%% file: ANA-HIGP-2024-08-PAPER.tex
\pdfoutput=1
\documentclass[PAPER, atlasdraft=false, UKenglish, cernpreprint,texlive=2025, texmf, orcidlogo]{atlasdoc}
\usepackage[full]{atlaspackage}
\usepackage{atlasbiblatex}

\usepackage{atlasphysics}
\graphicspath{{logos/}{figures/}}

\usepackage{ANA-HIGP-2024-08-PAPER-defs}

\newcolumntype{~}{>{\global\let\currentrowstyle\relax}}
\newcolumntype{^}{>{\currentrowstyle}}

\AtlasTitle{Measurement of the Higgs boson production in association with top quarks in multilepton final states in $pp$ collisions at  $\sqrt{s}=13$~TeV with the ATLAS detector}

\AtlasAbstract{%
A measurement of the associated production of a top-quark pair with
the Higgs boson ($t\bar{t}H$) in multilepton final states is presented.
The analysis is based on a data sample of proton--proton collisions at
$\sqrt{s}=13$~TeV recorded with the ATLAS detector at the CERN Large
Hadron Collider and corresponding to an integrated luminosity of 140
fb$^{-1}$.
Six final states defined by the number and flavour of reconstructed charged leptons are combined in
a simultaneous likelihood fit to extract the $t\bar{t}H$ signal and constrain the most relevant backgrounds.
The measured $t\bar{t}H$ cross-section normalised to Standard Model (SM) prediction is
$\sigma_{t\bar{t}H}/\sigma^{\text{SM}}=0.63^{+0.20}_{-0.19}$.
This result corresponds to an observed (expected) significance of
3.3$\sigma$ (5.3$\sigma$). Additionally, two other fits are used to measure the $t\bar{t}H$ cross-section
differentially in bins of the Higgs boson transverse momentum in the
simplified template cross-section framework, and to extract the associated production cross-section of a single top-quark with the Higgs boson ($tH$) together
with the $t\bar{t}H$ one.
The $CP$ structure of the top quark--Higgs boson Yukawa coupling is probed through an analysis
of $t\bar{t}H$ and $tH$ events.
The results are compatible with the SM hypothesis, and values of the mixing angle between \(CP\)-even and \(CP\)-odd top-Higgs Yukawa couplings of $| \alpha | > 62^\circ$ are excluded at 68\% confidence level.
}

\author{The ATLAS Collaboration}

\AtlasRefCode{HIGP-2024-08}

\PreprintIdNumber{CERN-EP-2025-239}

\AtlasDate{\today}

\arXivId{arXiv:2510.23755}

\AtlasJournalRef{\JHEP 05 (2026) 183}
\AtlasDOI{10.1007/JHEP05(2026)183}

\hypersetup{pdftitle={ATLAS document},pdfauthor={The ATLAS Collaboration}}

\begin{document}

\maketitle

\tableofcontents

\section{Introduction}
\label{sec:intro}
%


Since the discovery of the Higgs boson with a mass of approximately 125~\GeV by the ATLAS~\cite{HIGG-2012-27} and CMS~\cite{CMS-HIG-12-028} Collaborations in 2012,
the measurement of its properties has been one of the primary goals of the physics programme at the Large Hadron Collider
(LHC)~\cite{Evans:2008zzb}. So far no significant deviation from the Standard Model (SM) has been observed~\cite{HIGG-2021-23,CMS-HIG-23-013}.
The interaction between the Higgs boson and the top quark is of particular interest.
The large mass of the top quark of approximately 173~\GeV requires that it couples strongly to the Higgs boson in the SM.
For this reason, the top quark could play a special role in the electroweak symmetry breaking.
The measurement of the inclusive cross-section of the associated production of the Higgs boson with a top-quark pair (\ttH)
allows a more direct determination of the top-quark Yukawa coupling, relying on fewer theoretical assumptions than the loop-induced gluon--gluon fusion process.
The measurement of the Higgs boson production cross-section in association with a single top quark  ($tH$) and the kinematic properties of $tH$ and \ttH events can also provide information about
the $CP$ nature of the coupling~\cite{Gunion:1996xu,Ellis:2014hw,He:2015gu,Boudjema:2015nda}.
In the SM the Higgs boson is a scalar and its interactions are $CP$-even. The pure pseudoscalar hypothesis having $CP$-odd interactions with weak vector bosons and fermions has been excluded~\cite{HIGG-2013-17,CMS-HIG-14-018,HIGG-2019-10,CMS-HIG-20-007}.
However, it remains experimentally allowed that the Higgs boson is a $CP$-mixed state, which arises in extended Higgs sectors,
and would provide a new source of $CP$-violation beyond the SM, which is necessary to explain baryogenesis~\cite{Fontes:2015,Gunion:1996xu,Ellis:2014hw,He:2015gu,Boudjema:2015nda}.

The \ttH process was observed with a partial dataset of the LHC Run~2 by both ATLAS~\cite{HIGG-2018-13} and
CMS~\cite{CMS-HIG-17-035} using a combination of different channels.
Measurements of \ttH at $\sqrt{s}=13$~\TeV with the full Run-2 dataset, including $CP$ interpretations,
exist from ATLAS and CMS in the $H \to \gamma\gamma$ channel~\cite{HIGG-2019-01,HIGG-2020-16,CMS-HIG-19-015,CMS-HIG-19-013},
the $\hzzstar \to 4\ell$ channel~\cite{HIGG-2018-28,CMS-HIG-19-009}, the \htt channel~\cite{HIGG-2022-07},
and the $H \to b\bar{b}$ channel~\cite{HIGG-2020-03,HIGG-2020-24,CMS-HIG-19-011}.
In multilepton final states (with several light or $\tau$-leptons originating from both the decays of the Higgs boson and the top quarks), measurements of the inclusive \ttH production \xs have been performed
at $\sqrt{s}=13$~\tev\ by ATLAS~\cite{HIGG-2017-02} and CMS~\cite{CMS-HIG-19-008}
using datasets corresponding to a luminosity of 36~fb$^{-1}$ and 137~fb$^{-1}$, and reporting a signal strength\footnote{The signal strength is defined as the ratio of the measured cross-section to the SM cross-section.
Uncertainties on both the measurement and the prediction are included.} of $\mu_{\ttH} = 1.6^{+0.5}_{-0.4}$ and $\mu_{\ttH} = 0.92^{+0.26}_{-0.23}$ respectively.
A search for $CP$ violation in \ttH and \tH was also performed by CMS using the same dataset~\cite{CMS-HIG-21-006}.
The \tH process has been searched for by both ATLAS~\cite{HIGG-2020-16,HIGP-2024-03} and CMS~\cite{CMS-HIG-19-011,CMS-HIG-19-008} in the $H \to \gamma\gamma$,  $H \to b\bar{b}$
and multilepton channels with the full Run-2 dataset.
The measured signal strengths by ATLAS are $\mu_{\tH} = 3^{+4}_{-3}$ in $H \to \gamma\gamma$,  $\mu_{\tH} = 8.1\pm3.3$ combining the $H \to b\bar{b}$ and multilepton channels,
and $\mu_{\tH} = 5.7^{+4.1}_{-4.0}$ by CMS in the multilepton channel.
All those results are compatible with, but slightly above the SM prediction.

\begin{figure}[htb]
\centering
\subfloat[]{\includegraphics[height=3cm]{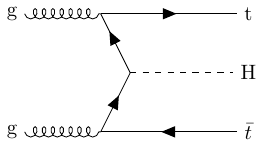}}\quad
\subfloat[]{\includegraphics[height=3cm]{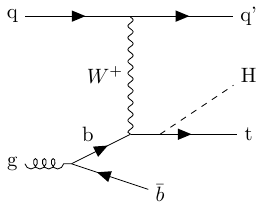}}\quad
\subfloat[]{\includegraphics[height=3cm]{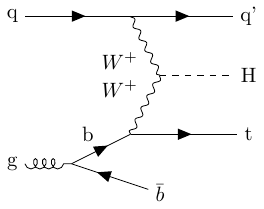}}
\caption{Representative Feynman diagrams for the signal processes considered in this analysis: (a) \ttH production and (b,c) $tHqb$ production.}
\label{fig:susyfeynman}
\end{figure}

This paper describes measurements of the inclusive \ttH production \xs, and of
the differential \ttH production cross-section as a function of the Higgs boson transverse momentum (\pTH),
as well as an interpretation of the $CP$ nature of the Higgs boson,
using a dataset corresponding to an integrated luminosity of 140~fb$^{-1}$ collected with
the ATLAS detector at a centre-of-mass energy $\sqrt{s}$=13~\TeV during the LHC Run~2 (2015--2018).
A simultaneous measurement of the \ttH and \tH signal strengths is also presented.
At the LHC, \tH production occurs via three modes: $tHqb$ ($t$-channel), $tWH$, and $s$-channel.
This analysis focuses on $tHqb$, which has a larger cross-section and is easier to distinguish from \ttH than $tWH$.
The $s$-channel has a very low cross-section and is neglected.
Examples of Feynman diagrams for both \ttH and $tHqb$ production are shown in Figure~\ref{fig:susyfeynman}.

In addition to the increased size of the analysed dataset, this measurement features significant improvements over
the previous publication based on 36~fb$^{-1}$~\cite{HIGG-2017-02}, benefitting from improved jet flavour tagging and lepton identification, a better understanding of the dominant
backgrounds \ttW and \ttZ from dedicated measurements~\cite{TOPQ-2019-30,TOPQ-2020-20}, a new simulation and a new systematic uncertainty model; besides,
an improved photon conversion rejection is used and the main backgrounds estimation relies on \textit{in situ} normalisation.

The differential \pTH measurement is done in the simplified template cross-section (STXS)~\cite{Andersen:2016qtm,Berger:2019wnu} formalism.
Since the leptons and missing transverse momentum in these final states can arise from either the Higgs boson or the top-quark decays, the Higgs system cannot be fully reconstructed unambiguously. To overcome this, a Graph Neural Network (GNN) is trained to estimate \pTH using the full event information.

\(CP\)-violation in the top--Higgs interaction is searched for using the same region definitions and background estimates,
but with a fit parameterised to accommodate for a possible admixture of $CP$-odd and $CP$-even Higgs states, affecting \ttH, $tHqb$, and $tWH$ cross-sections and event kinematics.
Indirect model-dependent constraints on the $CP$ structure of the top-quark Yukawa coupling can be inferred from the measurements of electron electrical dipole moments, Higgs boson
gluon--gluon fusion production and $H \to \gamma\gamma$ decays rates~\cite{Bahl:2020}.
The top--Higgs interaction Lagrangian term can be extended beyond the SM as~\cite{Demartin:2014fia}:
\begin{equation*}{\cal L}_{\ttH}=-\kprime y_t\phi\bar{\psi_t}(\cos{\alpha} + i\gamma_5\sin{\alpha})\psi_t
\end{equation*}
where $y_t$ is the SM Yukawa coupling, modified by a coupling modifier $\kprime$; $\alpha$ is the \(CP\)-mixing angle;
$\phi$ is the Higgs field; $\psi_t$ and $\bar{\psi_t}$ are top-quark spinor fields and $\gamma_5$ is a Dirac matrix.
The term containing $\gamma_5$ corresponds to a pseudoscalar component. The above expression reduces to the SM case for $\kprime = 1$ and $\alpha=0$.
Combining multilepton, $\hzzstar \to 4\ell$, and $H \to \gamma\gamma$ final states, CMS~\cite{CMS-HIG-21-006} constrains the $CP$-mixing angle to $|\alpha | < 48^\circ$ at the 68\% confidence level.
ATLAS results in the $H \to \gamma\gamma$ channel~\cite{HIGG-2019-01} excluded values of $|\alpha | > 43^\circ$ at the 95\% confidence level.

The measurements are performed by combining a total of six channels, categorised by their number of
hadronically decaying $\tau$-leptons (\tauhad) and light leptons ($\ell=e$ or $\mu$).
Requirements on the $\tau_{\textrm{had}}$ and $\ell$ multiplicities ensure that all channels are statistically independent.
These signatures are targeting \ttH events in which one or both top quarks decay leptonically,
and the Higgs boson decays to \hwwstar, \htt or \hzzstar with subsequent decays involving $\ell$ and \tauhad.
The event selection also accepts a large fraction of \tH events.
There are two zero-\tauhad channels (two-lepton same-sign, \twoLss,  and three-lepton, \threeL),  one single-\tauhad channel (\twoLssTau),  two two-\tauhad channels ( \oneLtwoTau and \twoLtwoTau) and one channel with no \tauhad requirement (\fourL).
In the zero-\tauhad channels the strategy builds upon the recent ATLAS \ttW
measurement~\cite{TOPQ-2019-30} to make use of the improved understanding of the non-prompt lepton background and \ttW modelling derived in that
analysis.

The paper is structured as follows.
The ATLAS detector is presented in Section~\ref{sec:detector}.
The recorded data and the Monte Carlo (MC) simulation samples used in the analysis are described in Section~\ref{sec:dataMC}.
The definition of reconstructed objects is described in Section~\ref{sec:objects}.
The event selection in each channel is given in Section~\ref{sec:evtselection}.
The background estimation method and the systematic uncertainties are described in Sections~\ref{sec:bkg} and~\ref{sec:syst}, respectively.
The results are presented in Section~\ref{sec:result}.


%
\section{ATLAS detector}
\label{sec:detector}
%

%
\newcommand{\AtlasCoordFootnote}{%
ATLAS uses a right-handed coordinate system with its origin at the nominal interaction point (IP)
in the centre of the detector and the \(z\)-axis along the beam pipe.
The \(x\)-axis points from the IP to the centre of the LHC ring,
and the \(y\)-axis points upwards.
Polar coordinates \((r,\phi)\) are used in the transverse plane,
\(\phi\) being the azimuthal angle around the \(z\)-axis.
The pseudorapidity is defined in terms of the polar angle \(\theta\) as \(\eta = -\ln \tan(\theta/2)\) and is equal to the rapidity
$ y = \frac{1}{2} \ln \left( \frac{E + p_z}{E - p_z} \right) $ in the relativistic limit.
Angular distance is measured in units of \(\Delta R \equiv \sqrt{(\Delta \eta)^{2} + (\Delta\phi)^{2}}\).}

The ATLAS detector~\cite{PERF-2007-01} at the LHC covers nearly the entire solid angle around the collision point.\footnote{\AtlasCoordFootnote}
It consists of an inner tracking detector surrounded by a thin superconducting solenoid, electromagnetic and hadronic calorimeters,
and a muon spectrometer incorporating three large superconducting air-core toroidal magnets.

The inner-detector system (ID) is immersed in a \qty{2}{\tesla} axial magnetic field
and provides charged-particle tracking in the range \(|\eta| < 2.5\).
The high-granularity silicon pixel detector covers the vertex region and typically provides four measurements per track,
the first hit generally being in the insertable B-layer (IBL) installed before Run~2~\cite{ATLAS-TDR-19,PIX-2018-001}.
It is followed by the SemiConductor Tracker (SCT), which usually provides eight measurements per track.
These silicon detectors are complemented by the transition radiation tracker (TRT),
which enables radially extended track reconstruction up to \(|\eta| = 2.0\).
The TRT also provides electron identification information
based on the fraction of hits (typically 30 in total) above a higher energy-deposit threshold corresponding to transition radiation.

The calorimeter system covers the pseudorapidity range \(|\eta| < 4.9\).
Within the region \(|\eta|< 3.2\), electromagnetic calorimetry is provided by barrel and
endcap high-granularity lead/liquid-argon (LAr) calorimeters,
with an additional thin LAr presampler covering \(|\eta| < 1.8\)
to correct for energy loss in material upstream of the calorimeters.
Hadronic calorimetry is provided by the steel/scintillator-tile calorimeter,
segmented into three barrel structures within \(|\eta| < 1.7\), and two copper/LAr hadronic endcap calorimeters.
The solid angle coverage is completed with forward copper/LAr and tungsten/LAr calorimeter modules
optimised for electromagnetic and hadronic energy measurements respectively.

The muon spectrometer (MS) comprises separate trigger and
high-precision tracking chambers measuring the deflection of muons in a magnetic field generated by the superconducting air-core toroidal magnets.
The field integral of the toroids ranges between \num{2.0} and \qty{6.0}{\tesla\metre}
across most of the detector.
Three layers of precision chambers, each consisting of layers of monitored drift tubes, cover the region \(|\eta| < 2.7\),
complemented by cathode-strip chambers in the forward region, where the background is highest.
The muon trigger system covers the range \(|\eta| < 2.4\) with resistive-plate chambers in the barrel, and thin-gap chambers in the endcap regions.

The luminosity is measured mainly by the LUCID--2~\cite{LUCID2} detector that records Cherenkov light produced in the quartz windows of photomultipliers located close to the beampipe.

Events are selected by the first-level trigger system implemented in custom hardware,
followed by selections made by algorithms implemented in software in the high-level trigger~\cite{TRIG-2016-01}.
The first-level trigger accepts events from the \qty{40}{\MHz} bunch crossings at a rate below \qty{100}{\kHz},
which the high-level trigger further reduces in order to record complete events to disk at about \qty{1.25}{\kHz}.

A software suite~\cite{SOFT-2022-02} is used in data simulation, in the reconstruction
and analysis of real and simulated data, in detector operations, and in the trigger and data acquisition
systems of the experiment.


%
\section{Data and simulated event samples}
\label{sec:dataMC}
%


The data were collected by the ATLAS detector in $pp$ collisions at the LHC with a centre-of-mass energy of $\sqrt{s} = \SI{13}{\TeV}$ with a peak instantaneous luminosity of ${\cal L}=2.1\times 10^{34}$~cm$^{-2}$s$^{-1}$, and a mean number of $pp$ interactions per bunch crossing of $\langle \mu \rangle = 34$.
Data quality requirements are applied to ensure that all relevant components were in nominal operating conditions and that the LHC beams were in stable-collision mode~\cite{DAPR-2018-01}.
The integrated luminosity of the resulting data sample corresponds to \SI{140}{\ifb} \cite{DAPR-2021-01}.
The uncertainty in the combined 2015--2018 integrated luminosity is 0.83\% \cite{DAPR-2021-01}, obtained using the LUCID-2 detector \cite{LUCID2} for the primary luminosity measurements
with complementary measurements from the ID and calorimeters.

Events are selected based on a combination of single-lepton\footnote{Unless otherwise stated, the term `lepton' will be used to refer to light leptons ($e$ or $\mu$).} and dilepton triggers~\cite{TRIG-2018-05,TRIG-2018-01}.
The use of a given trigger depends on the flavour and the transverse momenta
($\pt$) of the leptons in the event, and on the data-taking period.
Single-lepton triggers have $\pt$ thresholds between \SI{20}{\GeV} and \SI{26}{\GeV} while dilepton triggers have $\pt$ thresholds as
low as $12 (8)\,\si{GeV}$ for the leading (subleading) lepton.
The requirements imposed on the lepton \pT in the analysis are tighter than those applied in the trigger-object reconstruction and trigger decision, in order to be on the trigger efficiency plateau.

Simulated MC event samples are used to model the $\ttH$ and $\tH$ signals and to estimate the
backgrounds from SM processes.  A summary of the generators used for the simulation is shown in Table~\ref{tab:MCSamples}, and further details are provided below for the most relevant samples in the analysis.

\begin{table}[tbh]
\begin{center}
\caption{
The configurations used for event generation of signal and background processes.
The matrix element order refers to the order in the strong coupling constant of the perturbative calculation.
The `$\ttbar W$ (EWK)' sample also includes next-to-leading-order electroweak corrections.
Tune refers to the underlying-event tune of the parton shower generator, being either A14~\cite{ATL-PHYS-PUB-2014-021} or the default in \SHERPA .
All samples include leading-logarithm photon emission, either modelled by the parton shower generator or by \textsc{Photos}~\cite{Golonka:2005pn}.
The mass of the top quark ($m_t$) and SM Higgs boson were set to $172.5~\gev$ and $125~\gev$, respectively.
}
\label{tab:MCSamples}
\vspace{0.25cm}
\small
\setlength\tabcolsep{5pt}
\scalebox{0.8}{
\begin{tabular}{~l^l^l^l^l^l}
\hline\hline
Physics process & Event generator & Matrix element order & Parton shower & PDF set & Tune  \\
\hline
\ttH                   & \POWHEGBOX & NLO & \PYTHIA[8]\ & \NNPDF[3.0nlo] & A14 \\
\ttH ($CP$ interpretation) & \MGNLO & NLO & \PYTHIA[8] & \NNPDF[3.0nnlo]  & A14  \\
\tHqb   & \MGNLO & NLO & \PYTHIA[8] & \NNPDF[3.0nnlo]  & A14                                    \\
\tHW    & \MGNLO & NLO & \PYTHIA[8] & \NNPDF[3.0nnlo] & A14                                    \\ \hline
$\ttbar$               & \POWHEGBOX & NLO & \PYTHIA[8] & \NNPDF[3.0nlo] & A14  \\
$\ttbar W$             & \SHERPA[2.2.10] & \MEPSatNLO & \SHERPA &  \NNPDF[3.0nnlo] & \SHERPA default \\
$\ttbar W$ (EWK)        & \SHERPA[2.2.10] & LO & \SHERPA &  \NNPDF[3.0nnlo] & \SHERPA default \\ %

$\ttbar Z/\gamma^\ast$       & \MGNLO & \textsc{NLO} & \PYTHIA[8]  &  \NNPDF[3.0nlo] &  A14 \\
$\ttbar Z/\gamma^\ast$ (rad. decay) & \MGNLO & LO & \PYTHIA[8] & \NNPDF[3.0lo] & A14   \\
$t\bar t t\bar t$      &\MGNLO & NLO & \PYTHIA[8] & \NNPDF[3.1nlo] & A14 \\
$t\bar t t$            & \MGNLO & LO & \PYTHIA[8] & \NNPDF[2.3lo] \cite{Ball:2012cx} & A14 \\
$VV$, $qqVV$, $VVV$    & \SHERPA[2.2.2(1)]  & \MEPSatNLO & \SHERPA & \NNPDF[3.0nnlo] & \SHERPA default  \\
Single top             & \POWHEGBOX & NLO & \PYTHIA[8] & \NNPDF[3.0nlo] & A14   \\
($t$-, $Wt$-, $s$-channel) &  &  &                                     &  &  \\
$Z \to \ell^{+}\ell^{-}$         & \SHERPA[2.2.1] & \MEPSatNLO & \SHERPA & \NNPDF[3.0nnlo] & \SHERPA default \\ %
$Z \to \ell^{+}\ell^{-} (\gamma \to e^{+}e^{-})$ & \POWHEGBOX & NLO & \PYTHIA[8] & \CTEQ[6L1nlo] \cite{Pumplin:2002vw} & A14 \\
$Z \to \ell^{+}\ell^{-} (\gamma* \to e^{+}e^{-})$ & \POWHEGBOX & NLO & \PYTHIA[8] & \CTEQ[6L1nlo] & A14 \\
$t\bar{t}WW$ & \MGNLO & LO & \PYTHIA[8] & \NNPDF[2.3lo] & A14 \\
$\tZ$, $\tWZ$ & \MGNLO & NLO & \PYTHIA[8] & \NNPDF[3.0nlo] & A14 \\
\hline\hline
\end{tabular}
}
\end{center}
\end{table}

\subsection{Signal MC event samples}

The signal from $\ttH$ production was simulated using \POWHEGBOX[v2]~\cite{Nason:2009ai,Alioli:2010xd,Nason:2004rx,Frixione:2007vw}.
The \hdamp\ parameter, which controls the \pt\ of the first additional emission beyond the Born configuration and therefore regulates the high-\pt\ emission,  was set to $3 \times (2 m_t + m_H) / 4=352.5$~\GeV.
The functional form of the renormalisation (\muR) and factorisation (\muF) scales was set to $\sqrt[3]{m_\text{T}(t)\cdot m_\text{T}(\bar{t}) \cdot m_\text{T}(H)}$, where
$m_\text{T}=\sqrt{m^2+p_\text{T}^2}$ is the transverse mass of a generated particle, $m$ is its mass, and $p_\text{T}$ is its transverse momentum.
For this sample, the \NNPDF[3.0nlo] parton distribution function (PDF) set~\cite{Ball:2014uwa} was used.
The Monte Carlo prediction was normalised to cross-sections calculated at next-to-leading order (NLO) in quantum chromodynamics (QCD) and electroweak (EWK) couplings~\cite{deFlorian:2016spz,Beenakker:2002nc,Dawson:2003zu,Yu:2014cka,Frixione:2014qaa}.
The \tHqb\ (\tHW) samples were produced with \MGNLO[2.6] (2.8)~\cite{Alwall:2014hca} in the four-flavour (five-flavour) scheme with the \NNPDF[3.0nnlo] PDF.
The same flavour scheme was used in the matrix element (ME) calculation and the PDF.
The top-quark and $W$ boson decays were handled by \MADSPIN~\cite{Artoisenet:2012st} to account for spin correlations in the decay products.
The overlap of the \tHW\ process with \ttH\ at NLO was removed by using a diagram removal technique~\cite{Frixione:2008yi,Demartin:2016axk}.

To study \(CP\)-violation in the top--Higgs interaction, signal samples for \ttH, \tHqb and \tHW were also produced using \MGNLO[2.6] for different values of  $\kprime$ and $\alpha$, with all other parameters fixed to the SM predictions.
Following the procedure described in Ref.~\cite{HIGG-2020-03},  the yields of \ttH\ and \tH\ signals in each analysis bin are parameterised as a function of the model parameters by smoothly interpolating between generated MC samples.
This parameterisation makes use of two \ttH, ten \tHqb and ten \tHW samples generated with different values of $\alpha$ (between 0$^{\circ}$ and 180$^{\circ}$) and \kprime\ (between 0.5 and 2), in addition to the samples corresponding to the SM case.

\subsection{Background MC event samples}

The production of \ttbar events was modelled using the
\POWHEGBOX[v2]
generator at NLO with the \NNPDF[3.0nlo] PDF set
and the \hdamp parameter set to 1.5\,\mtop~\cite{ATL-PHYS-PUB-2016-020}.
The production of top-quark pairs in association with $W$ bosons (\ttW) is one of the largest backgrounds in the analysis.
Similarly to Ref.~\cite{TOPQ-2019-30},  simulated event samples for this process were produced using \Sherpa[2.2.10]~\cite{Bothmann:2019yzt}
with the \NNPDF[3.0nnlo] PDF set.
The ME was calculated for up to one additional parton at NLO and
up to two partons at leading order (LO) using \textsc{Comix}~\cite{Gleisberg:2008fv}
and \OPENLOOPS~\cite{Cascioli:2011va,Denner:2016kdg,Buccioni:2019sur}, and merged with the \Sherpa
parton shower~\cite{Schumann:2007mg} using the \MEPSatNLO
prescription~\cite{Hoeche:2012yf} with a merging scale of 30~\GeV.
The \muR and \muF were chosen to be $\muR = \muF =
\HT/2$, where \HT is defined as the scalar sum of the transverse masses $\sqrt{\pTsq+m^2}$ of all final-state
particles with transverse momentum \pT and invariant mass $m$.
Higher-order corrections related to EWK contributions were also included to account for corrections of orders $\alpha^{2}\alphas^{2}$ and $\alpha^3$~\cite{Kallweit:2015dum,Bothmann:2019yzt,Gutschow:2018tuk}.
Real emission contributions of the sub-leading EWK corrections at order $\alpha^{3}\alphas$~\cite{Frederix:2017wme} were also included.
The combination of contributions from NLO QCD and NLO EWK effects taken from this \Sherpa configuration closely follows the strategy described in Ref.~\cite{Frederix:2021agh}.
This \ttW event sample is normalised to a total cross-section of $\sigma(\ttW) = 745.3$~fb which also includes corrections at next-to-next-to-leading order (NNLO) in QCD as computed in Ref.~\cite{PhysRevLett.131.231901}.
The production of $t\bar{t}Z/\gamma^\ast$ events was simulated using
the \MGNLO[2.8.1] generator at NLO in \alphas with the \NNPDF[3.0nlo] PDF set.
The functional form of the renormalisation and factorisation scales (\muR, \muF) was set to the default scale $0.5 \times \sum_i \sqrt{m^2_i+p^{2}_{\textrm{T},i}}$, where the sum runs over all the particles generated from the matrix element calculation.
Top quarks were decayed at LO using \MADSPIN~\cite{Frixione:2007zp,Artoisenet:2012st} to preserve all spin correlations.

A dedicated \ttbar sample including rare $t \to Wb\gamma^\ast(\to \ell^+\ell^-)$ radiative decays, i.e.\ $t\bar{t} \to W^+bW^-\bar{b}\ell^+\ell^-$, was
generated using a LO matrix element and requiring $m(\ell^+\ell^-) > 1$~\GeV. In this
sample the photon can be radiated from the top quark, the $W$ boson,
or the $b$-quark. The $t\bar{t}Z/\gamma^\ast$ and $t\bar{t} \to W^+bW^-\bar{b}\ell^+\ell^-$ samples were combined and together form the
\enquote{$t\bar{t}Z/\gamma^\ast$} sample. The contribution
from internal photon conversions $(\gamma^\ast \to \ell^+\ell^-)$ with
$m(\ell^+\ell^-) < 1$~\GeV was modelled with QED multiphoton radiation via the
parton shower in an inclusive \ttbar sample.
These three components were combined to estimate the total $t\bar{t}Z/\gamma^\ast$ production.
Care was taken to avoid both double-counting of contributions and uncovered regions of phase
space when combining the different simulated event samples.  The LO cross-section for the $t\bar{t} \to W^+bW^-\bar{b}\ell^+\ell^-$ sample was scaled to
match the higher-order cross-section used for \ttbar production in the
phase space they overlap, and was assigned a 50\% normalisation
uncertainty.

The production of SM \tttt events was modelled using the \MGNLO[v2.6.2] generator that provides matrix elements at NLO in QCD with the \NNPDF[3.1nlo] PDF set.
The functional form of the renormalisation and factorisation scales was set to \(0.25\times \sum_i \sqrt{m^2_i+p^2_{\text{T},i}}\),
where the sum runs over all the particles generated from the matrix element calculation, following Ref.~\cite{Frederix:2017wme}.
Top quarks were decayed at LO using \MADSPIN to preserve all spin correlations.
The production of \tttt\  events is normalised to a cross-section of 13.37~fb computed at NLO in QCD including EWK corrections~\cite{Frederix:2017wme}.
Samples of diboson final states ($WW$, $WZ$ and $ZZ$, collectively referred to as \VV) were simulated with the
\SHERPA[2.2.1] or 2.2.2 generator depending on the process,
including off-shell effects and Higgs boson contributions, where appropriate.
Fully leptonic final states and semileptonic final states, where one boson
decays leptonically and the other hadronically, were generated using
matrix elements at NLO accuracy in QCD for up to one additional parton
and at LO accuracy for up to three additional parton
emissions. Samples for the loop-induced processes \(gg \to VV\) were
generated using LO-accurate matrix elements for up to one
additional parton emission for both the cases of fully leptonic and
semileptonic final states. The matrix element calculations were matched
and merged with the \SHERPA parton shower based on Catani--Seymour
dipole factorisation~\cite{Gleisberg:2008fv,Schumann:2007mg} using the MEPS@NLO
prescription~\cite{Hoeche:2012yf,Catani:2001cc,Hoeche:2009rj}.
The virtual QCD corrections were provided by the
\OPENLOOPS library. The
\NNPDF[3.0nnlo] set of PDFs was used, along with the
dedicated set of tuned parton-shower parameters developed by the
\SHERPA authors.

Additional MC samples are used to model processes with smaller impact in the analysis (such as single top, $VVV$, $t\bar t t$,  \ttWW,  \tZ, \tW, \tWZ or $Z \to \ell^{+}\ell^{-}$),  whose generator and settings are shown in Table~\ref{tab:MCSamples}.

All MC samples produced with \POWHEGBOX and \MGNLO are interfaced with \PYTHIA[8.2]~\cite{Sjostrand:2014zea} for parton shower, hadronisation, and underlying event, with parameters set according to the A14 tune~\cite{ATL-PHYS-PUB-2014-021}.
For all MC samples except those produced with \SHERPA, the decays of bottom and charm hadrons
were performed by \EVTGEN~\cite{Lange:2001uf}.

To simulate the effects of additional $pp$ collisions in the same, following or previous bunch crossings (pile-up), additional interactions were generated using the soft QCD processes provided by \PYTHIA[8.186]~\cite{Sjostrand:2007gs} with the A3 tune~\cite{ATL-PHYS-PUB-2016-017} and the MSTW2008LO PDF set~\cite{Martin:2009iq}, and overlaid onto each simulated hard-scatter event.
The MC samples were reweighted so that the pile-up distribution matches the one observed in the data.
All MC samples were processed through an ATLAS detector simulation~\cite{SOFT-2010-01} based on \GEANT~\cite{Agostinelli:2002hh}
or a fast simulation using a parameterisation of the calorimeter response and \GEANT for the other parts of the detector.  All MC samples were reconstructed in the same manner as the data.


%
\section{Object reconstruction and selection}
\label{sec:objects}
%


Interaction vertices from the $pp$ collisions are reconstructed from at least two tracks with $\pt$ larger than $500~\mev$ that are consistent with originating from the beam collision region in the $x$--$y$ plane. If more than one primary vertex candidate is found in the event, the candidate for which the associated tracks form the largest sum of squared $\pt$ is selected as the hard-scatter primary vertex~\cite{ATL-PHYS-PUB-2015-026}.

Electron candidates are reconstructed from energy clusters in the electromagnetic calorimeter matched to a track in the ID~\cite{EGAM-2021-01}. They are required to satisfy $\pt>10$~\GeV and $|\eta_{\mathrm{cluster}}|<2.47$, excluding the transition region between the endcap and barrel calorimeters ($1.37<|\eta_{\mathrm{cluster}}|<1.52$).
\enquote{\textit{Loose}} and \enquote{\textit{Tight}} electron identification working points (WPs) are used, based on a likelihood discriminant employing calorimeter and tracking variables that provide separation between electrons and jets, and between electrons and photons.

Muon candidates are reconstructed by performing a combined fit of the track information from the ID and MS~\cite{MUON-2018-03}. They are required to satisfy $\pt > 10~\gev$ and $|\eta| < 2.5$. \enquote{\textit{Loose}} and \enquote{\textit{Medium}} muon identification WPs are used. In the case of \enquote{\textit{Loose}} identification, muon candidates in the region $|\eta| < 0.1$, where muon spectrometer coverage is reduced, are also reconstructed from ID tracks matched to isolated energy deposits in the calorimeters consistent with the passage of a minimum-ionising particle.

Electron (muon) candidates are matched to the primary vertex by
requiring that the significance of their transverse impact parameter,
$d_0$,\footnote{The transverse impact parameter, $d_0$, is
defined in the $x$--$y$~plane as the distance of closest approach of the
track to the beamline.} satisfies $|d_0/\sigma(d_0)|<5\,(3)$, where
$\sigma(d_0)$ is the measured uncertainty in $d_0$, and by requiring
that their longitudinal impact parameter, $z_0$,\footnote{The longitudinal impact parameter, $z_0$, is defined as the distance in $z$ between the primary vertex and the point on the track used to evaluate $d_0$.}
satisfies
$|z_0 \sin\theta|<0.5$~mm.

Trigger leptons must be matched to a reconstructed lepton. The matched lepton transverse momentum must be at least 1~\GeV above the trigger threshold.

Electron and muon candidates arising from photon conversions, hadron decays,
or mis-reconstruction, are collectively referred to as non-prompt leptons.
To further suppress this source of background, lepton candidates are required to be isolated in the tracker and in the calorimeter. The track-based lepton isolation criterion is based on the quantity $I_R = \sum \pt^{\textrm{trk}}$, where the scalar sum includes all tracks (excluding the lepton candidate itself) compatible with the primary vertex, with $\pt^{\textrm{trk}}> 1$~\GeV, and within a cone of size $\Delta R=R_{\textrm{cut}}$ around the direction of the lepton.
The value of $R_{\textrm{cut}}$ is the smaller of 0.3 and $10~\gev/\pt^\ell$, where $\pt^\ell$ is the lepton $\pt$. All lepton candidates must satisfy $I_R/\pt^\ell < 0.15$.
They are also required to satisfy a calorimeter-based isolation criterion: the sum of the transverse energy within a cone of size $\Delta R=0.2$ around the lepton, after subtracting contributions from \pileup and the energy deposit of the lepton itself, is required to be less than 20\% (30\%) of the electron (muon) $\pt^\ell$.

The \enquote{\textit{Loose}} electron and muon identification and the impact parameter and isolation criteria described above constitute the \enquote{loose} lepton ($L$) definition.
It is used (together with the $\tauhad$ definition described later) to pre-select events and categorise them into the different channels, as described in Section~\ref{sec:evtselection},
ensuring that the channel event selections are mutually exclusive.
The \enquote{\textit{Tight}} electron and \enquote{\textit{Medium}} muon identification WPs and additional lepton criteria, described below, are used to reject further non-prompt leptons
and electrons with an incorrect charge assignment or coming from a photon conversion. In total five different lepton definitions are used in this analysis:
loose inclusive ($L$, and $L'$, further described below), medium inclusive ($M$), medium exclusive ($M_{\mathrm{ex}}$), and tight ($T$).
The various lepton definitions are summarised in Table~\ref{tbl:tightleps} and their use in the different channels is described in Sections~\ref{sec:evtselection} and~\ref{sec:bkg}.

Non-prompt leptons from hadron decays that contain bottom- or charm-quarks (referred to as \enquote{heavy-flavour (HF) non-prompt leptons}) are further rejected using a boosted decision tree (BDT) discriminant, referred to as the non-prompt-lepton BDT~\cite{HIGG-2017-02,TOPQ-2019-30}, based on isolation and lifetime information about a track-jet that matches the selected lepton.  %
Three WPs based on the non-prompt-lepton BDT are used: \textit{Tight}, \textit{VeryTight}, and \textit{Tight--not-VeryTight}.
The \textit{Tight} WP allows prompt muons (barrel/endcap electrons) satisfying the calorimeter- and track-based isolation criteria to be selected with an efficiency that is about 60\% ($60\%/70\%$) for $\pt \sim 20~\gev$ and reaches a plateau of 95\% ($95\%/90\%$) for $\pt \sim 40\ (40/65)~\gev$.
The prompt-lepton efficiency of the \textit{VeryTight} WP for muons (barrel/endcap electrons) that satisfy the calorimeter- and track-based isolation criteria is about 55\% ($55\%/60\%$) for $\pt \sim 20~\gev$ and reaches a plateau of 90\% ($85\%/83\%$) for $\pt \sim 40\ (40/65)~\gev$. The corresponding rejection factor\footnote{The rejection factor for a background is defined as the reciprocal of the efficiency evaluated for that background.} for muons (electrons) from the decay of $b$-hadrons ranges from 33 to 50 (20 to 50) for the \textit{Tight} WP, and from 50 to 100 (33 to 66) for the \textit{VeryTight} WP, depending on \pt and $\eta$, after resolving ambiguities between overlapping reconstructed objects.
The \textit{Tight--not-VeryTight} WP selects leptons that pass the \textit{Tight} WP but not the \textit{VeryTight} one; it enhances the selection of non-prompt leptons and is part of the event selection for control regions (CRs) enriched in HF non-prompt-lepton background, as described in Section~\ref{sec:bkg}.

In order to further suppress electrons with incorrect charge assignment, a BDT discriminant based on calorimeter and tracking quantities~\cite{EGAM-2018-01} is used.
Rejection factors of 19 (40) in the barrel (endcap) region are obtained in $\ttbar$ events,  for an efficiency of approximately 96\% (81\%).

Most electrons arising from material conversions, i.e.\ from photon conversions in the detector material, are rejected by the standard electron identification selection, but additional requirements are imposed to remove residual material-conversion candidates.
These candidates have a reconstructed displaced vertex with radius $r > 20$~mm that includes the track associated with the electron.\footnote{The beampipe and IBL inner radii are 23.5~mm and 33~mm, respectively.} The invariant mass of the associated track and the closest (in $\Delta\eta$) opposite-charge track reconstructed in the silicon detector, calculated at the conversion vertex, is required to be less than $100~\mev$. Internal conversion candidates, which correspond to low-mass virtual photon conversions (see Section~\ref{sec:dataMC}), must fail the requirements for material conversions, and the di-track invariant mass, calculated here at the primary vertex, is also required to be less than $100~\mev$.
Material and internal conversion candidates thus reconstructed are vetoed except in the $L$ definition (no veto applied), and in the electron selection of the CR enriched in conversions described in Section~\ref{sec:bkg}
(the conversion candidate $e^*$ passes the $M$ selection except that the conversion veto is reversed).

A looser lepton definition, denoted $L'$, which omits the non-prompt-lepton BDT requirement, while retaining the tighter identification, charge mis-assignment, and conversion vetos, is used to increase the acceptance of backgrounds in corresponding control regions.

\begin{table}
\begin{center}
\caption{\label{tbl:tightleps} Description of the loose inclusive ($L$ and $L'$), medium inclusive ($M$), medium exclusive ($M_{\mathrm{ex}}$), and tight ($T$) lepton definitions. The electron $e^{*}$ is required to fulfil, in addition to the corresponding lepton definition requirements, those corresponding to an internal or material conversion candidate.
The symbol `--' denotes that no requirement is applied for a given lepton definition.}
\resizebox{\textwidth}{!}{
\begin{tabular}{l||c|c|c|c|c||c|c|c|c|c}
\hline\hline
& \multicolumn{5}{c||}{Electron} & \multicolumn{5}{c}{Muon} \\
\hline
Lepton definition                  & $L$ & $L'$ & $M$ & $M_{\mathrm{ex}}$ & $T$ & $L$ & $L'$ & $M$ & $M_{\mathrm{ex}}$ & $T$ \\
\hline
Identification                     & \textit{Loose} & \multicolumn{4}{c||}{\textit{Tight}} & \textit{Loose} & \multicolumn{4}{c}{\textit{Medium}} \\
\hline
Transverse impact parameter        &  \multicolumn{5}{c||}{$<5$} & \multicolumn{5}{c}{$<3$ } \\
significance $|d_0|/\sigma_{d_0}$  & \multicolumn{5}{c||}{} &  \multicolumn{5}{c}{}  \\
\hline
Longitudinal impact parameter $z_0$     &  \multicolumn{10}{c}{$|z_0 \sin \theta|< 0.5$~mm}  \\
\hline
Isolation                          &  \multicolumn{5}{c||}{Yes} & \multicolumn{5}{c}{Yes} \\
\hline
Non-prompt lepton WP               &  \multicolumn{2}{c|}{--} & \textit{Tight} & \textit{Tight--not-}  & \textit{VeryTight} &  \multicolumn{2}{c|}{--} & \textit{Tight} & \textit{Tight--not-}  & \textit{VeryTight}  \\
&  \multicolumn{2}{c|}{}   &  & \textit{VeryTight} & & \multicolumn{2}{c|}{} &  & \textit{VeryTight} &  \\
\hline
Electron charge-misassignment veto &  -- & \multicolumn{4}{c||}{Yes} & \multicolumn{5}{c}{--} \\
Electron conversion candidate veto &  -- & \multicolumn{4}{c||}{Yes (except $e^{*}$)} & \multicolumn{5}{c}{--} \\
\hline\hline
\end{tabular}
}
\end{center}
\end{table}

The constituents for jet reconstruction are identified by combining measurements from both the ID and the calorimeter using a particle flow (PFlow) algorithm~\cite{PERF-2015-09}. Jet candidates are reconstructed from these PFlow objects using the \antikt~algorithm~\cite{Cacciari:2008gp, Fastjet} with a radius parameter of $R=0.4$. They are calibrated using simulation with corrections obtained by using \textit{in situ} techniques in data~\cite{JETM-2018-05}. Jet candidates with $\pT> 25$~\GeV\ and within $|\eta|<2.5$ are used to define the event selection. In order to reduce the effect of \pileup, each jet with $\pT<60$~\GeV\ and $|\eta|<2.4$ must satisfy the \enquote{\textit{Tight}} WP of the jet-vertex tagger (JVT)~\cite{PERF-2014-03} criteria used to identify jets that originate from the primary vertex.
Forward jets with $\pT> 25$~\GeV\ and $2.5<|\eta|<4.5$ are not used in the event selection but in some kinematic variables. They must satisfy the \enquote{\textit{Tight}} WP of the forward jet-vertex-tagger (fJVT)~\cite{ATL-PHYS-PUB-2019-026}.
A set of quality criteria is also applied to reject events containing at least one jet arising from non-collision sources or detector noise~\cite{ATLAS-CONF-2015-029}.

Jets containing $b$-hadrons are identified ($b$-tagged) via the DL1r algorithm~\cite{FTAG-2019-07} that uses a deep-learning neural network based on the distinctive features of $b$-hadron decays, primarily the impact parameters of tracks and the displaced vertices reconstructed in the ID. Additional input to this network is provided by discriminating variables constructed by a recurrent neutral network (RNN)~\cite{FTAG-2019-07}, which exploits the spatial and kinematic correlations between tracks originating from the same $b$-hadron. A multivariate $b$-tagging discriminant value is calculated for each jet.
In this article, three different WPs corresponding to 85\%, 77\%, and 70\% average expected efficiency to tag a $b$-quark jet are used, with light-jet\footnote{`Light-jet' refers to a jet originating from the hadronisation of a light quark ($u$, $d$, $s$) or a gluon.} rejection factors of about 40 to 600, and charm-jet ($c$-jet) rejection factors of about 3 to 12, as determined for jets with $\pt >20~\gev$ and $|\eta|<2.5$ in simulated $\ttbar$ events~\cite{FTAG-2018-01,FTAG-2020-08,FTAG-2019-02}.
Only the 85\% and 77\% WPs are used for event selection, while the 70\% WP is used for the overlap removal described below.
The notations \bveryloose, \bloose, and \bmedium are used to denote a $b$-tagged jet (\bjet) that passed the corresponding WP.
Correction factors derived from dedicated calibration samples enriched in \bjet{}s, $c$-tagged jets, or light-tagged jets, are applied to the simulated event samples.

Hadronic $\tau$-lepton decays produce a neutrino and visible decay products, mostly one or three charged pions and up to two neutral pions.
The reconstruction of the visible products of hadronically decaying $\tau$-leptons (\tauhadvis) is seeded by jets reconstructed using the \antikt~algorithm~\cite{Cacciari:2008gp,  Fastjet} where topological clusters of calorimeter energy deposits are used as inputs with a distance parameter of $R=0.4$.  BDTs are used to determine if tracks originate from a \tauhadvis, and one (1-prong) or three (3-prong) tracks with a total charge of $\pm 1$ are required.
An RNN algorithm ~\cite{ATL-PHYS-PUB-2019-033} using calorimeter and tracking-based variables is employed to identify \tauhadvis candidates and reject quark or gluon jets.  The chosen \enquote{\textit{Medium}} identification WP has an efficiency of 75\% (60\%) for 1-(3-) prong \tauhad decays. A separate BDT is then used to reject \tauhadvis candidates with one associated track that originate from electrons,  with a rejection factor of $\sim30-100$ depending on $\eta$ and $\pT$ while reaching an efficiency of about 95\% for real \tauhadvis.
The \tauhadvis energy scale is determined by combining information from the associated tracks, calorimeter energy clusters and reconstructed neutral pions~\cite{PERF-2014-06} using a multivariate regression technique~\cite{ATLAS-CONF-2017-029} trained in MC samples.
The \tauhadvis candidates with  $\pT>20$~\GeV and $|\eta|<2.5$,  excluding the electromagnetic calorimeter transition region, are selected.

Ambiguities between independently reconstructed electrons, muons, \tauhadvis and jets can arise. A sequential \enquote{overlap removal} procedure is performed to resolve these ambiguities and thus avoid double counting of detector signatures.
This procedure is applied to leptons satisfying the $L$ criteria.
If two electrons share the same track, only the one with the higher \pt\ is kept.
A \tauhadvis candidate within $\Delta R<0.2$ of an electron or a muon is removed.
If an electron and a muon share the same ID track, the muon is removed if it is reconstructed only from an ID track and calorimeter energy deposits consistent with a minimum-ionising particle (i.e.\ if it is \enquote{calo-tagged}), otherwise the electron is removed.
If an electron and a selected jet are found within $\Delta R < 0.2$, the jet is removed if it is not a \bmedium jet or if it has $\pt > 200~\gev$.
Muons ghost-associated~\cite{Cacciari:2008gn} with an $R=0.4$ jet must satisfy a jet--muon separation of $\Delta R < 0.4$.
If the overlapping jet is not a \bmedium jet and contains less than three tracks with $\pt > 500~\mev$, the jet is removed, otherwise the muon is removed.
If a \tauhadvis and a jet are found within $\Delta R < 0.2$ the object is interpreted as a \tauhadvis and the matched jet candidate is removed.
For each lepton in the event a lepton-\pt-dependent variable-size cone of maximum size $\Delta R=0.4$ is defined. If a selected jet, surviving all previous overlap criteria, is found in this cone, the lepton is removed.

The missing transverse momentum $\vec{p}_{\mathrm{T}}^{\mathrm{miss}}$ (with magnitude $\met$) is defined as the negative vector sum of the $\vec{p}_{\mathrm{T}}$ of all selected and calibrated objects in the event, including a term to account for the momenta of soft particles that are not associated with any of the selected objects~\cite{JETM-2020-03}.
This soft term is calculated from ID tracks matched to the primary vertex, which makes it more resilient to contamination from \pileup interactions.


%
\section{Event selection and categorisation}
\label{sec:evtselection}
%

\begin{figure}[tbh]
\centering
\subfloat[]{\includegraphics[height=5cm]{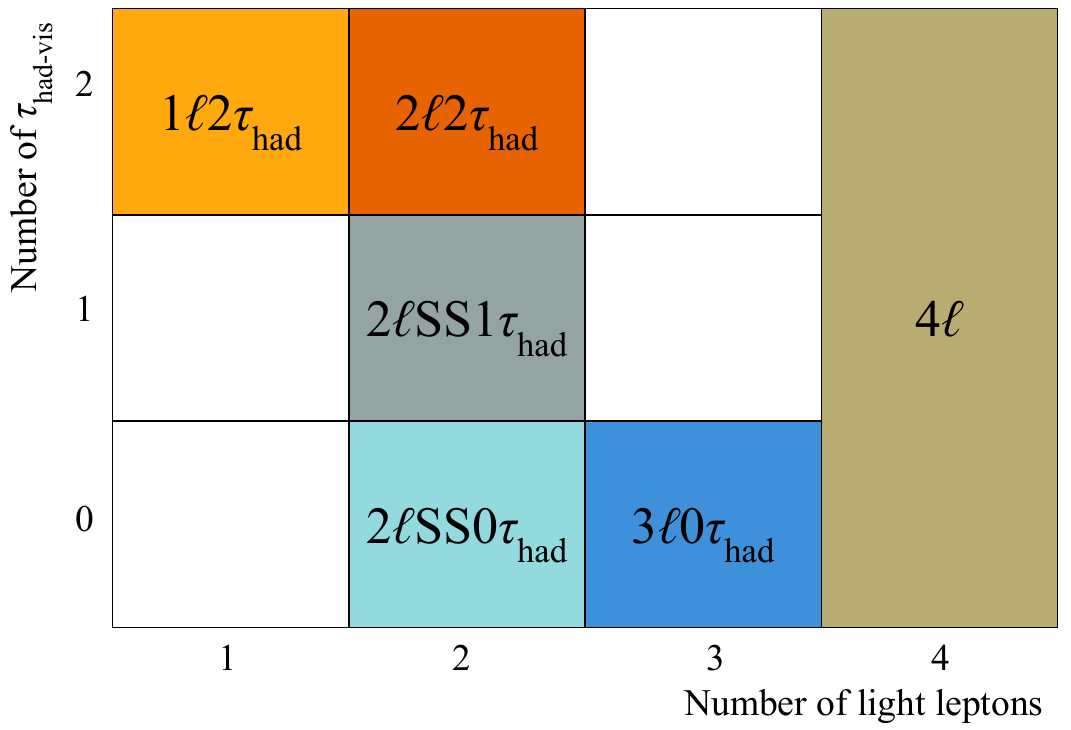}}
\subfloat[]{\includegraphics[height=5.5cm]{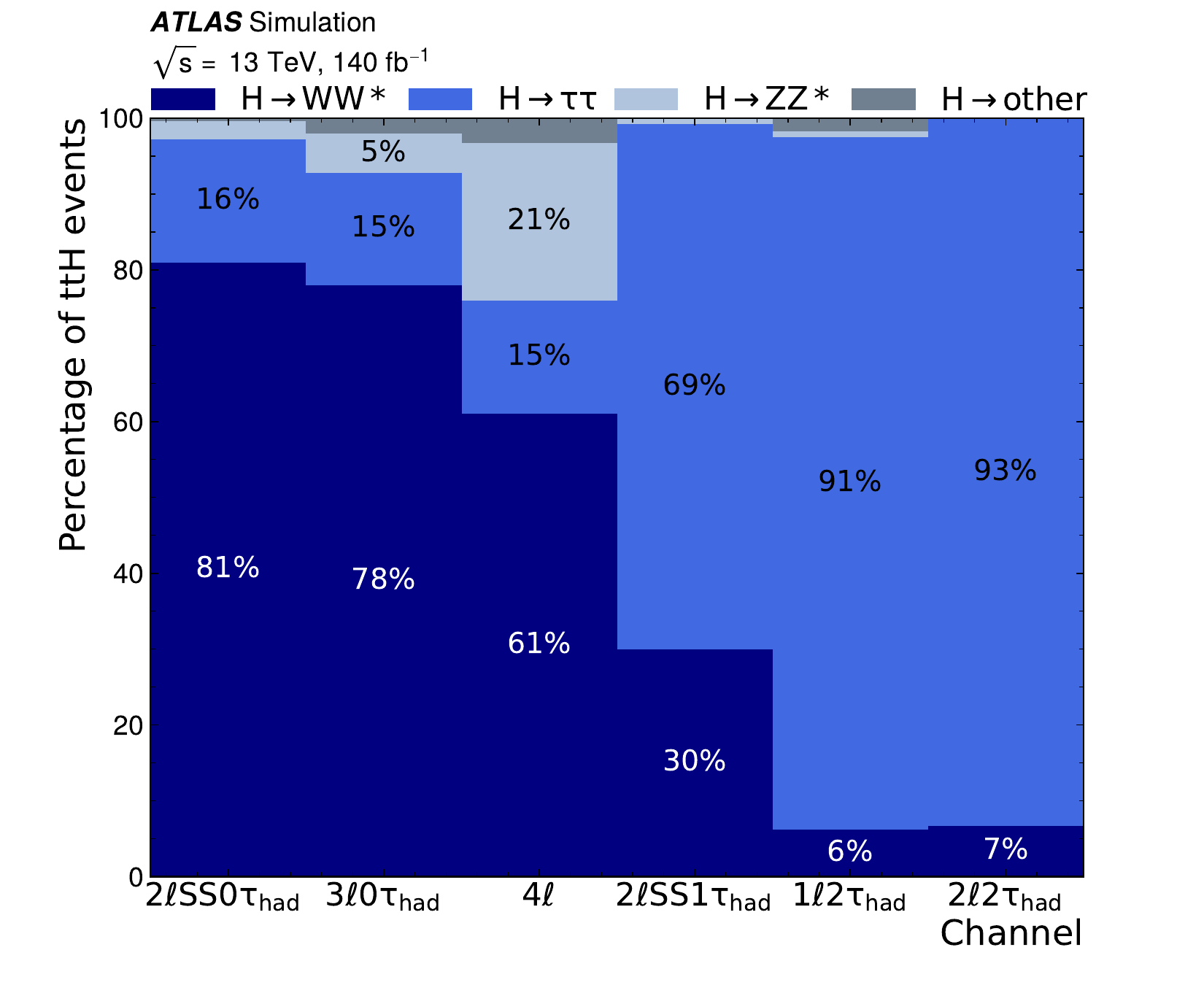}}
\caption{(a) Analysis channel definition in terms of light lepton and \tauhadvis multiplicities. (b) Fraction of the expected $\ttH$ signal arising from different Higgs boson decay modes in each signal channel. }
\label{fig:ttH_Channels}
\end{figure}

The analysis strategy is based on six different channels, corresponding to final states differentiated by the number and charge of loose light leptons of each flavour, and number of \tauhadvis candidates.
The considered channels are displayed in Figure~\ref{fig:ttH_Channels}(a) and are:
\begin{itemize}
\item \twoLss: two same-charge light leptons and no \tauhadvis candidates;
\item \threeL: three  light leptons and no \tauhadvis candidates;
\item \fourL: four  light leptons regardless of the number of \tauhadvis candidates;
\item \twoLssTau: two same-charge light leptons and one \tauhadvis candidate;
\item \oneLtwoTau: one light lepton and two \tauhadvis candidates;
\item \twoLtwoTau: two light leptons and two \tauhadvis candidates.
\end{itemize}

\begin{figure}[!phtb]
\centering
\subfloat[\twoLss]{\includegraphics[width=0.37\textwidth]{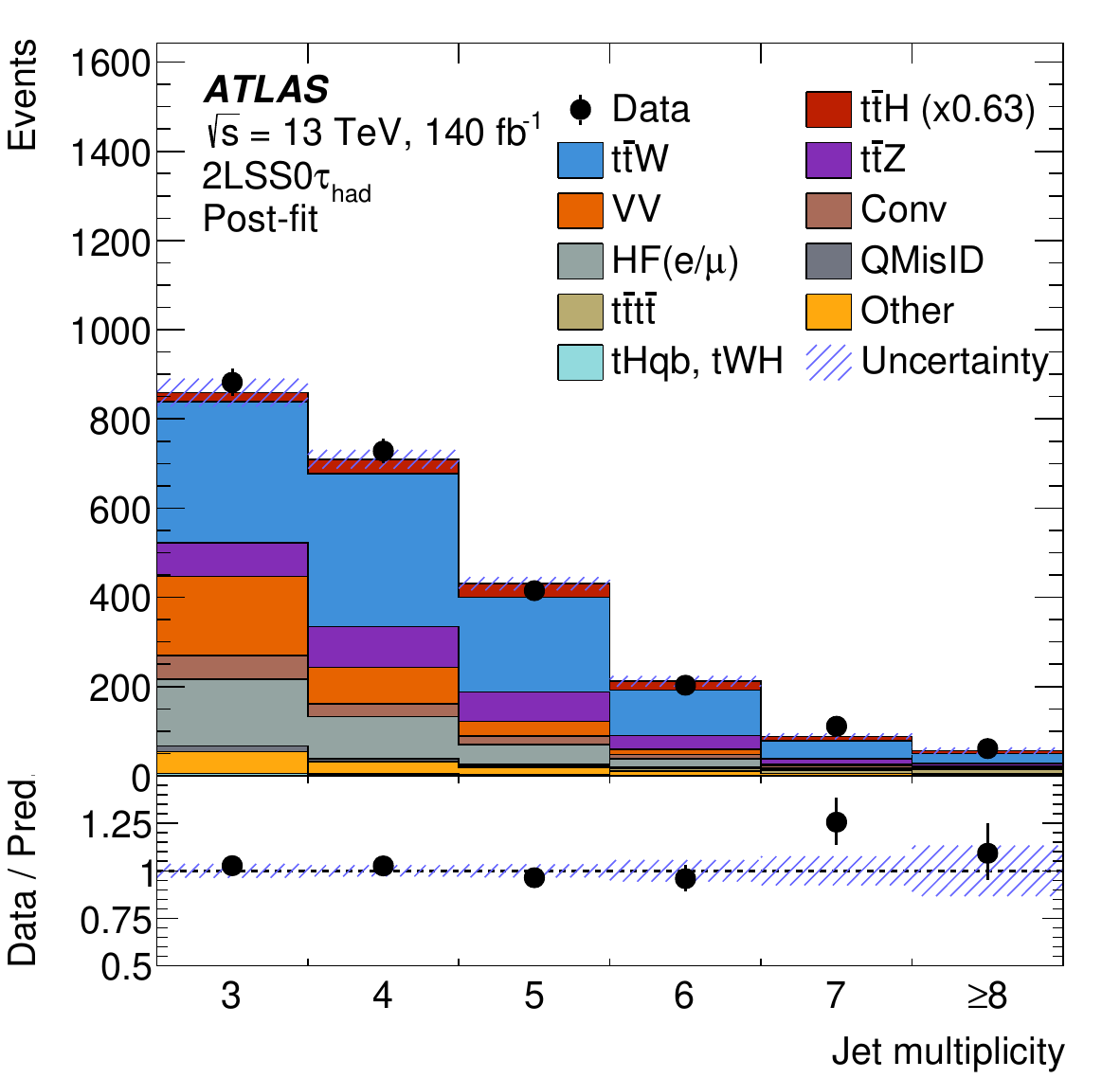}}
\subfloat[\threeL]{\includegraphics[width=0.37\textwidth]{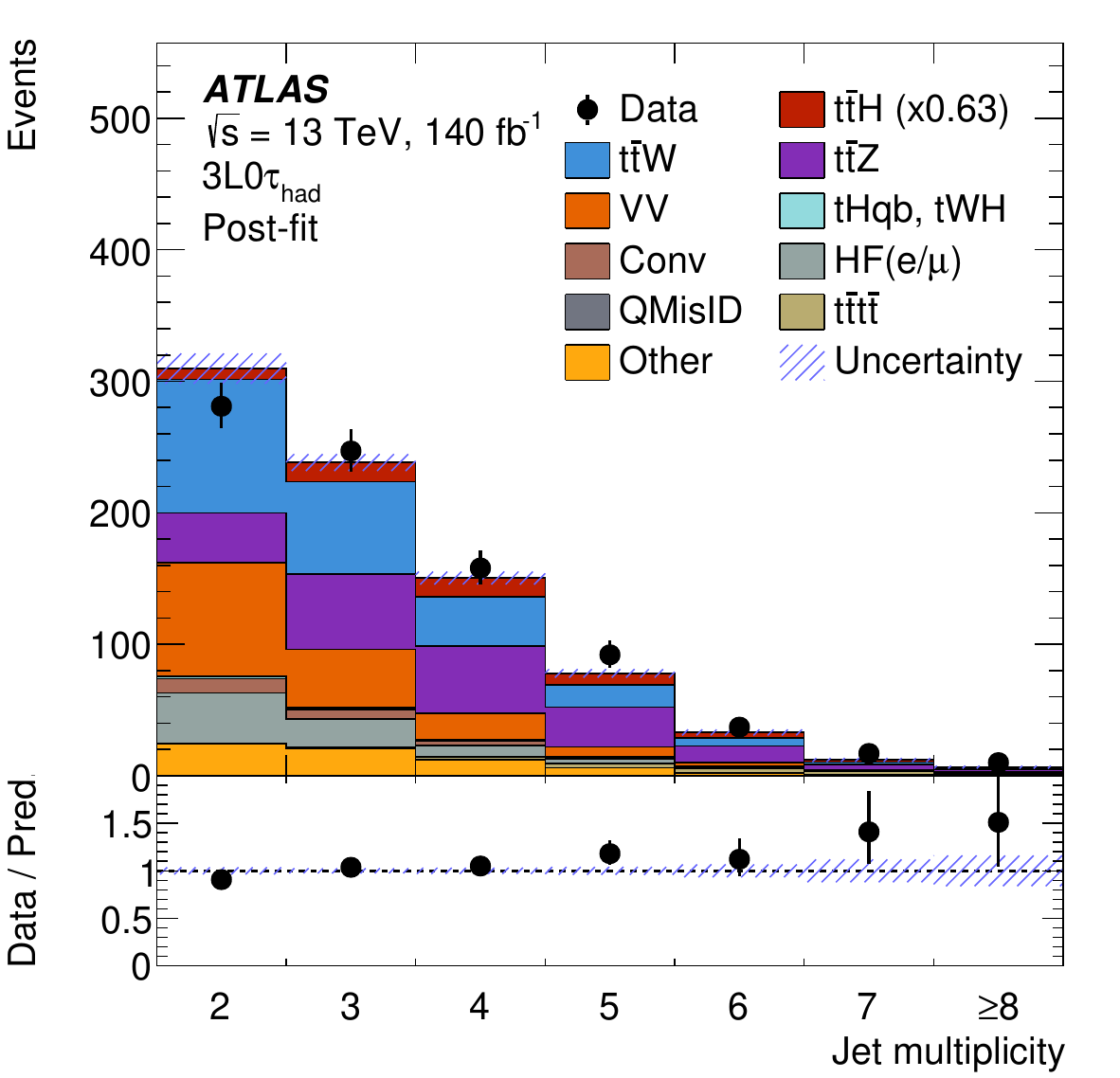}}\\
\subfloat[\fourL]{\includegraphics[width=0.37\textwidth]{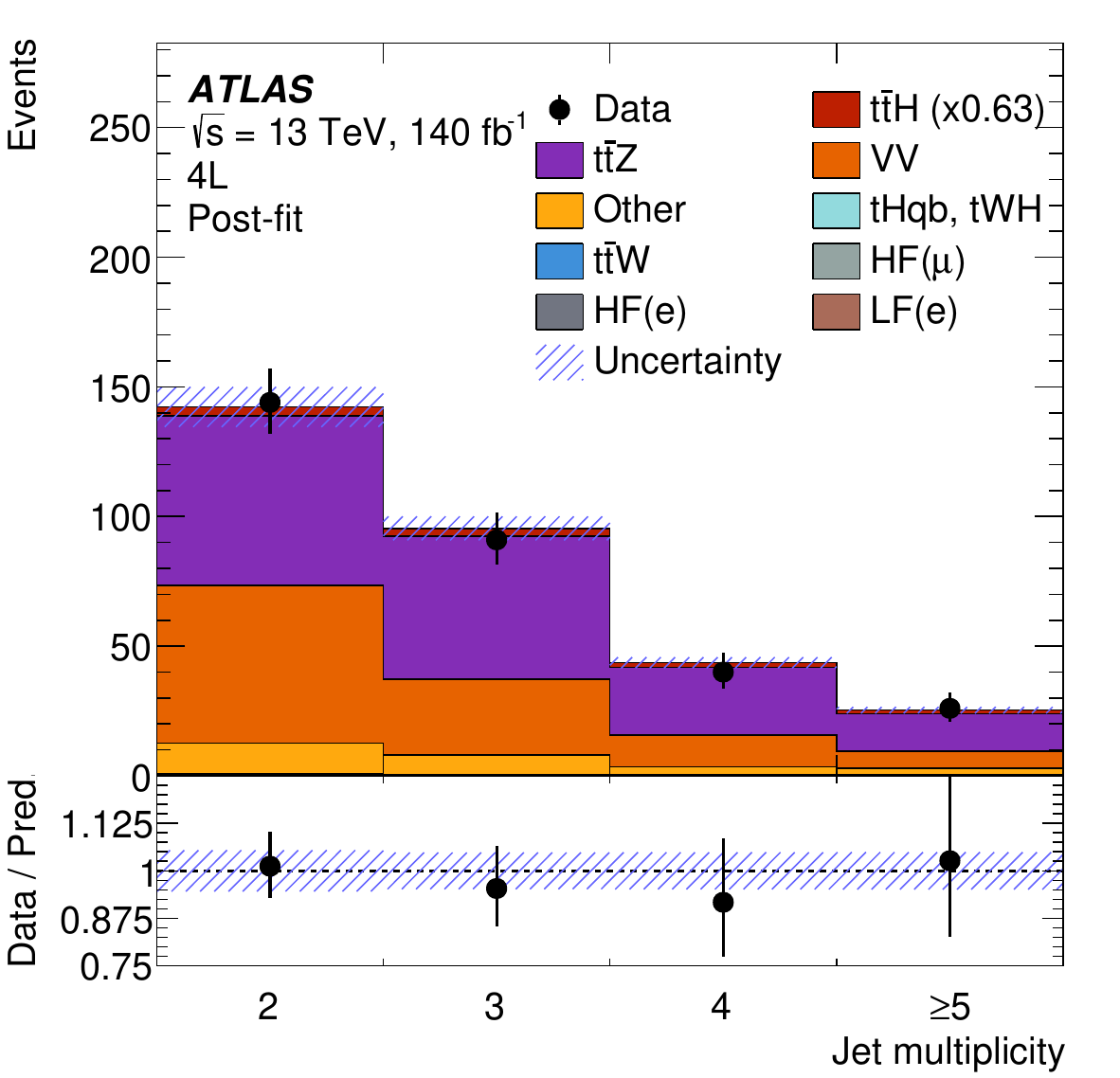}}
\subfloat[\twoLssTau]{\includegraphics[width=0.37\textwidth]{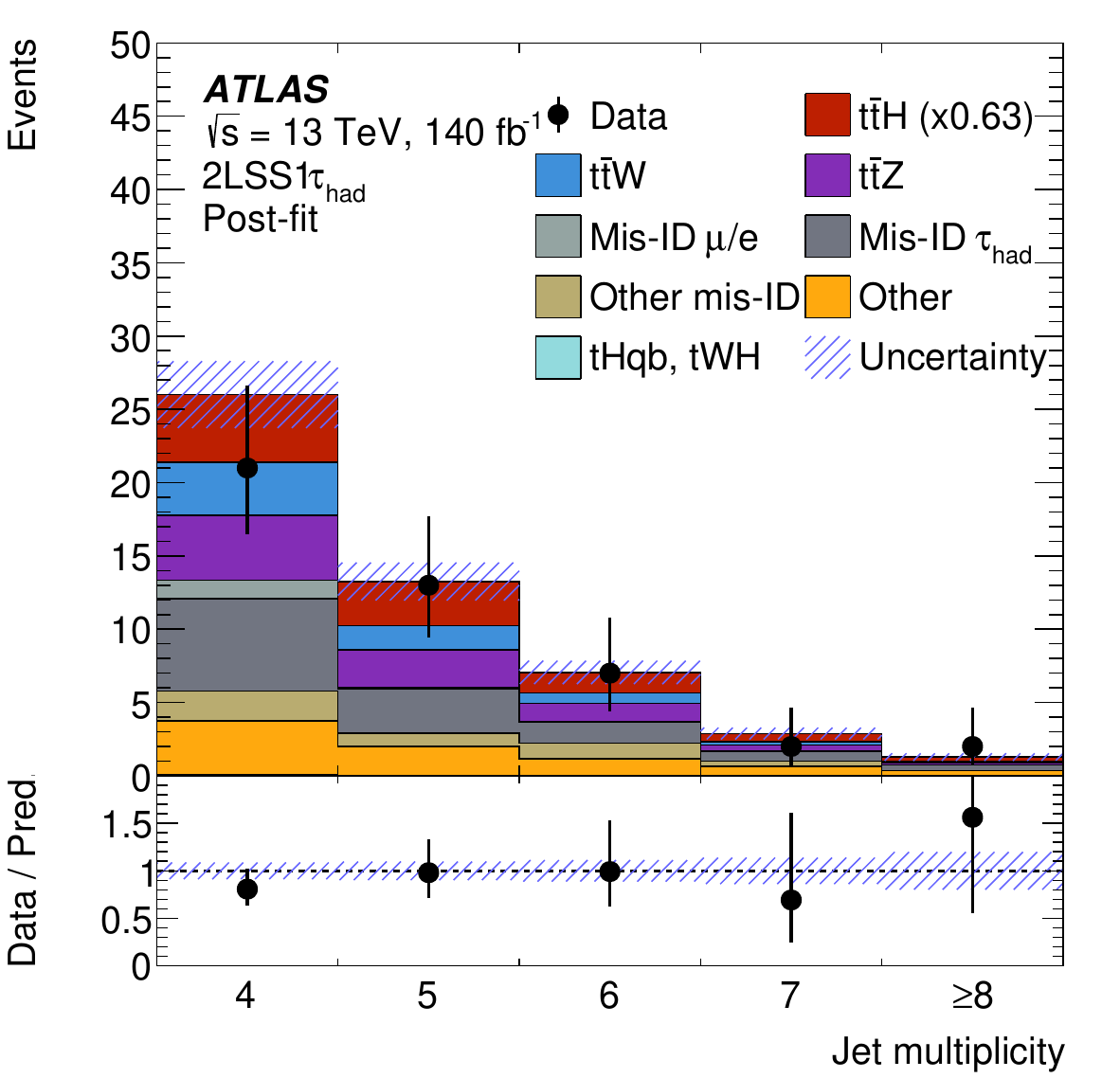}}\\
\subfloat[\oneLtwoTau]{\includegraphics[width=0.37\textwidth]{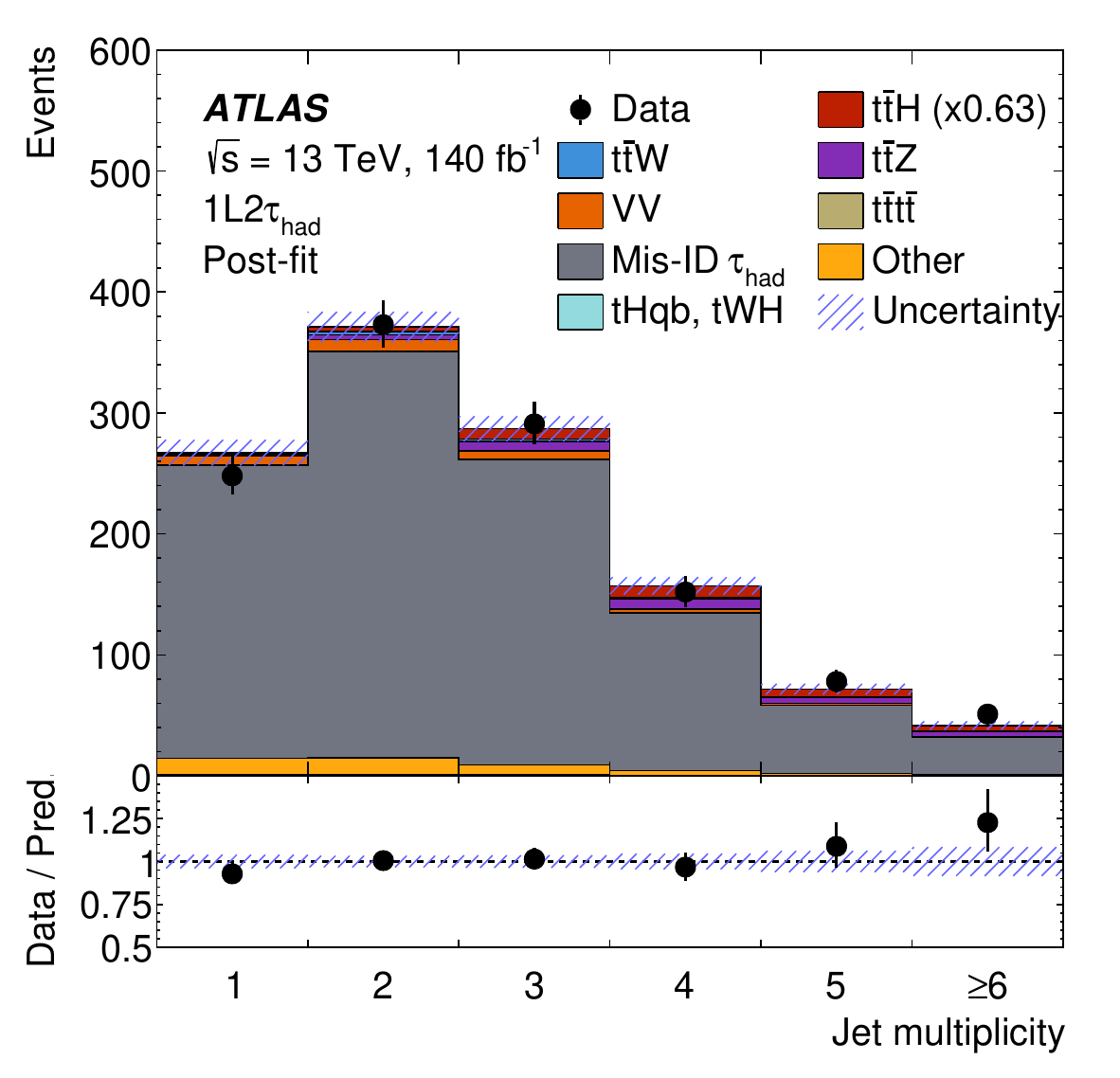}}
\subfloat[\twoLtwoTau]{\includegraphics[width=0.37\textwidth]{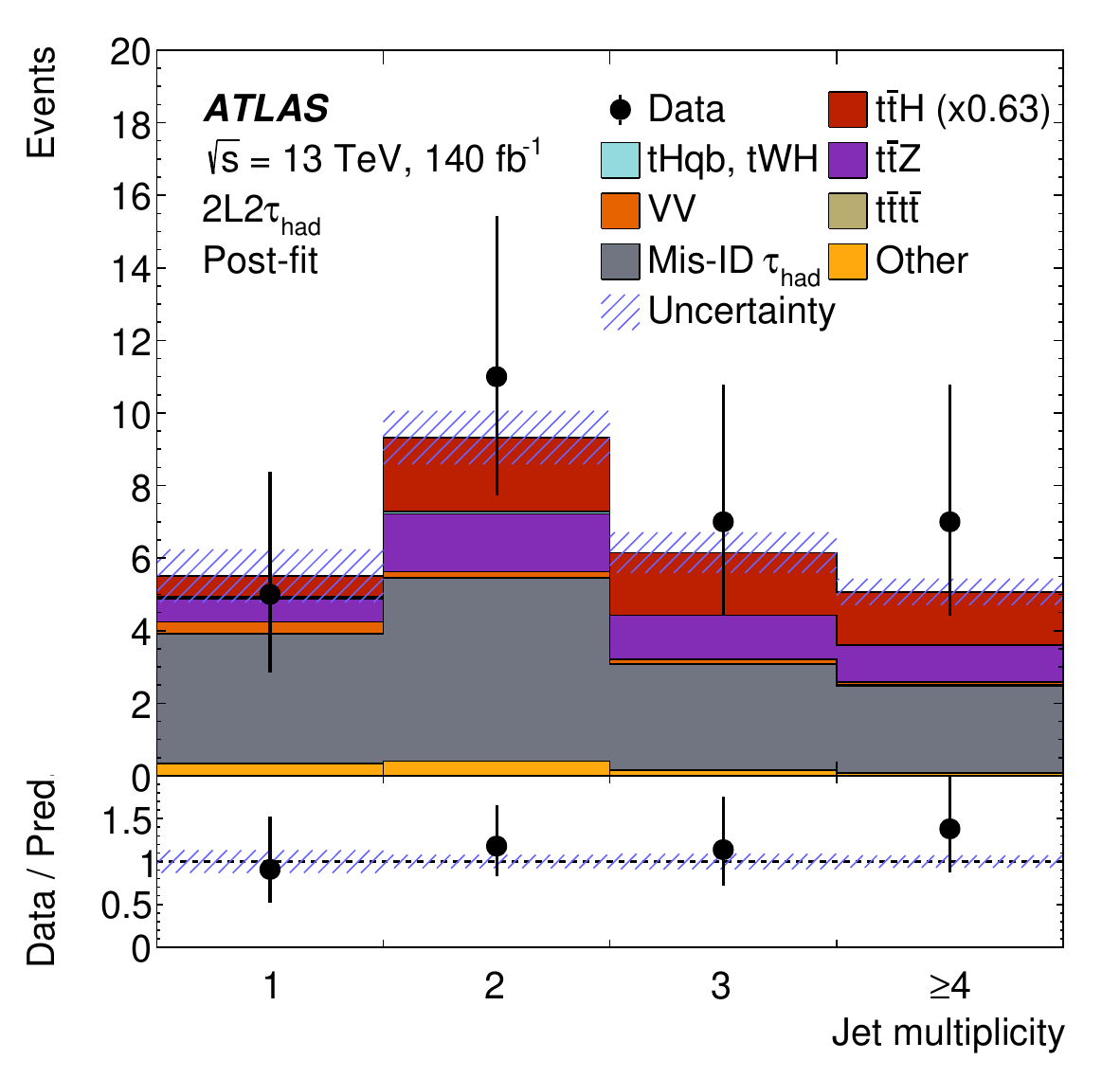}} \\
\caption{Comparison between data and the signal-plus-background prediction for the distribution of jet multiplicity in the (a) $\twoLss$, (b) $\threeL$, (c) $\fourL$, (d)  $\twoLssTau$, (e) $\oneLtwoTau$,  and (f) $\twoLtwoTau$ channels.
The signal and background normalisation factors and the nuisance parameters are set to their best-fit values as described in Section~\ref{sec:result}.
The shaded band indicates the total uncertainty, including correlations.
}
\label{fig:Njet_channels}
\end{figure}

These lepton selection criteria ensure orthogonality so that no event is classified in multiple channels.
To increase the signal sensitivity,  further requirements on the lepton,  \tauhadvis and jet properties are imposed in each channel, as shown in Table~\ref{tab:selection_channels} and detailed below.
Figure~\ref{fig:ttH_Channels}(b) shows the expected contribution in each channel from the different Higgs boson decay modes in the $\ttH$ signal.
The light leptons and \tauhadvis, as well as the jets in the event, can originate either from the Higgs boson decay products ($\tau$-lepton, $W$ or $Z$ bosons) or from the $W$ bosons in top-quark decays.
In addition,  multiple neutrinos can be produced in these decays together with the light leptons and \tauhadvis, thus an attempt to fully reconstruct the Higgs boson or top quarks is difficult.
Instead, the properties of the objects (such as multiplicity,  $\pT$, angular separation or invariant mass) in the event are exploited to separate the signal from the main backgrounds.
The jet multiplicity distribution for each channel is shown in Figure~\ref{fig:Njet_channels}, illustrating the background composition and signal contribution before any categorisation is applied.
The background processes are described in Section~\ref{sec:bkg} and, as detailed there,  some of the event selection requirements in Table~\ref{tab:selection_channels} are loosened or inverted to define background control regions.

Multivariate analysis (MVA) algorithms,  either BDTs or deep neural networks (DNNs), are used to further separate the signal from the relevant backgrounds.
The modelling of the input variables used in the training of the MVAs was studied and no significant discrepancies between data and simulation were observed.
The output scores from these BDTs or DNNs are used to define separate event categories in the  \twoLss, \threeL and \fourL\ channels,  and are the final discriminants in most categories.
The signal-enriched categories are the signal regions (SRs) in the analysis, while the categories enriched in the different background processes are used as CRs for their estimation, as detailed in Section~\ref{sec:bkg}.

\begin{figure}[!htbp]
\centering
\includegraphics[width=\linewidth]{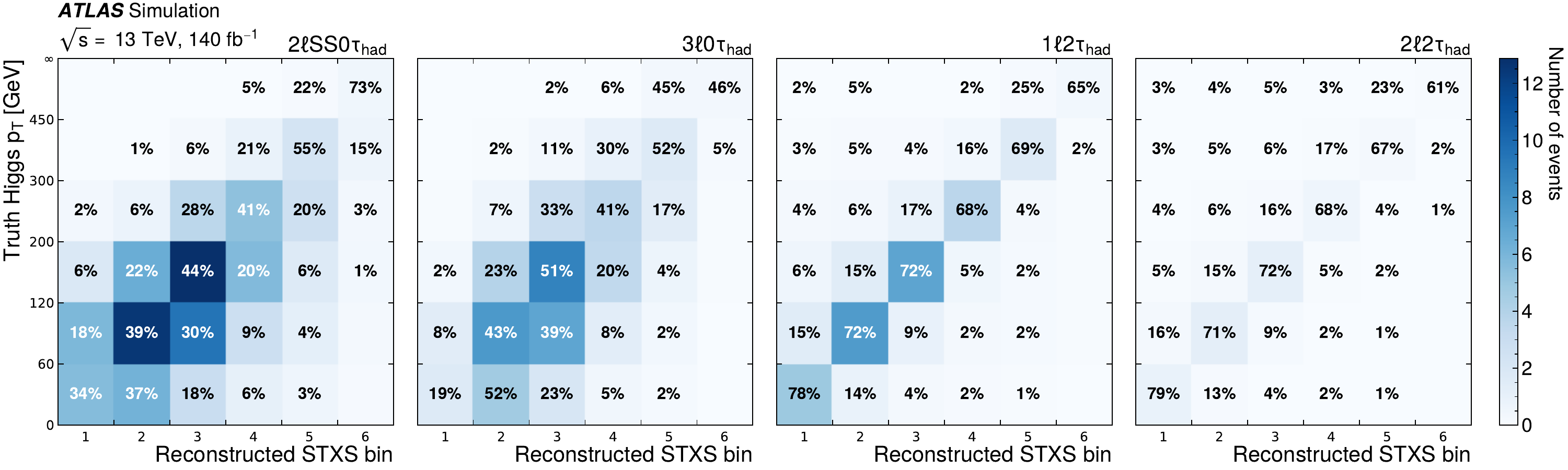}
\caption{
STXS migration matrices evaluated in the \ttH MC sample for the analysis channels featuring a \pTH estimation.
Colours denote the overall fraction of events in each STXS bin,  while numbers display the predicted yield fractions normalised to each truth STXS bin.
}
\label{fig:pTH_STXS}
\end{figure}

In the STXS stage-1.2 formalism, $\ttH$ production is measured in bins of the generator-level \pTH prior to Higgs-boson decay.
These bins are denoted as STXS 1--6 and correspond to \pTH values in the range 0--60~\GeV, 60--120~\GeV, 120--200~\GeV, 200--300~\GeV, 300--450~\GeV and $>$450~\GeV.
In the \twoLss and \threeL channels,  a GNN is trained to estimate \pTH and is used to further categorise the events according to the \ttH STXS regions. The input features for this GNN include: the four-momenta  of leptons,  jets, and \met;  the jet $b$-tagging information; the angular separation, invariant mass and \pT for all combinations of leptons,  jets, and \met; the jet multiplicity,  and the scalar sum of the \pT of all combinations of leptons and jets.   In the \oneLtwoTau and \twoLtwoTau channels,  \pTH is estimated using a BDT that takes as input the \pT of the two \tauhadvis in the event as well as their angular separation and invariant mass. Figure~\ref{fig:pTH_STXS} shows the migration across STXS bins for the different channels estimated in the \ttH MC sample.
No \pTH reconstruction is performed in the \fourL\ and  \twoLssTau\ channels.
Due to significant migration effects and limited statistics, some reconstructed  \pTH bins are merged depending on the channel:
in the \twoLss, \threeL, and \oneLtwoTau channels, bins 5 and 6 are merged together; in the \twoLtwoTau channels, bins 1 and 2 on one hand, and bins 3, 4, 5, and 6 on the other hand, are merged together.
As described in Section~\ref{sec:result}, some truth \pTH bins are also merged, such that the measurement is done with the following three bins: 0--120~\GeV, 120--200~\GeV, and $>$200~\GeV.

\begin{table}[tbhp]
\caption{
Summary of the channel definitions, with the requirements placed on the number of light leptons ($N_{\ell}$),  $\tauhadvis$ ($N_{\tauhadvis}$) and jets ($N_{\text{jets}}$), as well as the lepton \pT, invariant masses and total charge of the light leptons ($\sum q_{\ell}$) or $\tauhadvis$ ($\sum q_{\tauhadvis}$).
Some of the dilepton invariant mass requirements are only applied for pairs of same-flavour (SF) or SF opposite-sign (SFOS) leptons.
The requirements on `lepton definition' are summarised in Table~\ref{tbl:tightleps}.
The symbol `--' denotes that no requirement is applied for a given variable and channel.
}
\label{tab:selection_channels}
\begin{center}
\scalebox{0.9}{
\begin{tabular}{r|c|c|c|c|c|c}
\toprule
& \twoLss & \threeL & \fourL & \twoLssTau  & \oneLtwoTau &  \twoLtwoTau\\ \hline\hline
\multicolumn{7}{c}{Light leptons}\\ \hline\hline
$N_{\ell}$              & 2   & 3      & 4  & 2   & 1 & 2  \\

Lepton definition  &  $T$ & $\ell_0,\ell_1,\ell_2$: $L$,$T$,$T$ & $L$
&  $M$ & $L$  & $L$  \\
Lepton $\pT$ [\GeV]  &  $> 15$ & $\ell_0,\ell_1,\ell_2$: $>$ 10, 15, 15 & $> 10$
&  $> 15$ & $> 27$  & $> 10$  \\
$\sum q_{\ell}$     &  $\pm$2   & $\pm$1  & 0
&  $\pm$2  & $\pm$1 &  0 \\
$|m_{\ell\ell}-m_Z|$ [\GeV] &  --  & $>10$ (SFOS) & --
& $>10$ (SF) & -- & $>10$\\
$m_{\ell\ell}$ [\GeV] &  --  & $>12$ (SFOS) & $>12$ (SFOS)
& -- & -- & $>12$\\
$m(4\ell)$ [\GeV] & -- & -- & <115 or >130 & -- & -- & -- \\\hline\hline
\multicolumn{7}{c}{\tauhadvis and jets}\\ \hline\hline
$N_{\tauhadvis}$      & 0  & 0   & --  & 1 & 2 & 2 \\
$\sum q_{\tauhadvis}$     &  --   & --  & --
&  $\pm$1  & 0 &  0 \\ \hline
$N_{\text{jets}}$     &  $\geq$ 3 & $\geq$ 2 & $\geq$ 2 &  $\geq$ 4 & $\geq$ 1 & $\geq$ 1 \\\hline
$N_{\bveryloose}$   & $\geq$ 1 & $\geq$ 1 & $\geq$ 1& $\geq$ 1& -- & -- \\
$N_{\bloose}$          &  -- & --  & -- & -- & $\geq$ 1& $\geq$ 1 \\ \hline\hline
Number of categories & 14 & 11 & 2 & 1 & 6 & 2 \\
\bottomrule
\end{tabular}
}
\end{center}
\end{table}


\subsection{\twoLss channel}

Events in this channel must contain two same-charge light leptons with $\pT>15$~\GeV satisfying the $T$ requirements in Table~\ref{tbl:tightleps}.
To reject backgrounds like diboson or $\ttW$ production, selected events contain at least three jets with at least one of them being $b$-tagged with the 85\% $b$-tagging WP and no \tauhadvis candidates.
Categories in this channel are defined using a multi-class BDT considering separate classes for the $\ttH$ and $tHqb$ signal processes and the main $\ttW$ background.
A fourth class is included in the training containing the contributions from all other background processes, such as $\ttZ$, $\ttbar$ and $\tttt$.
This BDT uses 20 input variables describing the kinematics of the leptons, jets and $\met$.
The most discriminating variables\footnote{In the MVA algorithms in this paper,  the most discriminating variables are determined using the permutation importance method~\cite{breiman2001random,altmann2010permutation}} are  the jet multiplicity,  the invariant mass of the dilepton pair, and the average angular separation between the jets in the event.

\begin{figure}[!htb]
\centering
\includegraphics[height=9cm]{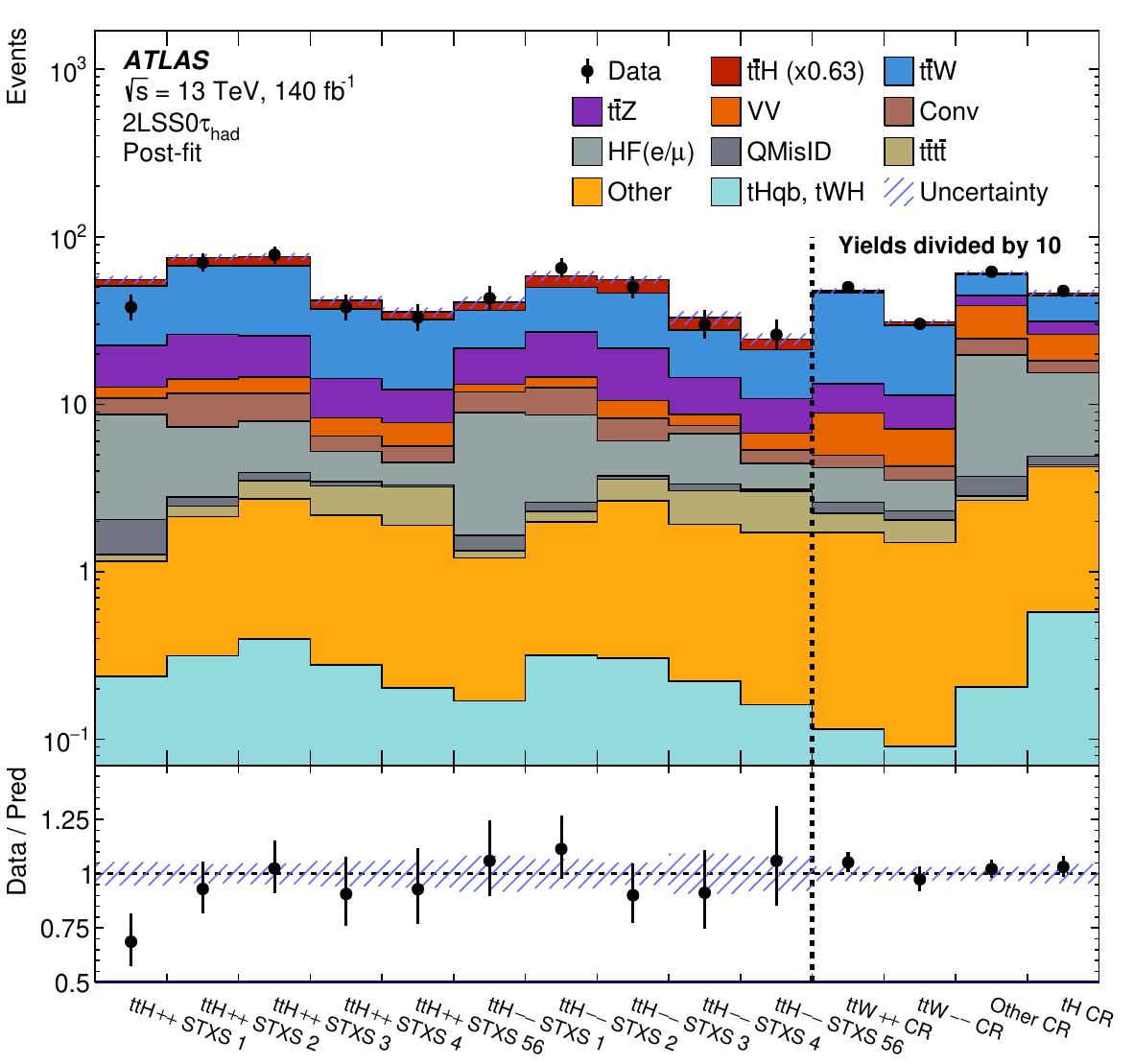}
\caption{Observed and expected event yields in all the categories in the \twoLss channel.
The signal and background normalisation factors and the nuisance parameters are set to their best-fit values as described in Section~\ref{sec:result}.
The categories denoted as `STXS $n$' correspond to the $n$-th bin in the STXS formalism described in Section~\ref{sec:evtselection}.  When several STXS bins are merged together,  all numbers are quoted (e.g. `STXS 56' corresponds to STXS bins 5 and 6).
The notation `++' or `$- -$' correspond to the total charge of the leptons in the category.
The shaded band indicates the total uncertainty, including correlations.
The vertical line indicates the bins in which the event yields have been scaled down by a factor 10.
}
\label{fig:2LSS_cat}
\end{figure}

Events are classified into categories according to the highest BDT output score and,  in the case of $\ttH$ and $\ttW$,  the total charge of the leptons (denoted as `++' or `$- -$') to exploit the relative charge asymmetry between $\ttW^+$ and $\ttW^-$ production.
To reduce the contamination from $\tttt$ production,  events with at least six jets and with at least three $b$-tagged jets at the 77\% working point are discarded from the $\ttH$ categories.
Furthermore, the $\ttH$ categories are split into five categories each according to the reconstructed $\pTH$ values.
In total,  14 categories are considered in this channel, whose observed and expected yields are shown in Figure~\ref{fig:2LSS_cat}.

\subsection{\threeL channel}

Events in this channel must contain three light leptons with total a charge of $\pm 1$.  The lepton with charge opposite to the other two is denoted as $\ell_0$, while the two leptons in the event with the same electric charge (denoted $\ell_1$ and $\ell_2$, with $\ell_1$ being the closest to $\ell_0$ in $\Delta R$) are required to fulfil the $T$ requirements in Table~\ref{tbl:tightleps} and have $\pT>15$~\GeV.
Additional requirements on the invariant mass of same-flavour opposite-sign (SFOS) dilepton pairs are applied to suppress $Z$-boson and other resonant backgrounds, and events also should have at least two jets with at least one of them being $b$-tagged with the 85\% $b$-tagging WP to reject the diboson, $\ttW$ and $\ttZ$ backgrounds.

As in the previous channel, event categories are defined using a multi-class BDT with six output nodes for the $\ttH$ and $tHqb$ signals as well as the $\ttW$, $\ttZ$, diboson and $\ttbar$ backgrounds.
A total of 25 input variables related to the multiplicities and properties of leptons, jets, $b$-jets and $\met$ are employed in the BDT training.
The most discriminating variables include the jet multiplicity,  the invariant masses of the $\ell_1\ell_2$ and $\ell_1\ell_0$ systems, and the average angular separation between leptons and jets.

\begin{figure}[!htb]
\centering
\includegraphics[height=9cm]{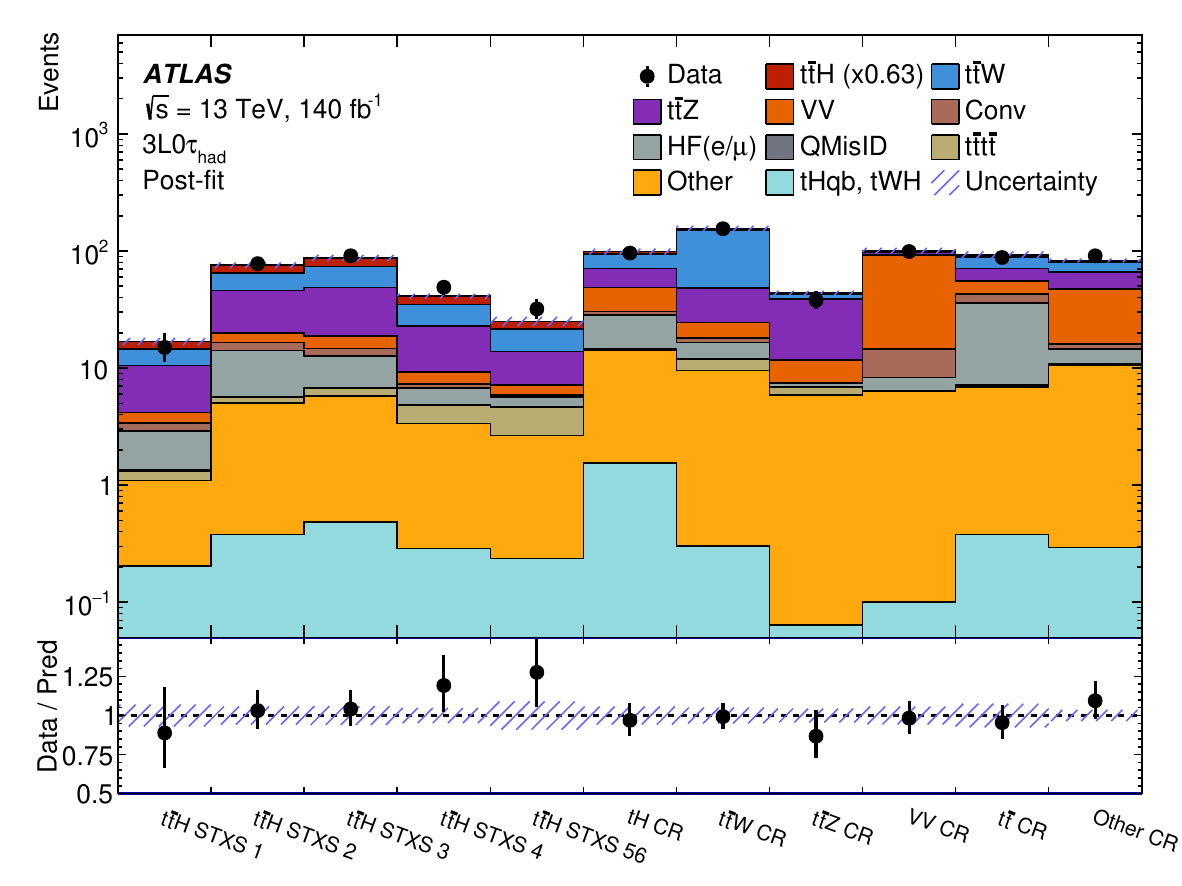}
\caption{Observed and expected event yields in all the categories in the \threeL channel.
The signal and background normalisation factors and the nuisance parameters are set to their best-fit values as described in Section~\ref{sec:result}.
The categories denoted as `STXS $n$' correspond to the $n$-th bin in the STXS formalism described in Section~\ref{sec:evtselection}.  When several STXS bins are merged together,  all numbers are quoted (e.g. `STXS 56' corresponds to STXS bins 5 and 6).
The shaded band indicates the total uncertainty, including correlations.
}
\label{fig:3L_cat}
\end{figure}

Seven categories, each dedicated to enhancing a given process, are built by applying
selections in the output scores to maximise its statistical significance,
with the remaining events included in a separate category.
The $\ttH$ category is further split into five categories according to the reconstructed $\pTH$ values.
In total,  11 categories are considered in the \threeL channel whose observed and expected yields are shown in Figure~\ref{fig:3L_cat}.

\subsection{\fourL\ channel}

Events in this channel are required to have four light leptons with total charge equal to zero, regardless of the number of \tauhadvis candidates.
Requirements are imposed in the invariant mass of any SFOS pair to reduce the contribution from low-mass resonances,
as well as a veto on the four-lepton invariant mass close to the Higgs boson mass to ensure orthogonality with the $H\to ZZ^*\to 4\ell$ analysis~\cite{HIGG-2018-28}.
Events are also required to have at least two jets with at least one of them being $b$-tagged with the 85\% $b$-tagging WP.

\begin{figure}[!htb]
\centering
\includegraphics[height=7cm]{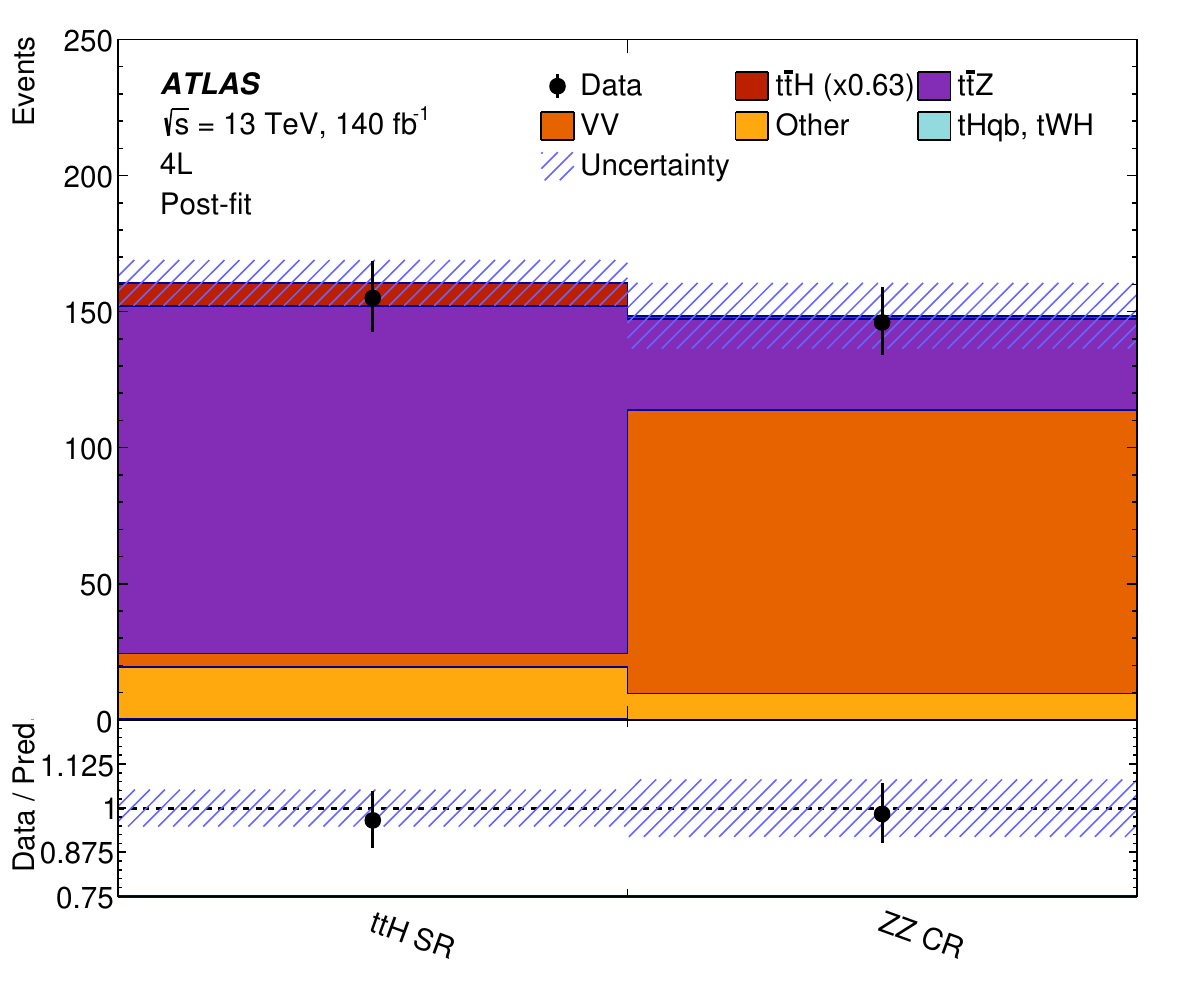}
\caption{Observed and expected event yields in all the categories in the \fourL\ channel.
The signal and background normalisation factors and the nuisance parameters are set to their best-fit values as described in Section~\ref{sec:result}.
The shaded band indicates the total uncertainty, including correlations.
}
\label{fig:4L_cat}
\end{figure}

This channel uses a DNN with three output nodes for the $\ttH$ signals as well as the $\ttZ$ and $ZZ$ backgrounds.
The DNN training is performed using 22 input variables, with the most discriminating ones including the $\met$ and the invariant masses of the different dilepton pairs in the event.
Using the DNN $ZZ$ output score, two separate categories are built enhanced in $\ttH$ and $ZZ$ production as shown in Figure~\ref{fig:4L_cat}.
Since the low part of the DNN distribution in the $\ttH$ category features a good purity in $\ttZ$ events,  no explicit $\ttZ$ category is defined.

\subsection{\twoLssTau channel}

Events in this channel are required to have two light leptons fulfilling the $M$ requirements in Table~\ref{tbl:tightleps}, with $\pT>15$~\GeV and the same electric charge, plus one \tauhadvis candidate with $\pT>20$~\GeV.
For the events where both light leptons have the same flavour,  their invariant mass must not be within 10~\GeV of the $Z$-boson mass to suppress backgrounds featuring $Z$-boson decays.
In addition,  further rejection of the main backgrounds is obtained by requiring events to have at least four jets and at least one of them must be $b$-tagged with the 85\% $b$-tagging WP.

A total of 17 input variables are used to train a multi-class BDT with three output nodes for the $\ttH$ signal as well as the $\ttW$ and $\ttbar$ backgrounds.
The BDT training is performed with a relaxed event selection accepting events with at least three jets to increase the available MC statistics.
The most discriminating variables in the BDT,  evaluated at the training selection level,  are the jet multiplicity,  the invariant mass of the system formed by the leptons and the $\met$,
and the scalar sum of the transverse momenta of jets, leptons and \tauhadvis candidates in the event.
Since it was found not to improve the sensitivity,  no further categorisation is performed in this channel but the BDT output score for the $\ttH$ class is used as discriminant variable.

\subsection{\oneLtwoTau channel}

Events in this channel are required to have one light lepton with $\pT>27$~\GeV fulfilling the $L$ requirements in Table~\ref{tbl:tightleps},  plus two \tauhadvis candidates with $\pT>20$~\GeV and opposite electric charge.
Events must also have at least one jet fulfilling the 77\% $b$-tagging WP requirements.
Two separate categories are built depending on the number of jets in the event: a first category enhanced in $tHqb$ signal contains all the events with one or two jets, while events with at least three jets constitute a second category enhanced in $\ttH$ signal events.
Similarly to the \twoLss channel, the $\ttH$ category is split into five categories each according to the reconstructed $\pTH$ values.
In total,  6 categories are considered in this channel, whose observed and expected yields are shown in Figure~\ref{fig:2tau_cat}.

Separate binary BDT classifiers are trained in each category to improve the signal separation from the background, and are used as discriminant variable in the analysis.
For the $tHqb$-enhanced category,  17 input variables are used for the BDT training with the most relevant being minimum angular distance between the jets and \tauhadvis, and the azimuthal angle between the top-quark candidate (reconstructed with the lepton, $\met$ and leading $b$-jet) and the di-\tauhadvis system.
For the $\ttH$-enhanced category,  12 input variables are used, the most relevant ones being the azimuthal separation between the di-\tauhadvis system and the $\met$, and the largest $|\eta|$ value of the two \tauhadvis candidates.

\subsection{\twoLtwoTau channel}

\begin{figure}[!htb]
\centering
\includegraphics[height=7cm]{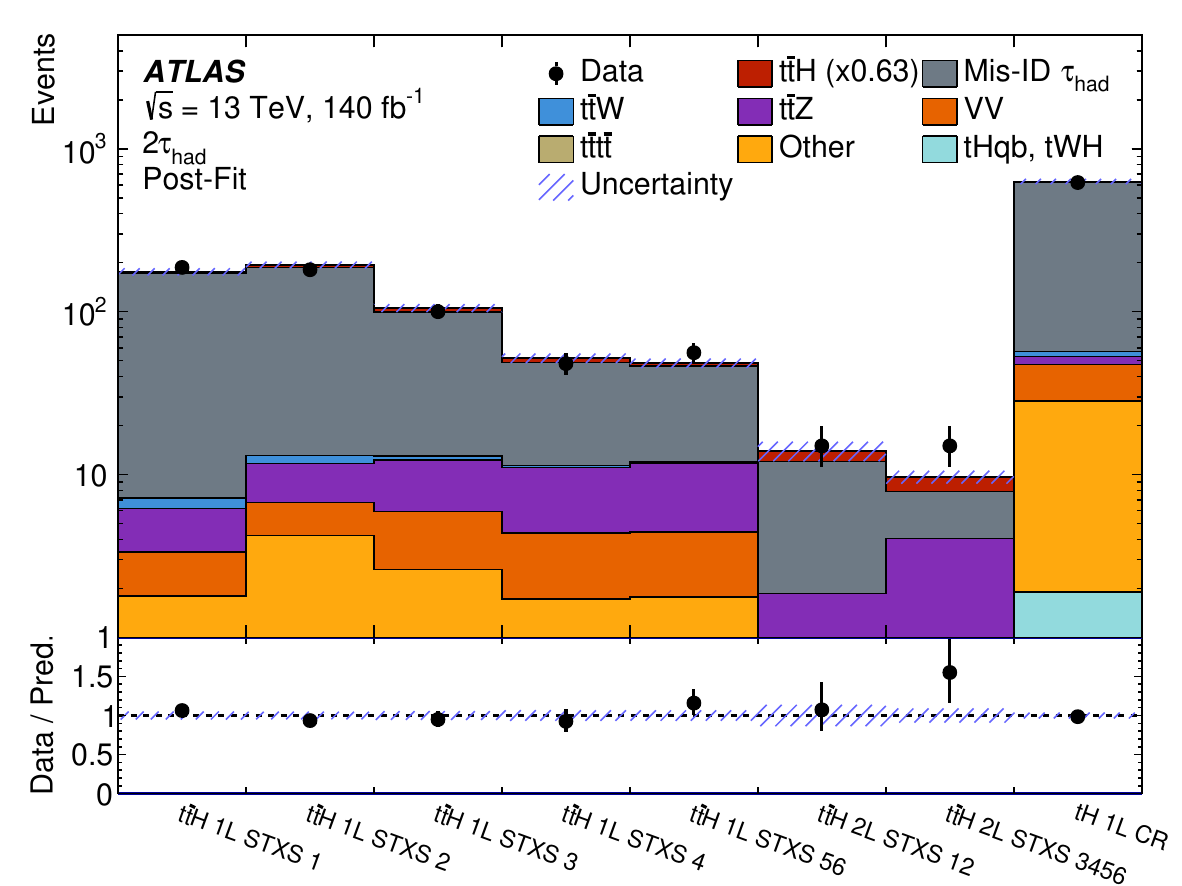}
\caption{Observed and expected event yields in all the categories in the \oneLtwoTau and \twoLtwoTau channels.
The signal and background normalisation factors and the nuisance parameters are set to their best-fit values as described in Section~\ref{sec:result}.
The shaded band indicates the total uncertainty, including correlations.
}
\label{fig:2tau_cat}
\end{figure}

Events in this channel are required to have two light leptons fulfilling the $L$ requirements in Table~\ref{tbl:tightleps} with opposite electric charge and two \tauhadvis candidates with $\pT>20$~\GeV and opposite electric charge.
Requirements are imposed on the dilepton invariant mass to reduce the contribution from $Z$ boson and other light resonances.
Events must also have at least one jet fulfilling the 77\% $b$-tagging WP requirements.
Figure~\ref{fig:2tau_cat} shows the observed and expected yields in this channel.
Two separate categories are built according to the reconstructed $\pTH$ value.

A binary BDT classifier is used to improve the channel sensitivity,  which is trained to separate the $\ttH$ signal from the background, which is dominated by events with misidentified $\tauhad$.
This BDT uses 10 input variables, the most discriminating ones being the largest $|\eta|$ value of the two \tauhadvis candidates, the azimuthal separation between the di-\tauhadvis system and the $\met$ and the angular separation between \tauhadvis and jets.
No further categorisation is performed in this channel although the BDT output score for the $\ttH$ class is used as discriminant variable in the analysis.

\FloatBarrier


%
\section{Background estimation}
\label{sec:bkg}

Background processes are categorised as either irreducible or reducible.
Irreducible backgrounds are defined as events in which all selected leptons are prompt, i.e.
are produced in $W$ or $Z$ boson decays, leptonic $\tau$-lepton decays, or photon internal conversions.
Events containing at least one selected non-prompt lepton, a misidentified $\tauhad$, or a light lepton with a mis-assigned charge (QMisID), are considered reducible.

The misidentified $\tauhad$ background is estimated with data-driven techniques. All other backgrounds, including non-prompt leptons,
are estimated from MC,
with the yields of $\ttW$, $\ttZ$, $WZ$, $WW/ZZ$, and non-prompt lepton backgrounds being adjusted via normalisation factors that are determined by performing a likelihood fit to data
across all event categories described in Section~\ref{sec:evtselection} as well as dedicated control regions defined specifically to help constraining a particular background.
Table~\ref{tab:selectionCR} shows the definition of the control regions used in the \twoLss, \threeL, and \llll channels.
Table~\ref{tab:selectionCRtaus} shows the definition of the control regions used in the \twoLssTau, \oneLtwoTau and \twoLtwoTau channels.
For some backgrounds,  like diboson production, the simulation is improved by applying additional corrections derived from data control samples before the simultaneous fit as detailed below.
The values of the normalisation factors given in this section are obtained from the combined fit for the inclusive \ttH cross-section measurement presented in Section~\ref{sec:result}.
For background sources normalised to data, uncertainties on their predicted cross-section are ignored.

\begin{table}[!htb]
\begin{center}
\caption{Event selection summary of the control regions in the \twoLss, \threeL, and \llll~channels. The lepton types $L$, $L'$, $M$, $M_{\mathrm{ex}}$, and $T$ correspond to the definitions described in Table~\ref{tbl:tightleps}. The $L'$ category is only used in the \lllVV and \lllttZ regions with an opposite-sign muon. The notation $e^*$ is used to denote material conversion or internal conversion candidates, as described in Section~\ref{sec:objects}.
}

\resizebox{\textwidth}{!}{
\def\arraystretch{1.4}
\begin{tabular}{l||c|c|c|c||c|c}
\hline\hline
& \multicolumn{4}{c||}{\twoLss and \threeL channels} &  \multicolumn{2}{c}{\llll channel} \\	\hline
Region                             & \lllVV  & \lllttZ & \lllintc,  & \llttetm, \llttemt   & \lll $\mu$ HF & \lll e LF \\
&         &         & \lllmatc   & \llttmutm, \llttmumt & \lll e HF   & \\
&         &         &            & \llttemm, \llttmumm  & & \\ \hline
Lepton requirement                     & \multicolumn{2}{c|}{\lll}         & $\mu\mu e^*$ & \llSS & $e^{\mp}\mu^{\pm}\mu^{\pm}$   & $\mu^{\pm}\mu^{\mp}e^{\mp}$ \\
& \multicolumn{2}{c|}{}             &              &       & $\mu^{\mp}e^{\pm}e^{\pm}$     &                           \\
Lepton definition                      & \multicolumn{2}{c|}{$(L$ or $L',M,M)$}   & $(L,M,M)$ & {\small$(T,M_{\mathrm{ex}})$ or $(M_{\mathrm{ex}},T)$ or $(M_{\mathrm{ex}},M_{\mathrm{ex}})$} & \multicolumn{2}{c}{$L$} \\
Lepton \pt[\GeV{}]                & \multicolumn{3}{c|}{(10, 15, 15)}                                 & (15, 15) & \multicolumn{2}{c}{(10,10,10)}\\
$|\mllOSSF - m_Z|$ [\GeV{}]            & \multicolumn{2}{c|}{$<$ 10} & $>$ 10 & -- &  -- & $<$ 10 \\
$|\mlll - m_Z|$ [\GeV{}]               & \multicolumn{2}{c|}{$>$ 10}  & $<$ 10 & -- & -- & $>$ 15 \\
$m_{\mathrm{T}}(\ell_0, \met)$ [\GeV{}]  & \multicolumn{3}{c|}{--}  & $<$ 250 for \small$TM_{\mathrm{ex}}$ and $M_{\mathrm{ex}}T$ pairs & \multicolumn{2}{c}{--}\\
$\met$ [\GeV{}]                   & \multicolumn{3}{c|}{--} & -- & -- & $<$ 20 \\
\njets                             & 2 or 3 & $\geq 4$ &  --  & $\geq 2$ & \multicolumn{2}{c}{$\ge$ 1} \\
\nbjets                            & 1 \bveryloose     & $\geq 1$ \bveryloose & 0 \bveryloose & 1 \bveryloose & $\geq 1$ \bveryloose & 0 \bveryloose \\
\tauhadvis candidates              &  \multicolumn{3}{c|}{0} & 0 & \multicolumn{2}{c}{--} \\
\threeL channel veto                  &  \multicolumn{3}{c|}{--} & -- & \multicolumn{2}{c}{Yes} \\
\hline\hline
Region split                           & -- & -- & internal,   & sub-leading $e/\mu$ $\times$                                                & $e^{\mp}\mu^{\pm}\mu^{\pm}$ & -- \\
&    &    & material    & {\small($TM_{\mathrm{ex}}$, $M_{\mathrm{ex}}T$, $M_{\mathrm{ex}}M_{\mathrm{ex}}$)} & $\mu^{\mp}e^{\pm}e^{\pm}$   & \\
\hline\hline
\end{tabular}
}
\label{tab:selectionCR}
\end{center}
\end{table}

\begin{table}[!tb]
\begin{center}
\caption{Event selection summary in the control and validation regions used in the \twoLssTau, \oneLtwoTau and \twoLtwoTau channels.
The lepton types correspond to the definitions in Table~\ref{tbl:tightleps}.
Regions with two leptons or \tauhadvis can require their charges to have opposite sign (OS) or same sign (SS).
The \tauhadvis definition denoted as $M$ is described in Section~\ref{sec:objects}, while the definition denoted as \textit{VL} employs the \enquote{\textit{Very loose}} ID WP of Ref.~\cite{ATL-PHYS-PUB-2019-033}.
}
\resizebox{\textwidth}{!}{
\def\arraystretch{1.2}
\begin{tabular}{l||c|c|c||c|c|c|c}
\hline\hline
& \multicolumn{3}{c||}{\twoLssTau channel} &  \multicolumn{4}{c}{\oneLtwoTau and \twoLtwoTau channels} \\	\hline
Region                  & CR-OS     & CR-SS     & VR$_{1\tauhad}$  & FF-Z & FF-tt & VR$^{1\ell}_{2\tauhad}$ & VR$^{2\ell}_{2\tauhad}$\\
\hline
Lepton requirement                 & 2, OS  & 2,  SS  &2,  OS  & 2, OS  & 2, OS & 1 & 2,  OS   \\
Lepton definition                  & $M$ & $L'$ & $M$ & $L$ & $L$ & $L$ & $L$ \\
\tauhadvis requirement        & 1 & 1 & 1 & 2, OS & 2, OS & 2, SS  & 2, SS     \\
\tauhadvis definition             & $M$   & $M$  & $M$ &\textit{VL} & \textit{VL} & $M$  & $M$  \\
$|\mllOSSF - m_Z|$ [\GeV{}]    & -- & -- & -- & $<$ 10 & $>$ 10 & --& --\\
\njets                             & 2 or 3 &  2 or 3 & $\geq 4$ & $\geq 1$ & $\geq 1$ & $\geq 1$ & $\geq 1$ \\
\nbjets       & $\geq 1$ \bveryloose  & $\geq 1$ \bveryloose & $\geq 1$   \bveryloose & 0 \bloose & $\geq 1$ \bloose & $\geq 1$ \bloose  & $\geq 1$ \bloose \\
\hline\hline
Region split                       & -- & $ee$, $e\mu$, $\mu e$, $\mu\mu$  & -- & -- & -- & -- & --  \\
\hline\hline
\end{tabular}
}
\label{tab:selectionCRtaus}
\end{center}
\end{table}


\subsection{Irreducible backgrounds}
\label{sec:bkg_irred}


Background contributions with prompt leptons originate from a wide range of physics processes with the relative importance of individual processes varying by channel.
The main irreducible backgrounds originate from \ttW, \ttZ/$\gamma^*$ and \VV. Smaller contributions
originate from the following rare processes: \tZ, \tW, \tWZ, \ttWW, \VVV, \tttt and \ttt production. Backgrounds with prompt leptons are estimated from simulation using the samples
described in Section~\ref{sec:dataMC}.

The \VV simulated event sample does not model the jet multiplicity spectrum in data well.  Therefore, the same \njets-dependent data-driven correction as
the one described in the recent ATLAS \ttW measurement~\cite{TOPQ-2019-30}
is applied to the \VV sample in the \twoLss, \threeL and \twoLssTau channels. This correction is
derived from an inclusive trilepton diboson-enriched region with zero $b$-tagged jets at the 85\% WP.

In addition to the MVA-based \VV and \ttZ CRs defined in Section~\ref{sec:evtselection}, two additional \lllVV and \lllttZ~CRs
are used in the likelihood fit to improve the prediction of the background contribution from the \VV and \ttZ /$\gamma^*$ processes, respectively, in the \twoLss and \threeL channels.
Both rely on the selection of an SFOS lepton pair compatible with the $Z$ mass.
The number of jets and \bjets provide good discrimination between \VV and \ttZ: events in the \lllVV CR are required to have 2 or 3 jets, exactly one of which
is $b$-tagged, whereas events in the \lllttZ~CR must have $\geq 4$ jets and at least one $b$-tagged jet. The three leptons must satisfy the ($L$, $M$, $M$) (resp.  $L'$, $M$, $M$) selection
when the opposite-sign lepton is an electron (resp. a muon).
The yield in the \lllVV CR, and the number of \bjets in the \lllttZ CR, are included in the likelihood fit. Two normalisation factors are used for the \VV processes: one for $WZ$, and
one for $WW/ZZ$. The measured normalisation factors are:
$\lambda_{WZ}= 1.06\pm 0.10$, $\lambda_{WW/ZZ}= 1.13 \pm 0.17$, and $\lambda_{\ttZ}= 1.09 \pm 0.09$.

The rate of the background from internal conversions with $\textrm{m}_{e^+e^-} < 1$~\GeV in the \twoLss and \threeL channels is estimated using the dedicated control region \lllintc,
which selects $Z \rightarrow \mu\mu\gamma$ events where the photon is reconstructed as an electron candidate falling into the internal conversion category.
The total yield in the region is used in the likelihood fit to determine the normalisation factor $\lambda_{\mathrm{IntC}}= 0.89 \pm 0.16$.
The purity of internal conversions in the IntC region reaches 93\%.

The \ttW process is the main irreducible background in the \twoLss and \threeL channels. Its normalisation is estimated in the likelihood fit thanks to the MVA-based CRs described in Section~\ref{sec:evtselection}.
No additional CR targeting \ttW is defined.
Its normalisation factor is found to be $\lambda_{\ttW}= 1.18 \pm 0.07$. Although it cannot be interpreted as an inclusive cross-section measurement because some acceptance uncertainties are not included
(see Section~\ref{sec:syst}), it is in agreement with the dedicated cross-section measurements of ATLAS~\cite{TOPQ-2019-30} and CMS~\cite{CMS-TOP-21-011}.
Fitting two distinct \ttW normalisation factors for the \twoLss and the \threeL channels leads to compatible values of the normalisation factors.
The other channels do not have any significant constraining power on \ttW.


\subsection{Reducible backgrounds}
\label{sec:bkg_red}

\subsubsection{Non-prompt leptons in the \twoLss, \threeL and \fourL~channels}
\label{sec:0tau_fakes}

The non-prompt lepton backgrounds are in general
small in all SRs and are estimated from simulation,
with the normalisation determined from data. The main
contribution to the non-prompt-lepton background is from $\ttbar$
production, followed by much smaller contributions from $V$+jets and
single-top-quark processes. The non-prompt leptons in the simulated event
samples are labelled according to whether they originate from HF or
light-flavour (LF) hadron decays, or from a material conversion
candidate. The HF category includes leptons from bottom and charm
decays, which is the dominant source of non-prompt leptons in the \twoLss and \threeL channels;
in the \llll channel, LF non-prompt electrons also contribute significantly.
Several dedicated control regions enriched in specific processes
are included in the simultaneous likelihood fit to data in order to constrain several normalisation factors.
Due to differences in the lepton and event selection, the nature of the non-prompt background is different in the \twoLss/\threeL channels
on the one hand and in the \fourL~channel on the other hand, thus different normalisation factors and control regions are used.

Several corrections are applied to the \ttbar and overall non-prompt-lepton background
simulations before the fit.
First, the \ttbar$+$~$\geq${}1~$b$-jet contribution is multiplied by a factor of 1.3 as previously measured by ATLAS~\cite{HIGG-2020-23}.
Second, the shape of the $b$-jet multiplicity distribution in the non-prompt-lepton background simulation is
corrected to match data in an orthogonal \twoLss validation region
enriched in non-prompt leptons, where one of the leptons must pass a
looser non-prompt-lepton BDT score but not pass the $M$ lepton WP.
In the~\fourL~channel, the non-prompt-electron candidate transverse momentum is reweighted in order to correct for a slight disagreement with data in the corresponding control region,
and the correction is taken as an additional systematic uncertainty.

In the~\twoLss and~\threeL channels, the normalisation factor for HF non-prompt leptons is estimated separately
for electrons and muons, and is denoted by $\lambda^\textrm{\ll/\lll}_{e HF}$ and
$\lambda^\textrm{\ll/\lll}_{\mu HF}$ respectively.
There are six \twoLss CRs enriched in contributions
from HF non-prompt leptons in \ttbar events,
i.e.\ \llttetm, \llttemt, \llttemm, \llttmutm, \llttmumt,
and \llttmumm. In these CRs, the distribution of the transverse momentum
$p_{\textrm{T}}^{\ell 1}$ of the sub-leading lepton is used in the fit.
Requiring at least one $M_{\textrm{ex}}$ lepton in these CR events
provides separation from the processes with prompt leptons, especially
$\ttW$, and thus optimises the sensitivity to the HF non-prompt
electron and muon contributions. The $\ttH$ signal contribution in
these control regions is at most 8\% of the total
expected yield. The measured normalisation factors for the \twoLss/\threeL HF non-prompt leptons are
$\lambda^\textrm{\ll/\lll}_{e HF} = 1.20 \pm 0.41$ and
$\lambda^\textrm{\ll/\lll}_{\mu HF} = 1.12 \pm 0.16$.

In the~\fourL~channel,  the reducible background is dominated by \ttbar events with two non-prompt leptons. The HF electron and muon components are estimated
with the normalisation factors $\lambda^\textrm{\llll}_{e HF}$ and $\lambda^\textrm{\llll}_{\mu HF}$, and a third one, $\lambda^\textrm{\llll}_{e LF}$,
corresponds to LF non-prompt electrons.
A CR enriched in $t\bar{t}+\mu^{\text{HF}}$ (resp. $t\bar{t}+e^{\text{HF}}$) is designed by selecting \lll events with three $L$ leptons with flavour
$e^{\mp}\mu^{\pm}\mu^{\pm}$ (resp. $\mu^{\mp}e^{\pm}e^{\pm}$) but not selected by the \threeL channel. The same-sign lepton with the lowest transverse momentum is labelled as the non-prompt-lepton candidate,
and its transverse momentum distribution is included in the combined fit.
A CR enriched in electron LF from $Z$+\text{jets} events is designed by selecting \lll events with one SFOS lepton pair with an invariant mass close to the $Z$ mass;
the \met distribution of this CR is included in the combined fit.
The measured normalisation factors for HF and LF non-prompt leptons in the \fourL channel are
$\lambda^\textrm{\llll}_{\mu HF} = 0.91 \pm 0.06$,
$\lambda^\textrm{\llll}_{e HF} = 0.93 \pm 0.13$, and
$\lambda^\textrm{\llll}_{e LF} = 1.17 \pm 0.22$.

Finally, a normalisation factor $\lambda^\textrm{MatC}_e$ is determined for the background from material conversions. The contribution from material conversions in the~\fourL\ channel is negligible,
so this only applies to the \twoLss/\threeL channels.
Similarly to \lllintc, the \lllmatc CR is enriched in material conversions using $Z \rightarrow \mu\mu\gamma$ events
and only the total event yield is used in the fit.
The normalisation factor is measured to be $\lambda^\textrm{MatC}_e = 1.13 \pm 0.21$.

In all non-prompt normalisation factors the uncertainties are mostly statistical. %

\subsubsection{Electron charge mis-identification}

Backgrounds due to electron QMisID
primarily affect the \twoLss and \twoLssTau channels and arise mainly from $t\bar{t}$
production, with one electron either undergoing hard bremsstrahlung
followed by an asymmetric conversion ($e^\pm \to e^\pm \gamma^* \to
e^\pm e^+ e^-$) or having mismeasured track curvature.  The muon charge
misassignment rate is negligible in the $\pt$ range relevant to this
analysis.

The QmisID background is estimated using MC simulation after applying a correction derived by comparing the predicted yield with the observed data in a sample of $Z \to e^+e^-$ events reconstructed as SS di-electron pairs.
This correction decreases the MC yields by 10\%, and a conservative 50\% uncertainty is assigned to the QmisID background estimates.
The contribution from QmisID is less than 2\% of the total background in the \twoLss and \twoLssTau channels.


\subsubsection{Non-prompt leptons and misidentified $\tau_{\text{had}}$ backgrounds in the \oneLtwoTau, \twoLtwoTau, and \twoLssTau channels}

\begin{figure}[!htb]
\centering
\subfloat[]{\includegraphics[height=7cm]{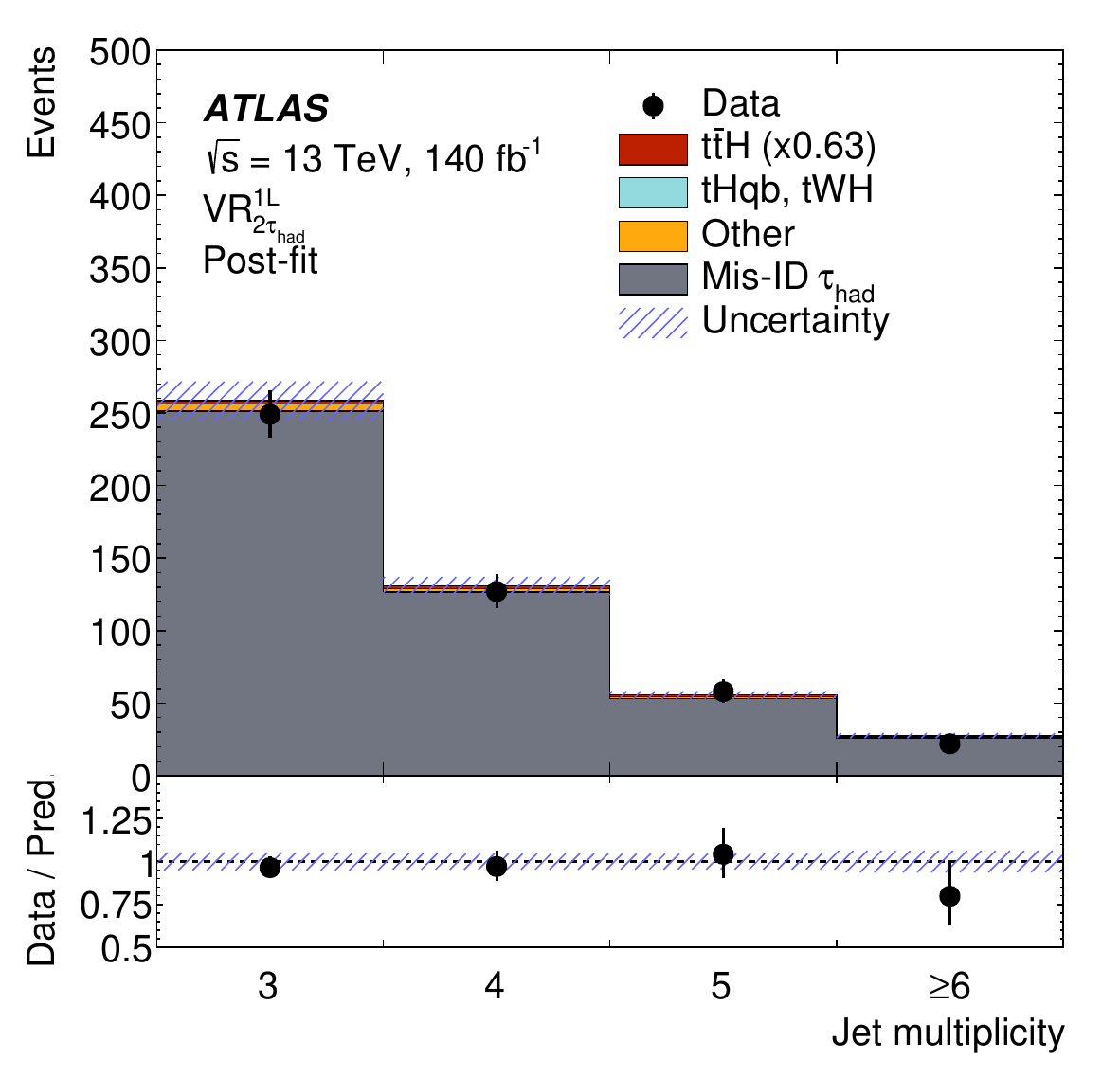}}
\subfloat[]{\includegraphics[height=7cm]{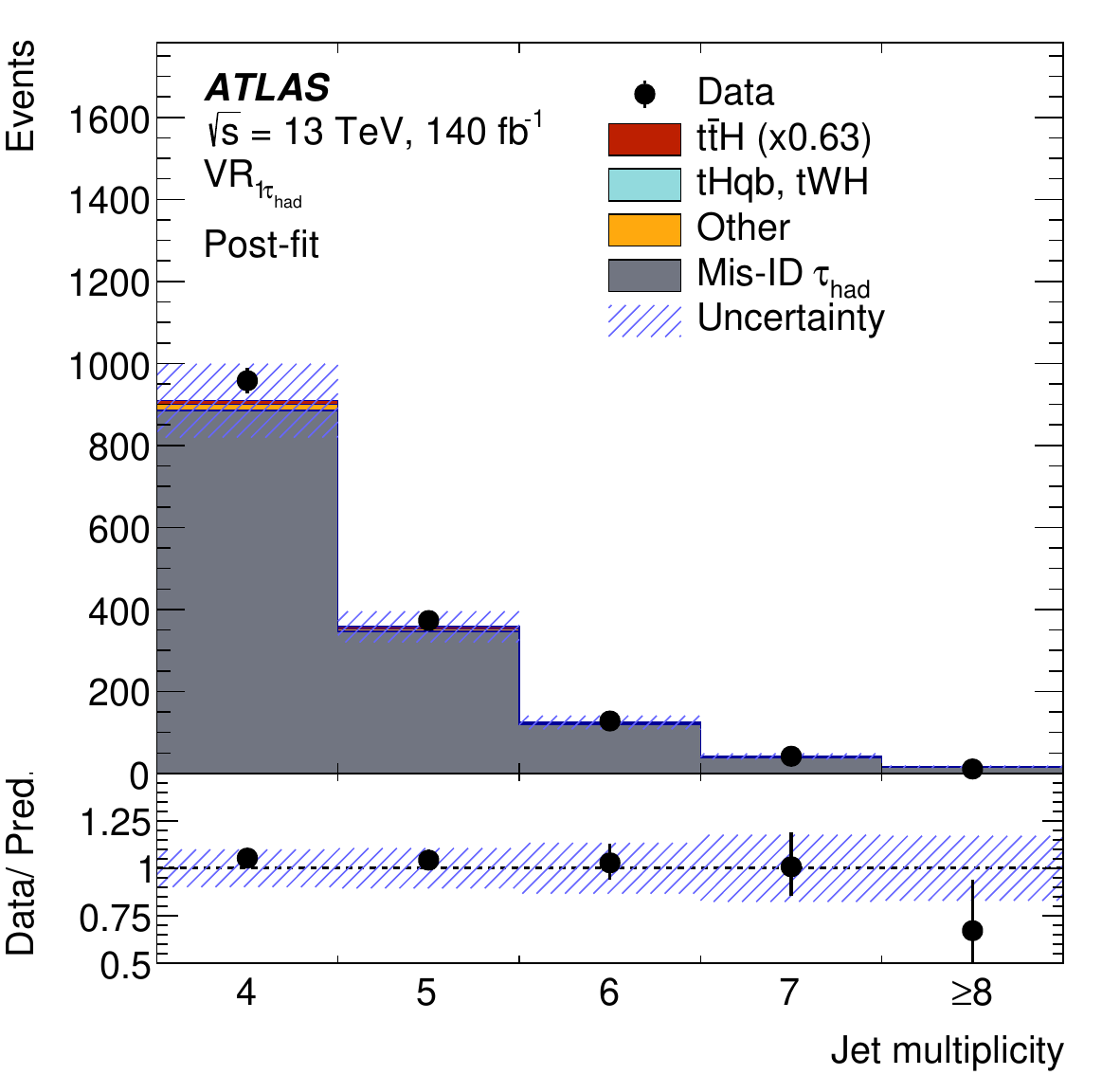}}
\caption{Distribution of the jet multiplicity in the validation regions of the (a) \oneLtwoTau and (b) \twoLssTau channels, obtained by reverting the charge requirements on the \tauhadvis and leptons, respectively.
The signal and background normalisation factors and the nuisance parameters are set to their best-fit values as described in Section~\ref{sec:result}.
The shaded band indicates the total uncertainty, including correlations.
}
\label{fig:tauBkg_VR}
\end{figure}

In the \oneLtwoTau and \twoLtwoTau channels,  the main background arises from $\ttbar$+jets production where one jet is misidentified as a \tauhadvis.
The dominant source of misidentified \tauhadvis is light-quark jets,  with sizeable contributions from gluon-initiated jets and heavy-flavour jets.
This background contribution  is evaluated using the fake-factor (FF) technique~\cite{EGAM-2019-01}.
Data events are selected if they satisfy a very loose requirement on the \tauhadvis identification score but do not satisfy the \enquote{Medium} WP requirements.
Residual contributions from processes with real \tauhadvis satisfying this requirement are evaluated using simulations and subtracted.
The distribution of the misidentified \tauhad background is obtained by multiplying the contribution of data events selected with the inverted identification criterion by a fake-factor defined as the ratio of misidentified \tauhad that respectively pass or fail the \enquote{Medium} identification WP.
The fake-factors are measured in control regions with two light leptons with opposite charge and two \tauhadvis, which are dominated by $Z$+jets and \ttbar production.
These fake-factors are binned in ($\pT$, $\eta$) and their values are in the range 0.20--0.45 (0.04--0.20) for 1(3)-prong \tauhadvis.
The correctness of the background estimates is assessed at low BDT score and, as shown in Figure~\ref{fig:tauBkg_VR}(a), also in the VR$^{1\ell}_{2\tauhad}$ and VR$^{2\ell}_{2\tauhad}$ (Table~\ref{tab:selectionCRtaus}) validation regions (VRs) requiring two \tauhadvis with the same electric charge.

In the \twoLssTau channel,  the method described in Section~\ref{sec:0tau_fakes} is extended to account for both the non-prompt lepton and the misidentified \tauhad background contributions.
Dedicated CRs are defined relaxing the jet multiplicity requirements to include only events with two or three jets.
The first CR is obtained by requiring the light leptons in the event to have the same electric charge and it is dominated by the misidentified \tauhad background.
Additional CRs dominated by non-prompt light leptons are defined by loosening the lepton selections requirements and splitting according to the flavour composition ($ee$, $e\mu$, $\mu e$ and $\mu\mu$).
In the simultaneous fit, normalisation factors are obtained for the misidentified $e$, $\mu$ and \tauhad contributions predicted in the MC.
These normalisation factors are measure to be $\lambda^{\twoLssTau}_{eHF/LF} = 0.68 \pm 0.21$,  $\lambda^{\twoLssTau}_{\mu HF/LF} = 1.11 \pm 0.15$ and $\lambda^{\twoLssTau}_{\textrm{Mis-ID} \tau} = 1.01 \pm 0.07$.
The results are validated in the VR$_{1\tauhad}$ region (Table~\ref{tab:selectionCRtaus}) with the same kinematic selection as the \twoLssTau except for the lepton charge requirements, as shown in Figure~\ref{fig:tauBkg_VR}(b).

%

%
\section{Systematic uncertainties}
\label{sec:syst}
%


Systematic uncertainties in the results arise from uncertainties in the experimental procedures, signal and
background cross-sections, and the modelling of the relevant physics processes.
In the different regions, each uncertainty can affect the normalisations and/or the shape of the discriminants used in the fit.

The dominant experimental uncertainties arise from the calibration of the jet energy scale and resolution~\cite{JETM-2018-05}.
The calibration of the $b$-tagging efficiencies and $c$- and light-jet rejection rates have
smaller effects among the experimental uncertainties~\cite{ATL-PHYS-PUB-2017-003,FTAG-2020-08, FTAG-2019-02}.
Uncertainties are considered for the JVT selection efficiency~\cite{PERF-2014-03} and for the correction of the pile-up profile in the simulations to match that in data.
Other detector-related systematic uncertainties include the light lepton and $\tauhadvis$ reconstruction and identification efficiency, energy scale and energy resolution,  $\met$ modelling, light lepton trigger efficiency and luminosity.

Several sources of uncertainties related to the signal modelling are considered.
Except for \ttH cross-section measurements, two cross-section uncertainties are applied accounting for the effect of varying the PDF and $\alpha_S$ and for missing higher-order terms in the fixed order perturbative QCD calculations. They amount to $\pm 3.6\%$ and $\pm 9.2\%$, respectively~\cite{deFlorian:2016spz}.
Systematic uncertainties related to the migrations of events between the Higgs boson \pt boundaries in the STXS scheme~\cite{ATL-PHYS-PUB-2023-031} are also included.
Uncertainties in the Higgs boson decay branching fractions are also applied following the recommendation in Ref.~\cite{deFlorian:2016spz}.
The impact of using a different parton shower and hadronisation model is evaluated by comparing the nominal \POWPY[8] sample with an alternative \POWHER[7]~\cite{Bellm:2015jjp} sample.
Uncertainties in the matching between the matrix element and parton shower are estimated by comparing the nominal event sample with an alternative \POWPY[8] sample where the definition of the hardness of the \POWHEG emission, calculated via the parameter $\pt^{\textrm{hard}}$~\cite{Hoche:2021mkv}, is varied following Ref.~\cite{ATL-PHYS-PUB-2023-029}.
An uncertainty related to the amount of initial-state radiation (ISR) is estimated by varying independently the factorisation and renormalisation scales in the matrix element as well as considering ISR \alphas variations taken from the A14 tune~\cite{ATL-PHYS-PUB-2020-023}.

Since normalisation factors are extracted from the fit for the \ttW, \ttZ and \VV backgrounds, cross-section uncertainties for these processes are omitted.
Moreover the modelling variations on \ttW, \ttZ and \VV described below are all normalised to the yield of selected events of the nominal prediction, such as not to include overall acceptance uncertainties.
As a result the normalisation factors cannot be interpreted as cross-section measurements.

Systematic uncertainties in the modelling of the \ttW, \ttZ and \VV backgrounds due to missing higher-order QCD corrections are estimated by varying independently the factorisation and renormalisation scales by a factor of 0.5 and 2 with respect to the central value.

For systematic uncertainties in the modelling of the \ttW background due to the matrix element and parton shower algorithm and parameter choices, the nominal \Sherpa prediction is compared to a sample generated with \MGNLO{}+\PYTHIA[8] using the FxFx NLO multĳet merging~\cite{Frederix:2021agh}.
In addition, an uncertainty related to the choice of parton shower and hadronisation model is derived by comparing the relative difference between \POWPY[8] and \POWHER[7] and applying that variation to the nominal \ttW prediction.
Uncertainties in the PDF modelling are also included.

Uncertainties in the modelling of additional jets for \ttZ production are applied by varying the ISR \alphas variations taken from the A14 tune.
In addition, parton shower, hadronisation and underlying event modelling uncertainties are derived by comparing the nominal \MGNLO{}+\PYTHIA[8] sample with an alternative \MGNLO{}+\HERWIG[7] sample.

Systematic uncertainties associated to the data-driven $N_\mathrm{jets}$-dependent correction factors applied to the \VV production are derived based on the $N_\mathrm{jets}$ fit function parameter uncertainties. These are applied only in the channels where the correction factors are used (\twoLss, \threeL and \twoLssTau).

The uncertainties in higher-order \xs{}s are taken into account for the rare background processes  (\tZ, \tttt, \ttWW, $VH$, $tWZ$, \VVV, \ttt, \tHjb and \tHW) except for those whose normalisation is measured in the fit.
A $^{+70}_{-15}$\% normalisation uncertainty is assigned to \tttt\ production to account for the 15\% theory uncertainty in this process ~\cite{Frederix:2017wme} as well as the 70\% difference of the ATLAS measurement~\cite{TOPQ-2021-08} of $22.5^{+6.6}_{-5.5}$~fb with respect to the SM prediction of $13.4^{+1.0}_{-1.8}$~fb~\cite{vanBeekveld:2022hty}.  This uncertainty treatment is consistent with the recent update of that calculation at next-to-leading logarithmic accuracy of $17.4^{+1.4}_{-3.0}$~fb in Ref.~\cite{vanbeekveld2025thresholdresummationproductionquarks}.
Additionally, uncertainties associated to the  \tttt\ matrix element and choice of parton shower are estimated by comparing the nominal sample to alternative \SHERPA and \MGNLO{}+\HERWIG[7] samples, respectively.

The systematic uncertainties in the estimation of the non-prompt light lepton background closely follow the strategy in Ref.~\cite{TOPQ-2019-30},  including uncertainties related to the non-prompt lepton BDT (applied in the \twoLss and \threeL channels only) and uncertainties related to the $\ttbar$ MC modelling.
Uncertainties are estimated for the modelling of the non-prompt-lepton BDT input variables,  namely the muon energy deposit in the calorimeter relative to the expected value ($E_\mathrm{cluster}/E_\mathrm{expected}$), the electron track \pt divided by the jet track \pt, and the secondary vertex's longitudinal significance, using tracks with $\pt > 500$~\MeV for non-prompt muons.
These uncertainties are included as variations that can affect the shape of distributions in each region, but not the normalisation.
An uncertainty of $\pm$20\% is considered to account for non-prompt-lepton rate differences between the \textit{Tight-not-VeryTight} and \textit{VeryTight} non-prompt lepton BDT working points.  Uncertainties in the $t\bar{t}$ parton shower are assessed by using a \POWHER[7] $t\bar{t}$ MC simulation as a variation of the nominal \POWPY[8] prediction, separately for the electron and muons channels.
The following uncertainties associated to the generation of the \ttbar MC sample are also taken into account: normalisation and factorisation scales,  initial- and final-state radiation, $p_{\mathrm{T}}^{\mathrm{hard}}$ and \hdamp\ parameters.

Uncertainties in the modelling of \ttbar production in association with heavy-flavour jets are accounted for by assigning an uncorrelated 50\% uncertainty on the $\ttbar+b$ and $\ttbar+c$ background processes.
In the \fourL\ channel,  due to the sizeable contribution from leptons produced in semileptonic decays of heavy-flavour jets in \ttbar\ production, an additional uncertainty comparing the predictions from \ttbar MC samples produced in the four-flavour and five-flavour schemes is included.

Uncertainties in the electron conversion background extrapolation from $Z$-enriched to \ttbar-enriched regions are derived from the residual data/simulation mismodelling in
\twoLss validation regions built by reversing the electron conversion veto.
Internal and material conversion extrapolation uncertainties of 50\% and 10\%, respectively, are applied.

A systematic uncertainty of 50\% is assigned to the background from electrons with misidentified charge in the \twoLss and \twoLssTau channels.

In the \twoLssTau channel, the same set of uncertainties related to the non-prompt light lepton background are applied.  Dedicated uncertainties of 20\%  for material conversions and 40\%  for the extrapolation between $L'$ and $M$ lepton definitions are also applied.
Additional extrapolation uncertainties related to the difference in misidentified \tauhad kinematics and composition between the signal and control regions are included.

The systematic uncertainties in the estimation of the misidentified \tauhad background in the \oneLtwoTau and  \twoLtwoTau channels include statistical uncertainties in the \tauhad fake-factors,  uncertainties due to the subtraction of processes with two real \tauhadvis in the control regions, and an overall 10\% uncertainty due to observed non-closure in the validation region with same-sign \tauhadvis pairs.  Additional uncertainties due to the composition differences between the signal and control regions are evaluated by comparing the fake-factors measured in data from $Z$+jets and $\ttbar$ control regions.

Table~\ref{tab:grouped_impacts} summarises the impact of the different uncertainty sources on the measurement of the \ttH production cross-section.

\begin{table}[tbh]
\centering
\caption{Summary of the different sources of uncertainty in the ratio of $\sigma_{\ttH}$ to the SM expectation.
The quoted values are obtained by repeating the fit, while fixing the set of nuisance parameters of the sources corresponding to each category to their best-fit values, and subtracting in quadrature the resulting uncertainty from the total uncertainty of the nominal fit presented in the first row.
Experimental uncertainties for reconstructed objects combine efficiency and energy/momentum scale and resolution uncertainties.
}
\label{tab:grouped_impacts}
\begin{tabular}{lc}
\toprule
& $\Delta \sigma_{\ttH}/\sigma^{\text{SM}}$       \\
\midrule
Total uncertainty                                 & $+0.20$ $-0.19$ \\
\midrule
Statistical uncertainty                           & $+0.13$ $-0.12$  \\
Total systematic uncertainty     & $\pm0.15$ \\
\midrule
$ttH$ modelling     & $+0.03$ $-0.02$ \\
$\ttW$ modelling    & $\pm0.05$ \\
$\ttbar$ modelling & $\pm0.04$ \\
$\tttt$ modelling   & $\pm0.03$ \\
$\ttZ$ modelling   & $\pm0.01$ \\
Diboson modelling  & $\pm 0.01$ \\
Other backgrounds modelling    & $\pm0.03$ \\
\hline
\ttW normalisation       & $+0.07$ $-0.06$ \\
\ttZ normalisation       & $+0.05$ $-0.05$ \\
Non-prompt normalisation & $+0.02$ $-0.02$ \\
Diboson normalisation    & $+0.01$ $-0.01$ \\
\hline
MC samples size                        & $\pm0.04$ \\
Electrons, muons and $\tauhadvis$          & $+0.04$ $-0.03$ \\
Jets                                       & $+0.04$ $-0.03$  \\
Flavour tagging                    & $<0.01$\\
$\met$                                  & $\pm 0.01$ \\
Misidentified $\tau$-lepton background             & $\pm 0.03$  \\
Luminosity                                         & $<0.01$ \\
\bottomrule
\end{tabular}
\end{table}

\newpage


%
\section{Results}
\label{sec:result}
%

\FloatBarrier

A combined profile-likelihood fit is performed on all bins in the signal and control regions to measure the $\ttH$ cross-section and the normalisation of the different background processes.
Similar fits are also performed to measure the $\ttH$ cross-section in STXS bins and to measure the $\ttH$ and $tHqb$ cross-sections simultaneously.
The statistical analysis is based on a likelihood function constructed as the product of Poisson
probability terms ($P$) over the bins of the input distributions:

\begin{equation}
\label{eqn:llh}
L\left(\vec{\mu},  \vec{\lambda}, \vec{\theta}, \text{data}\right) = \prod_{k\in\text{reg}}
\prod_{b\in\text{bins}}   P(n_{k,b} | n^{\text{signal}}_{k,b}(\vec{\mu}, \vec{\theta}) +
n^{\text{bkg}}_{k,b}(\vec{\lambda}, \vec{\theta})) \prod_{\theta\in\vec{\theta}} G(\theta),
\end{equation}

where $n_{k,b}$, $n^{\text{signal}}_{k,b}$ and $n^{\text{bkg}}_{k,b}$
stand for the number of observed events, the number of expected signal events and the
number of expected background events in bin $b$ of analysis region $k$, respectively.
The systematic uncertainties (described in Section~\ref{sec:syst}) are incorporated in the likelihood function through nuisance parameters, $\vec{\theta}$, which are constrained by Gaussian probability terms, $G$.
The statistical uncertainty due to the limited number of simulated events is included in the likelihood in the form of additional nuisance parameters with Poisson constraint terms.
Both $\vec{\mu}$ (signal-strength parameter,  defined as a multiplicative factor to the yield of expected signal events) and $\vec{\lambda}$ (normalisation factors for several backgrounds) are treated as free parameters in the likelihood fit and are estimated by maximising the likelihood.
The test statistic is constructed from the profile likelihood ratio, and the confidence intervals for the parameters of interest are
derived using the asymptotic approximation~\cite{Cowan:2010js}.

\subsection{$\ttH$ cross-section measurements}

\begin{figure}[!phtb]
\centering
\subfloat[]{\includegraphics[width=0.775\linewidth]{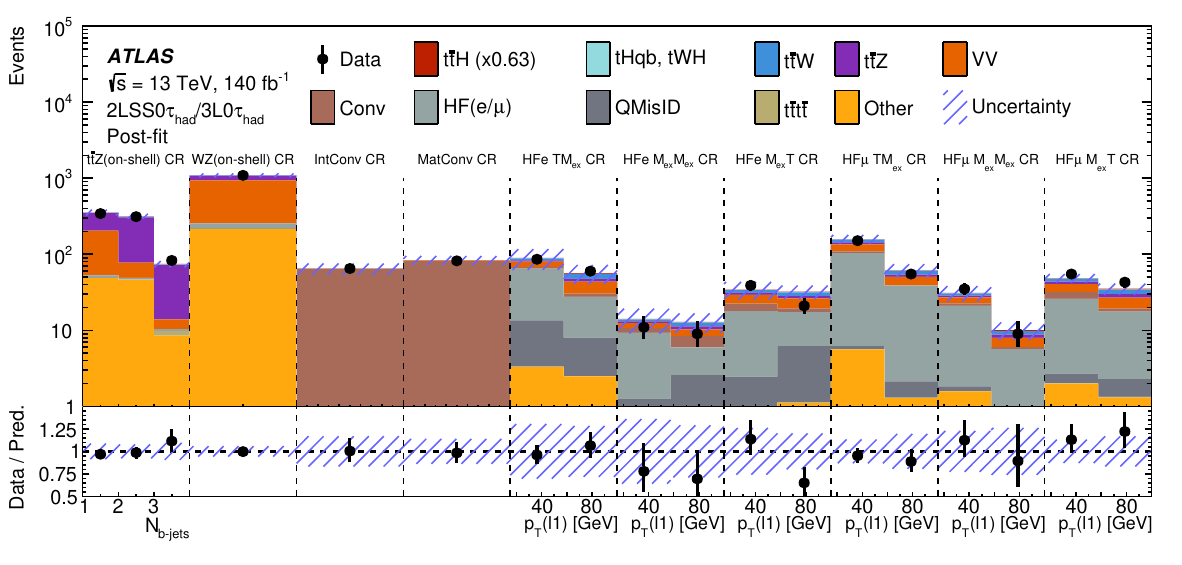}}\\
\subfloat[]{\includegraphics[width=0.775\linewidth]{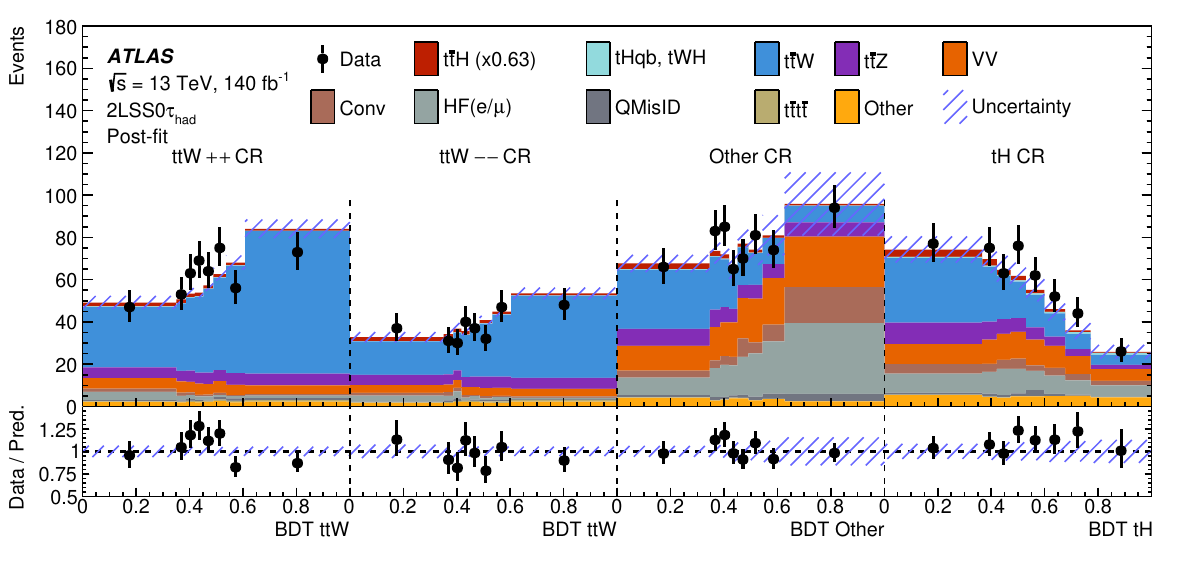}}\\
\subfloat[]{\includegraphics[width=0.775\linewidth]{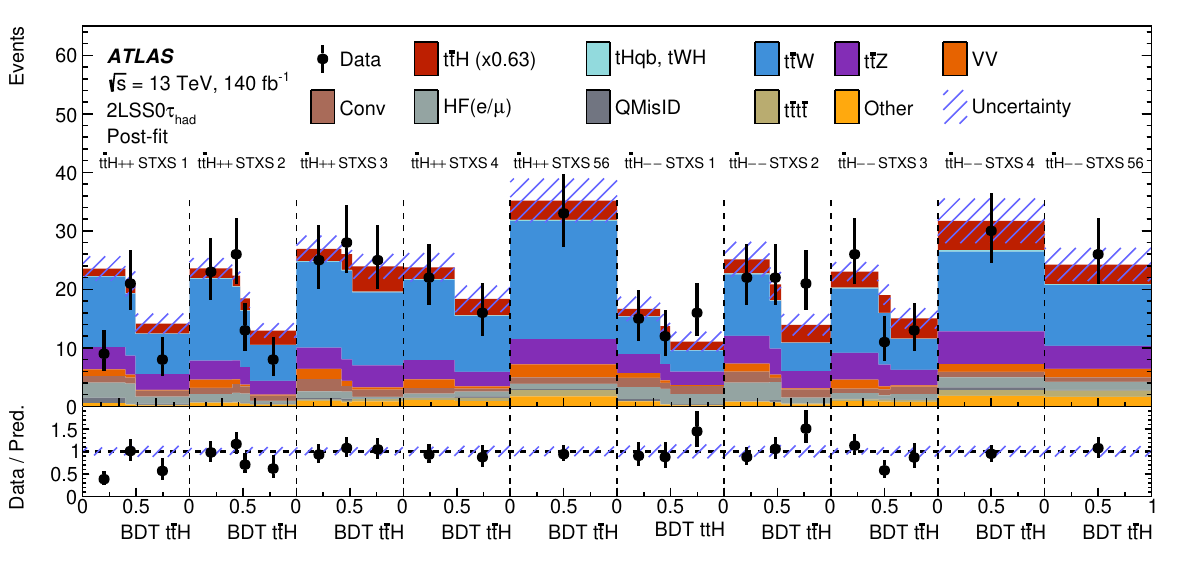}}
\caption{
Comparison of the observed and expected event yields in all the regions and bins used in the statistical analysis for (a) the \twoLss and \threeL control regions, and (b,c) the \twoLss channel.
The signal and background normalisation factors and the nuisance parameters are set to their best-fit values.
Regions labelled `IntConv' and `MatConv' stand for the control regions enhanced in internal an material conversions, respectively.
The regions denoted as `STXS $n$' correspond to the $n$-th bin in the STXS formalism described in Section~\ref{sec:evtselection}.  When several STXS bins are merged together,  all numbers are quoted (e.g. `STXS 56' corresponds to STXS bins 5 and 6). `++' and `$--$' refer to the charge of the two leptons.
The shaded band indicates the total uncertainty, including correlations.
}
\label{fig:postfit_2L}
\end{figure}

\begin{figure}[!phtb]
\centering
\subfloat[]{\includegraphics[width=0.775\linewidth]{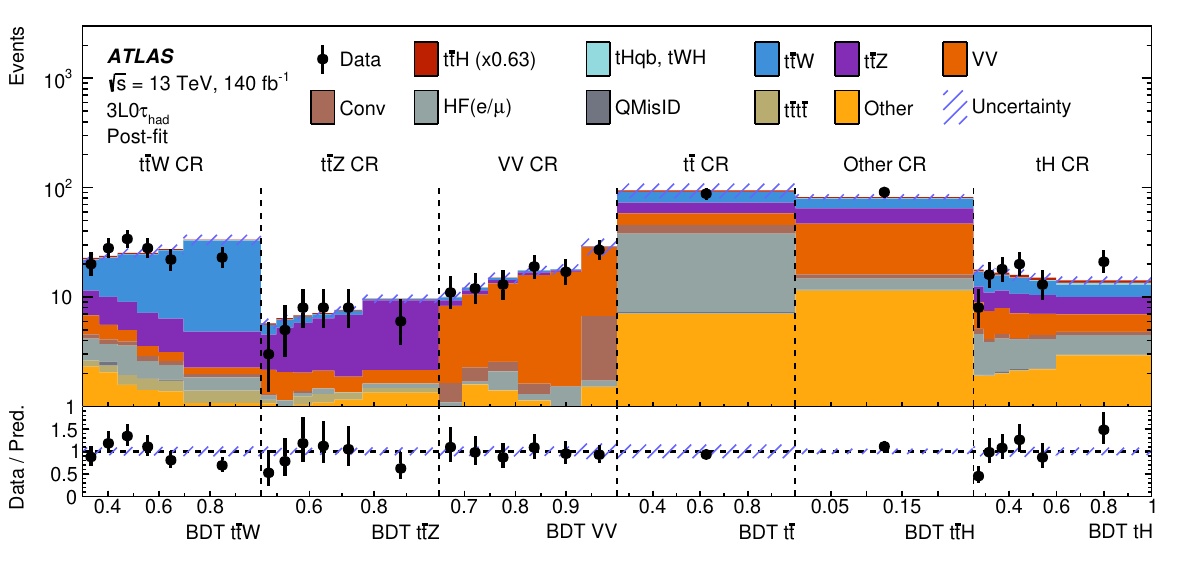}}\\
\subfloat[]{\includegraphics[width=0.775\linewidth]{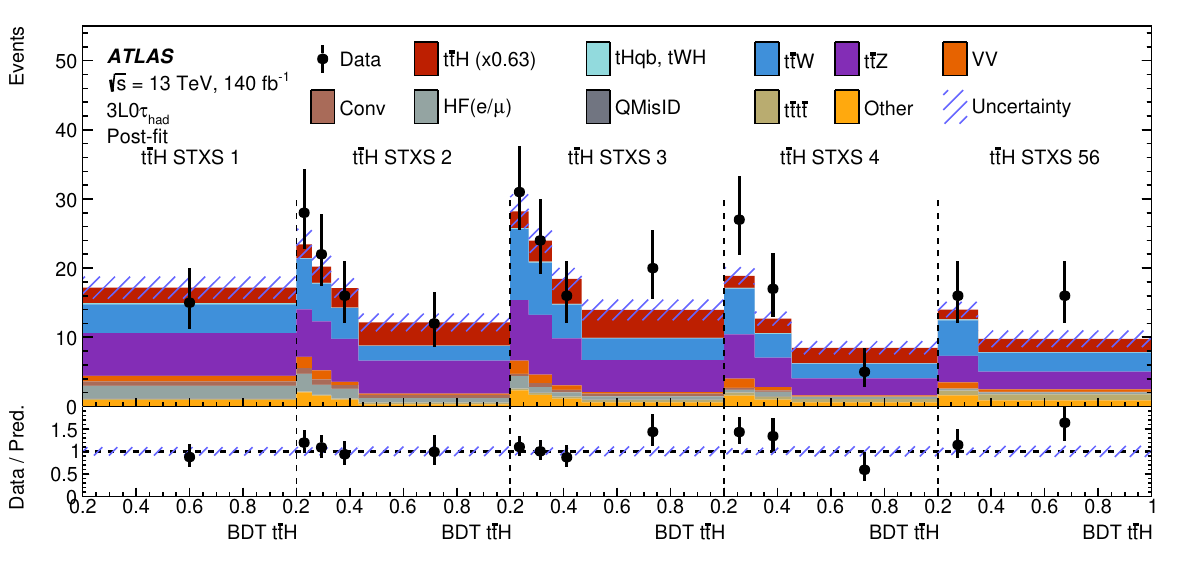}}\\
\subfloat[]{\includegraphics[width=0.775\linewidth]{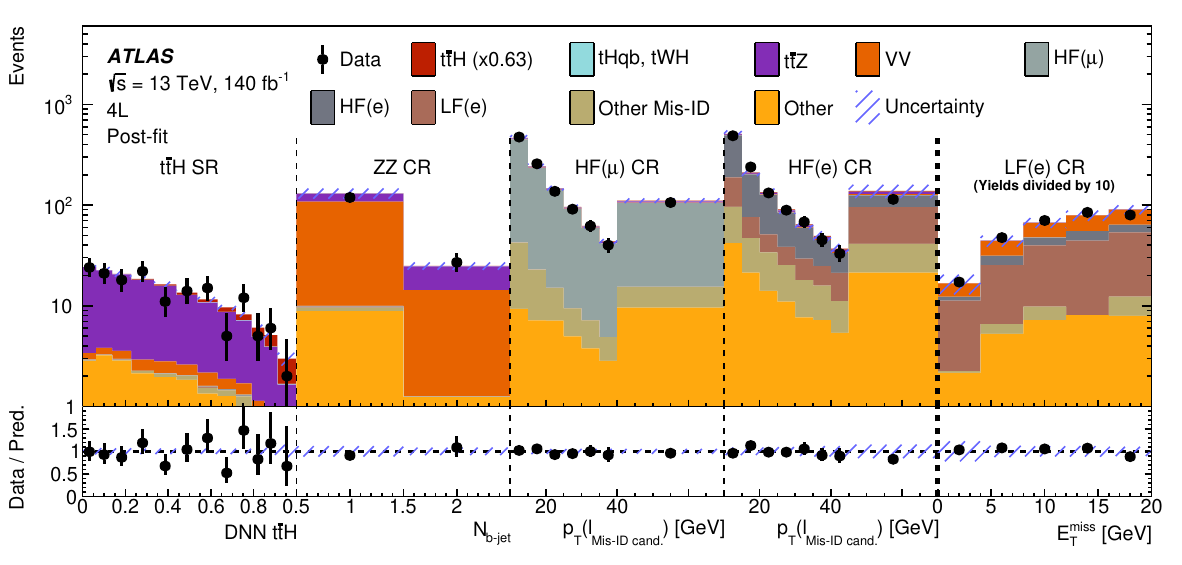}}
\caption{
Comparison of the observed and expected event yields in all the regions and bins used in the statistical analysis for the (a,b) \threeL\ and (c) \fourL\ channels .
The signal and background normalisation factors and the nuisance parameters are set to their best-fit values.
The regions denoted as `STXS $n$' correspond to the $n$-th bin in the STXS formalism described in Section~\ref{sec:evtselection}.  When several STXS bins are merged together,  all numbers are quoted (e.g. `STXS 56' corresponds to STXS bins 5 and 6).
The shaded band indicates the total uncertainty, including correlations.
}
\label{fig:postfit_3L_4L}
\end{figure}

\begin{figure}[!htbp]
\centering
\subfloat[]{\includegraphics[width=0.85\linewidth]{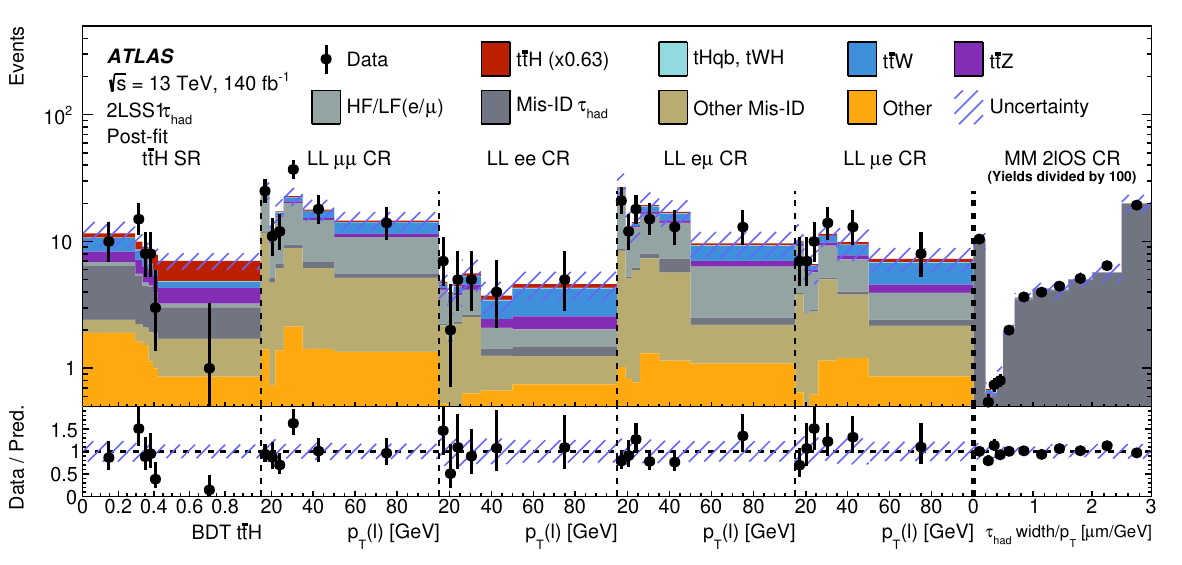}}\\
\subfloat[]{\includegraphics[width=0.85\linewidth]{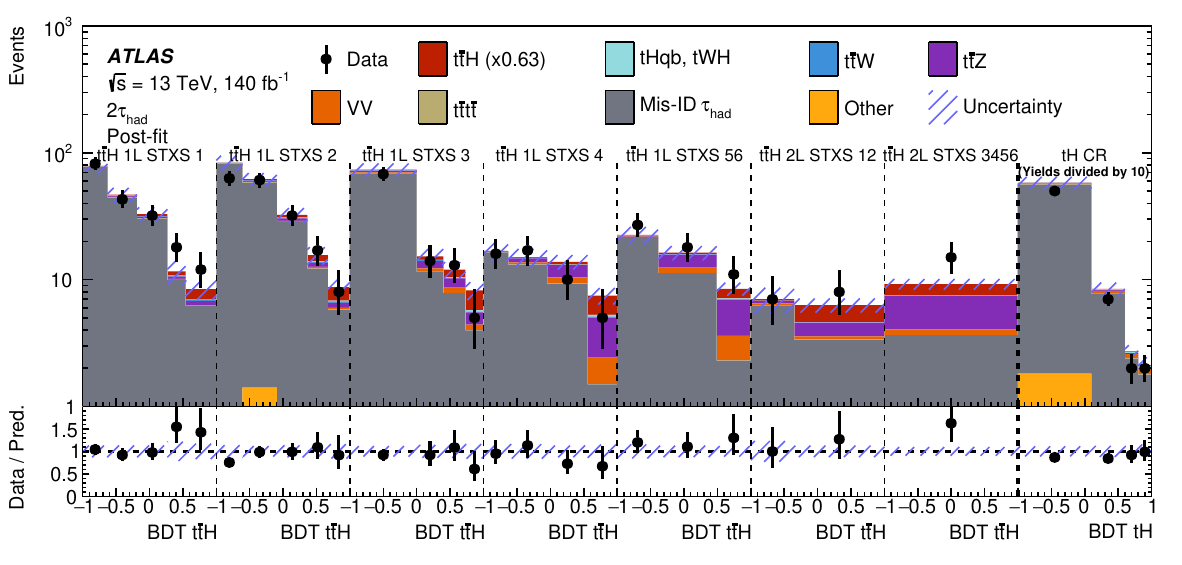}}
\caption{
Comparison of the observed and expected event yields in all the regions and bins used in the statistical analysis for the (a) \twoLssTau, and (b) \oneLtwoTau and \twoLtwoTau channels.
The signal and background normalisation factors and the nuisance parameters are set to their best-fit values.
The regions denoted as `STXS $n$' correspond to the $n$-th bin in the STXS formalism described in Section~\ref{sec:evtselection}.  When several STXS bins are merged together,  all numbers are quoted (e.g. `STXS 3456' corresponds to STXS bins from 3 to 6).
The shaded band indicates the total uncertainty, including correlations.
}
\label{fig:postfit_taus}
\end{figure}

Figures~\ref{fig:postfit_2L}-\ref{fig:postfit_taus} show the data and background predictions in each bin entering the statistical analysis.
An excess of events over the expected background is found with an observed (expected) significance of
3.3$\sigma$ (5.3$\sigma$).
The measured \ttH cross-section ($\sigma_{\ttH}$) is $321^{+102}_{-99}$~fb,  while the SM prediction ($\sigma^{\text{SM}}$) is $507^{+35}_{-50}$~fb.
The compatibility of this result with the SM is 7.2\%.

\begin{figure}[!htbp]
\centering
\includegraphics[width=0.8\linewidth]{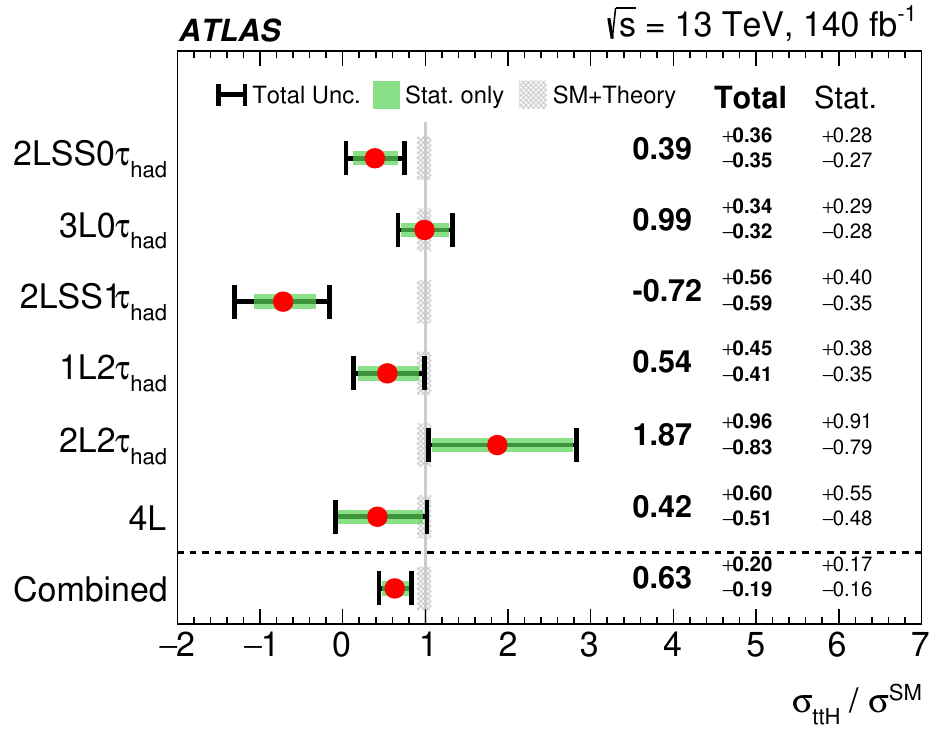}
\caption{The observed best-fit values of the \ttH cross-section relative to the SM expectation
and their uncertainties by analysis channel and
combined. The individual $\sigma_{\ttH}/\sigma^{\text{SM}}$
values for the channels are obtained from a simultaneous fit with the cross-section
parameter for each channel floating independently.
The inclusive theory systematics in $\sigma^{\text{SM}}$ are shown as a grey dashed band but are not included in the uncertainties of the measurement.
}
\label{fig:Resultscombinedfit:NormFactors_signal}
\end{figure}

The best-fit value of $\sigma_{\ttH}/\sigma^{\text{SM}}$\footnote{Results are quoted as $\sigma_{\ttH}/\sigma^{\text{SM}}$ for convenience,  but the inclusive theory systematics in $\sigma^{\text{SM}}$ are not included.} for each individual channel and the combination of all channels are shown in Figure~\ref{fig:Resultscombinedfit:NormFactors_signal}.
The individual channel results are extracted from the full fit but with a separate parameter of
interest for each channel.
The measured results are compatible with the SM prediction in most of the channels. The probability that the six fitted cross-sections are compatible with a single value is 12.4\%.
The largest deviation is observed in the \twoLssTau channel, with a fitted value of  $\sigma_{\ttH}/\sigma^{\text{SM}}=-0.72^{+0.56}_{-0.59}$.
As shown in Figure~\ref{fig:postfit_taus},  there is a downward fluctuation in data at high value of the \twoLssTau BDT.

\begin{figure}[!htbp]
\centering
\includegraphics[width=0.45\linewidth]{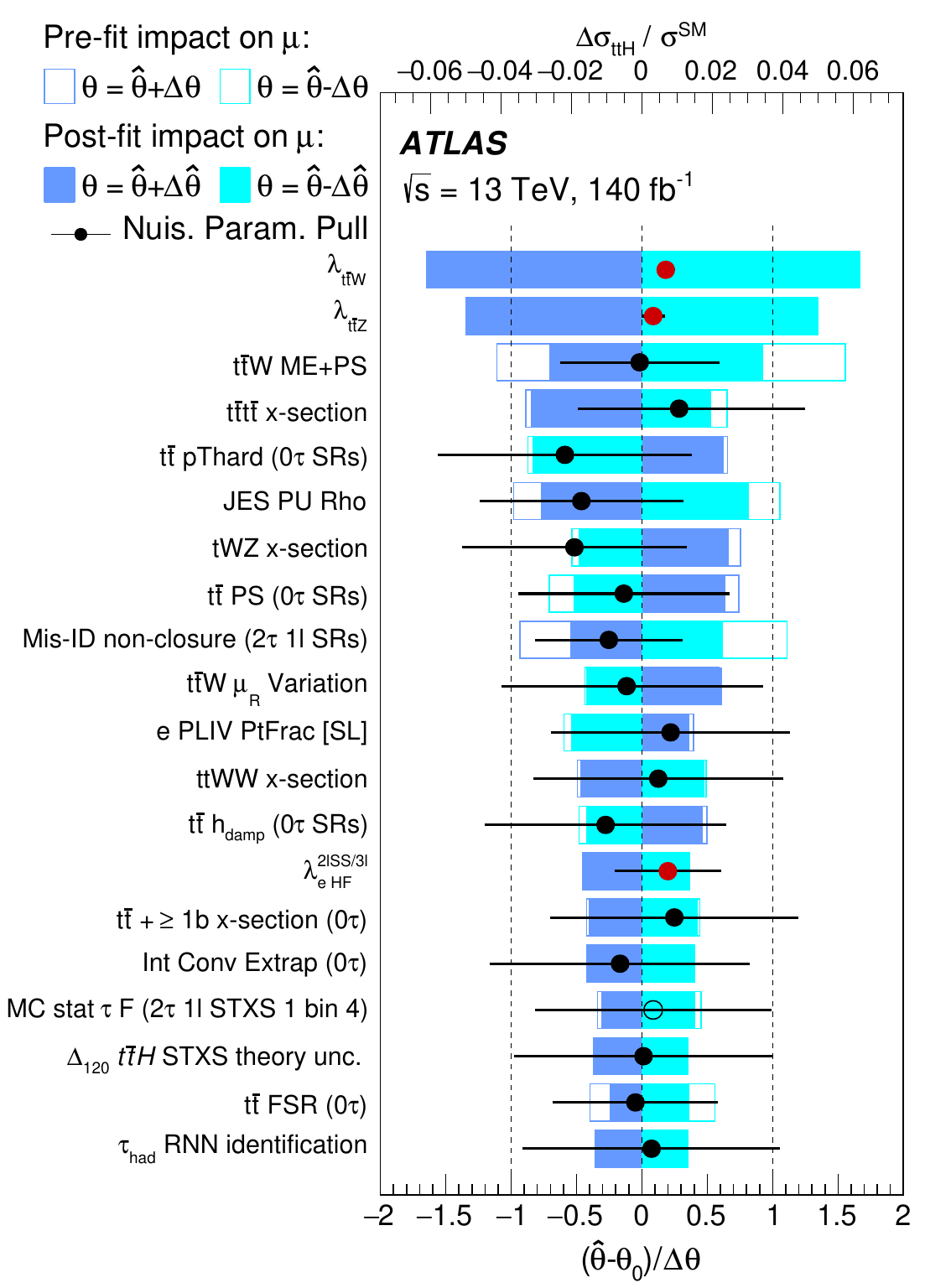}
\caption{Ranking of the 20 modelling and experimental systematic uncertainties with the largest post-fit impact on the inclusive cross-section.
The empty (filled) blue rectangles correspond to the pre-(post-)fit impact on $\sigma_{\ttH}/\sigma^{\text{SM}}$ and refer to the upper scale of the plot.
The impact of each nuisance parameter, $\Delta\sigma_{\ttH}/\sigma^{\text{SM}}$, is computed by comparing the nominal best-fit value of $\sigma_{\ttH}/\sigma^{\text{SM}}$ with the result of the fit when fixing the considered nuisance parameter to its best-fit value, $\hat{\theta}$, shifted by its pre-fit (post-fit) uncertainties $\pm\Delta\theta$ ($\pm\Delta\hat{\theta}$).
The black markers show the pulls of the nuisance parameters relative to their nominal values, $\theta_{0}=0$.
The red markers show the pulls of the normalisation factors relative to their nominal value, $\theta_{0}=1$.
No pre-fit uncertainty is defined for the normalisation factors as they are free parameters of the fit.
The pulls and their relative post-fit errors, $\Delta\hat{\theta}/\Delta\theta$, refer to the lower scale of the plot.
The inclusive theory systematics in $\sigma^{\text{SM}}$ are not included in the uncertainties of the measurement.
For brevity, the following acronyms are used in the figure: matrix element (ME), parton shower( PS), cross-section (x-section), pile-up density (PU Rho),  non-prompt-lepton BDT (PLIV),  final state radiation (FSR).
The STXS migration uncertainty for the bin boundary at $\pTH=120$~\GeV is denoted as $\Delta_{120}$.
}
\label{fig:Resultscombinedfit:Ranking}
\end{figure}

The highest ranked nuisance parameters are shown in Figure~\ref{fig:Resultscombinedfit:Ranking}.
The most relevant nuisance parameters correspond to the $\tttt$ background cross-section uncertainty as well as in the normalisation factors of $\ttW$ and $\ttZ$ production.
Mild constraints are observed for the nuisance parameters related to the $\ttW$ modelling uncertainties and
the non-closure uncertainties in the \oneLtwoTau channel.

\FloatBarrier

\begin{figure}[!htbp]
\centering
\includegraphics[width=0.65\linewidth]{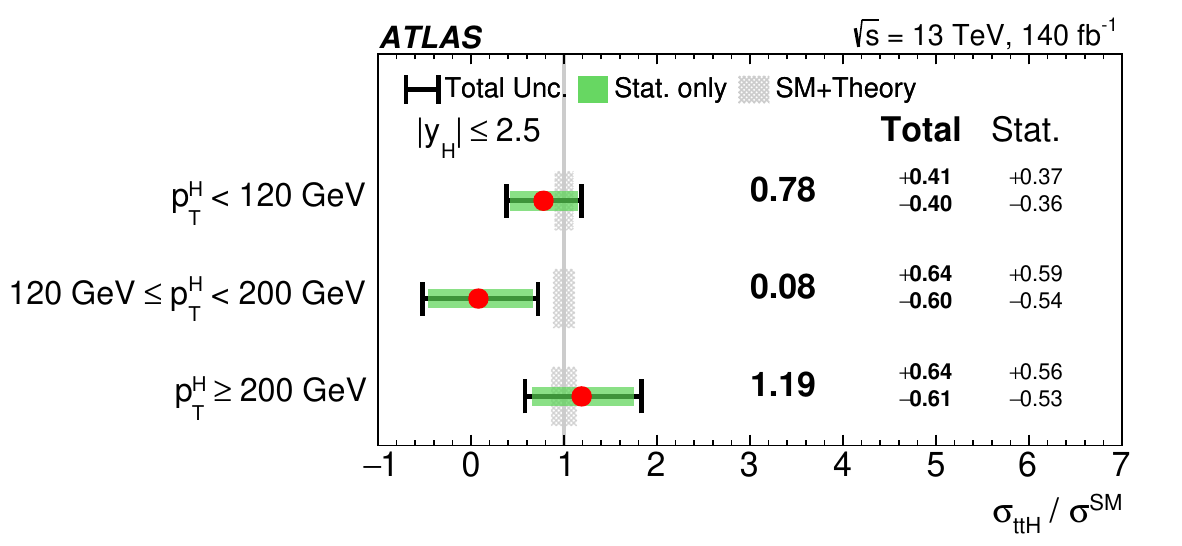}
\caption{The observed best-fit values of the
\ttH cross-section relative to the SM expectation
and their uncertainties in the simplified template cross-section measurement.
The inclusive theory systematics in $\sigma^{\text{SM}}$ are shown as a grey dashed band but are not included in the uncertainties of the measurement.
}
\label{fig:Resultscombinedfit:NormFactors_signal_STXS}
\end{figure}

Table~\ref{tab:STXS_result} and Figure~\ref{fig:Resultscombinedfit:NormFactors_signal_STXS} show the results of the STXS measurement of three separate \pTH bins.
These results are obtained by performing a fit using all the channels, although the sensitivity for different STXS bins comes mostly from the channels featuring \pTH reconstruction (i.e.  \twoLss, \threeL, \oneLtwoTau and \twoLtwoTau).
Excellent agreement with the SM is found for the $\pTH<120$~\GeV  and $\pTH>200$~\GeV bins, although the measured cross-section value in the $120<\pTH<200$~\GeV bin is lower than the SM prediction.
This is due to data underfluctuations in the STXS3 categories of the \twoLss and \oneLtwoTau channels, as well as the effect of anti-correlations between the measurement in the different STXS bins.
The compatibility of the STXS fit with the SM is 28\%. The compatibility of the STXS fit with the inclusive fit is 55\%.

\begin{table}[tbh!]
\caption{Observed best-fit values of the \ttH cross-section and their SM expectation in the simplified template cross-section measurement.
}
\label{tab:STXS_result}
\begin{center}
\renewcommand{\arraystretch}{1.4}{
\begin{tabular}{|l|c|c|}
\toprule
STXS bin  & $\sigma_{\ttH}$ [fb] & $\sigma^{\text{SM}}$ [fb] \\ \midrule
$\pTH < 120$~\GeV & $231^{+122}_{-119}$ & $297^{+20}_{-29}$ \\
$120 \leq \pTH < 200$~\GeV & $10^{+81}_{-76}$ & $127^{+9}_{-12}$ \\
$\pTH \geq 200$~\GeV & $92^{+49}_{-47}$ & $77^{+5}_{-8}$ \\
\bottomrule
\end{tabular}
}
\end{center}
\end{table}

\FloatBarrier
\subsection{$\ttH$ and $tHqb$ combined measurement}

\begin{figure}[!htbp]
\centering
\includegraphics[width=0.75\linewidth]{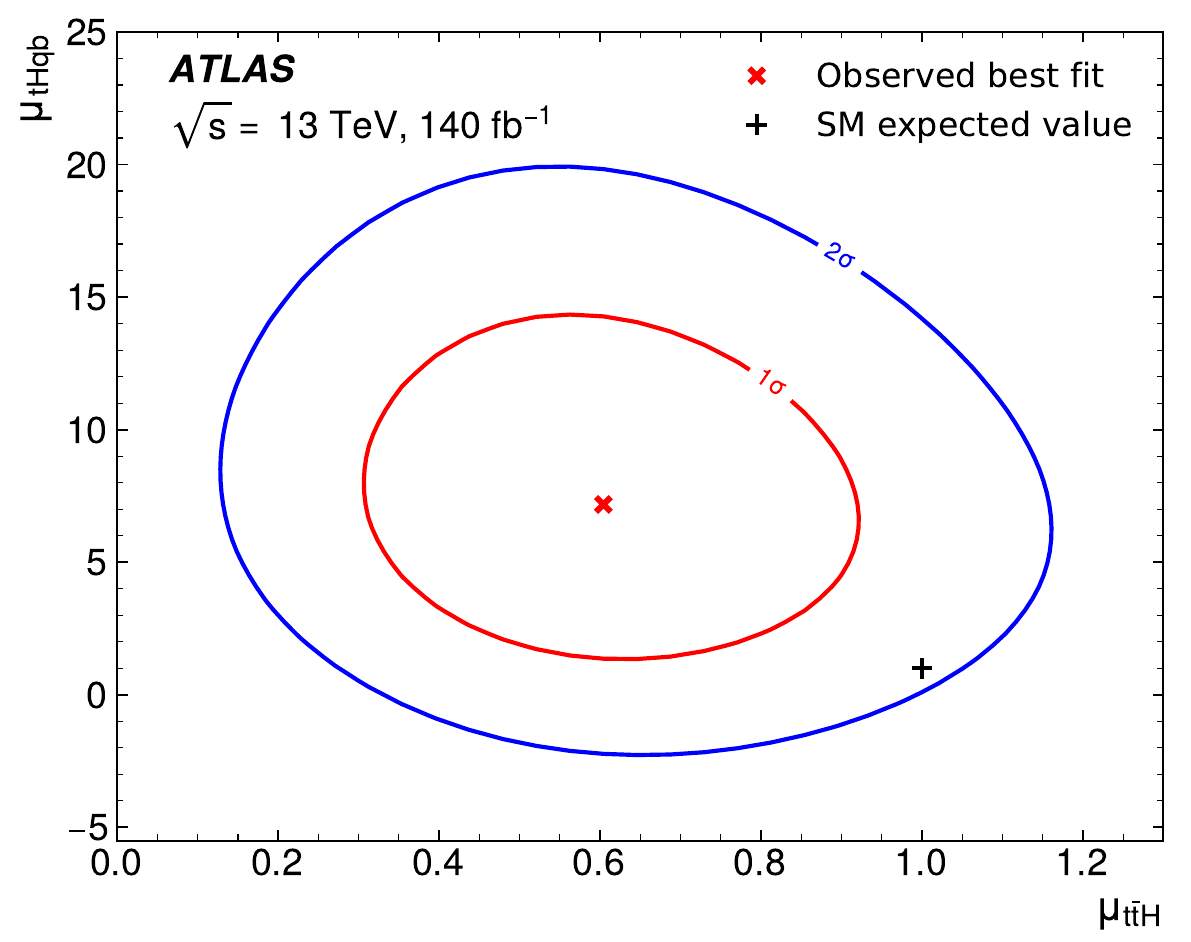}
\caption{Observed exclusion contours in the $(\mu_{\ttH},  \mu_{tHqb})$ plane,  with the best-fit values shown with the cross and the 1$\sigma$ and 2$\sigma$ contours shown with solid lines.
Uncertainties in both the measurement and the SM prediction are included.
}
\label{fig:Resultscombinedfit:NormFactors_signaltH}
\end{figure}

Sensitivity to $tHqb$ can also be achieved since the  \twoLss, \threeL and \oneLtwoTau channels have dedicated categories enhanced in this production process.
Figure~\ref{fig:Resultscombinedfit:NormFactors_signaltH} shows the results of a fit to all the channels with both $\mu_{\ttH}$ and $\mu_{tHqb}$ signal strengths floating freely,
while the $tWH$ process is fixed at its SM prediction.
The best-fit values of the signal strengths are $\mu_{\ttH}=0.59^{+0.22}_{-0.20}$ and $\mu_{tHqb}=7.2^{+4.6}_{-4.0}$ (including uncertainties in the SM prediction), with a $-11\%$ correlation between them.
The measured $\mu_{tHqb}$ is consistent with the dedicated \tH ATLAS result that combines multilepton and $H \to b\bar{b}$ channels~\cite{HIGP-2024-03}.

\FloatBarrier

\subsection{$CP$ interpretation}

To determine the $\alpha$ and \kprime\ $CP$ parameters, a binned profile likelihood fit is performed using only the \twoLss, \threeL, \oneLtwoTau and \twoLtwoTau channels.
In this case, the value of the likelihood in Eq.~(\ref{eqn:llh}) varies according to the expected \ttH, $tHqb$ and $tWH$ signal yields, which are functions of $\alpha$ and \kprime.
By scanning the value of the test statistic in grid points in  $\kappa'_t$ and $\alpha$, two-dimensional exclusion contours in the $(\kappa'_t , \alpha)$ plane are obtained.

\begin{figure}[!phtb]
\centering
\subfloat[]{\includegraphics[width=0.37\textwidth]{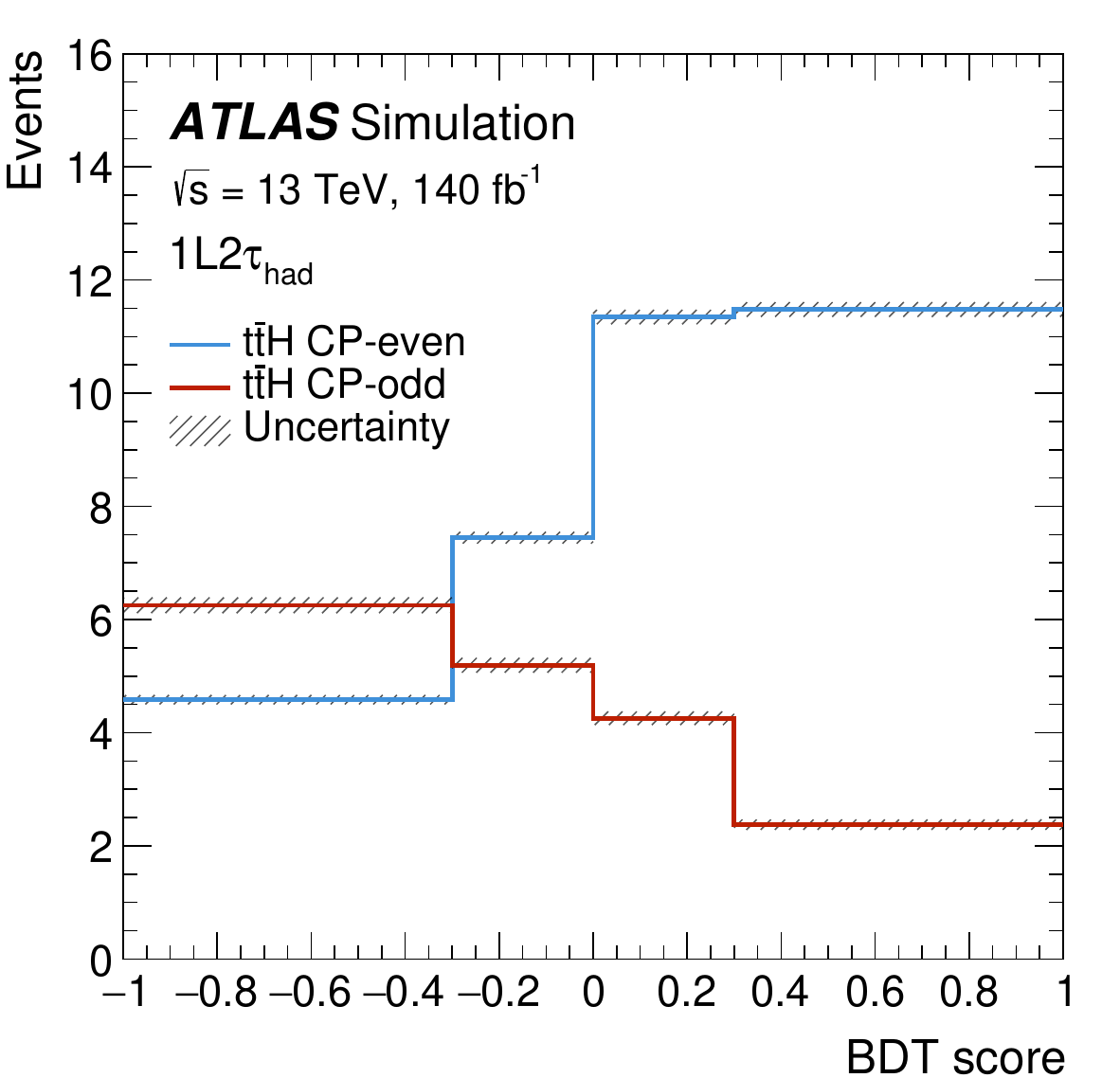}}
\subfloat[]{\includegraphics[width=0.37\textwidth]{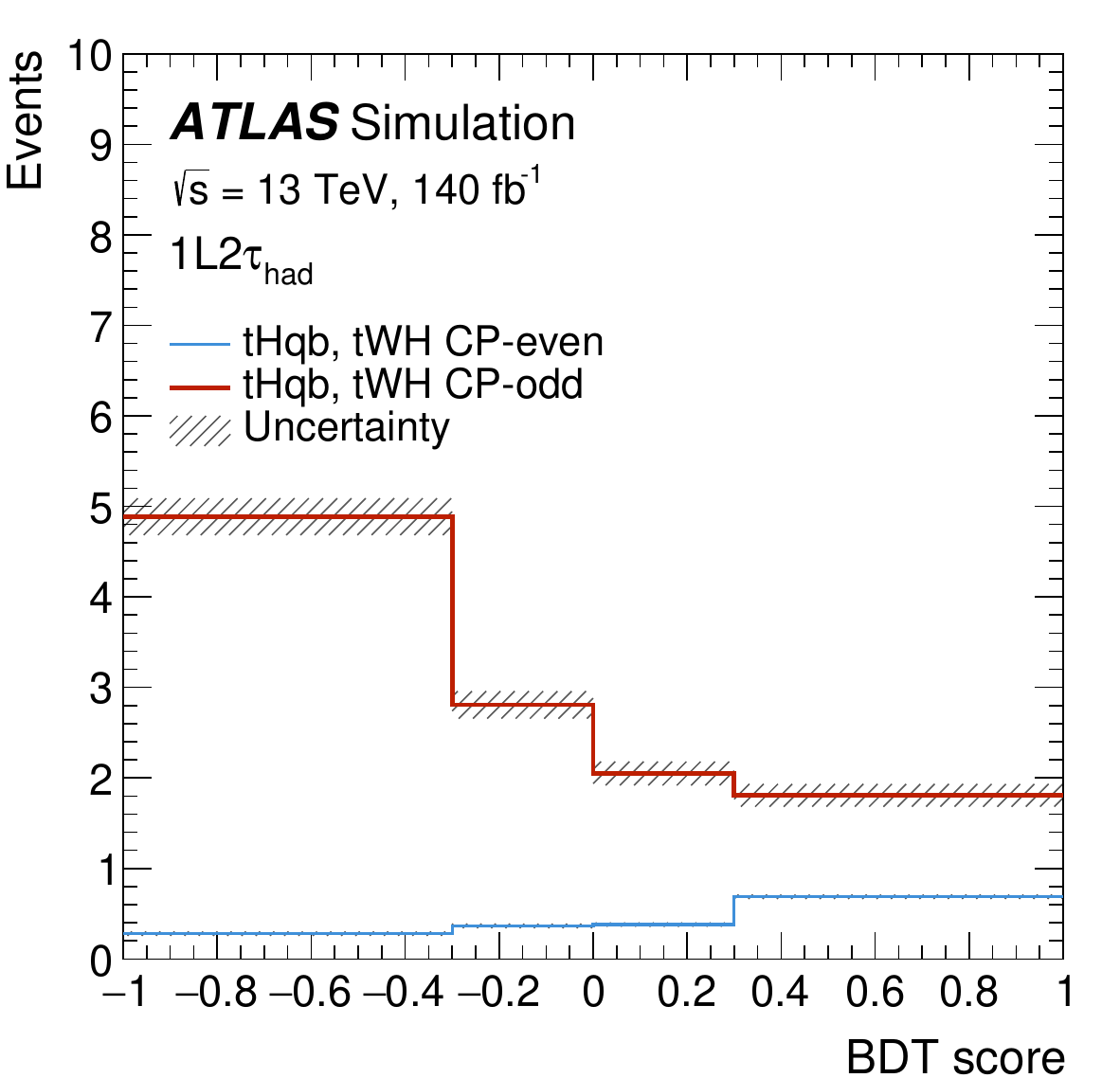}}
\caption{Distributions of the scores of the $CP$ BDT in the \oneLtwoTau channels for  (a) \ttH and (b) $tHqb$ signals under the SM ($CP$-even) and $CP$-odd hypotheses,
normalised to the expected event yields assuming the predicted cross-sections.
Only MC statistical uncertainties are shown.}
\label{fig:CP_BDT}
\end{figure}

In the  \twoLss and \threeL channels, the same discriminants used in the STXS measurement are also employed for the $CP$ interpretation.
In the \oneLtwoTau and \twoLtwoTau channels,  dedicated $CP$ BDTs are trained to separate between the $CP$-even and $CP$-odd signal hypotheses.
Separate BDTs are built for \ttH and $tHqb$, using 17 input variables, the most relevant ones being the angular separations between $\tauhad$ and jets,  jets and leptons,  and $\tauhad$ and leptons.
The performance of these BDTs is illustrated in Figure~\ref{fig:CP_BDT}.
The difference in $tHqb$ yields between the $CP$-odd and $CP$-even hypothesis are due to the change of the relative sign between the Higgs-boson couplings to the top quark and the $W$ boson in the diagrams contributing to this process.
The \ttH $CP$ BDT is used as discriminant variable for the \twoLtwoTau channel and for the \ttH-enhanced category in the \oneLtwoTau channel, while the $tHqb$ $CP$ BDT is employed in the \oneLtwoTau $tHqb$-enhanced category.

\begin{figure}[!htb]
\begin{center}
\includegraphics[height=0.55\textwidth]{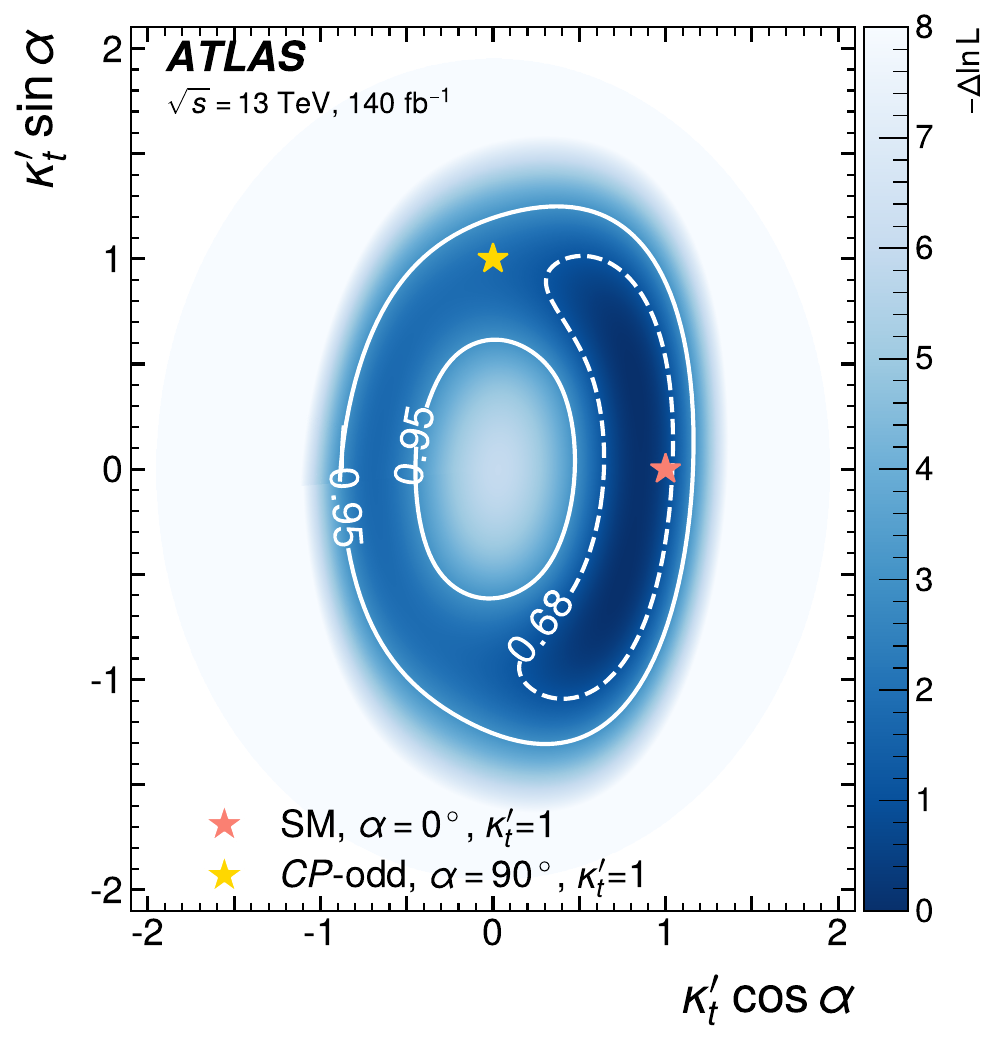}
\end{center}
\caption{The observed exclusion contours in the $(\kappa'_t\cos\alpha , \kappa'_t\sin\alpha)$ plane.
Regions contained in the dashed, and in the solid lines are compatible with the best-fit result at
68\% and 95\% confidence level,
with stars representing the \(CP\)-even and \(CP\)-odd results with \kprime=1.
}
\label{fig:cp_scan}
\end{figure}

The exclusion contours in the $(\kappa'_t , \alpha)$ plane are shown in Figure~\ref{fig:cp_scan}.  The results are compatible with the SM hypothesis, and the observed (expected) exclusion of the pure \(CP\)-odd hypothesis is at 1.8$\sigma$ (3.1$\sigma$).
Because the likelihood is very flat around the minimum, the best-fit value is sensitive to numerical instabilities.
The observed (expected) exclusion at 68\% confidence level on the  \(CP\)-mixing angle is $| \alpha | > 62^\circ$ ($43^\circ$).
The difference between the expected and observed exclusions is mainly coming from the low fitted cross-section in the  \twoLss channel.
The compatibility between the results from fits to the channels without \tauhad (\twoLss, \threeL) and those with \tauhad (\oneLtwoTau and \twoLtwoTau) is 34\%.

%
%
\FloatBarrier

\section{Conclusion}
\label{sec:conclusion}
%

Several measurements of the associated production of a top-quark pair with
the Higgs boson (\ttH) in multilepton final states
based on a dataset of proton--proton collisions at
$\sqrt{s}=13$~\TeV recorded with the ATLAS detector at the CERN Large
Hadron Collider and corresponding to an integrated luminosity of 140~fb$^{-1}$ are presented.
The inclusive \ttH production cross-section is measured to be $321^{+102}_{-99}$~fb.
This result is compatible with the Standard Model prediction of $507^{+35}_{-50}$~fb and corresponds to an observed (expected) significance of 3.3$\sigma$ (5.3$\sigma$). The \ttH cross-section is also measured differentially in bins of the Higgs boson transverse momentum in the
simplified template cross-section framework.
Another fit is done in which \ttH and $tHqb$ are left floating freely separately, leading to signal strengths of
$\mu_{\ttH}=0.59^{+0.22}_{-0.20}$ and $\mu_{tHqb}=7.2^{+4.6}_{-4.0}$.
Finally, the $CP$ structure of the top quark-Higgs boson Yukawa coupling is probed;
the results are compatible with the SM hypothesis, and values of the \(CP\)-mixing angle of $| \alpha | > 62^\circ$ are excluded at 68\% confidence level while the expected exclusion is $| \alpha | > 43^\circ$ .


%
\section*{Acknowledgements}
%

%
%

%
%

We thank CERN for the very successful operation of the LHC and its injectors, as well as the support staff at
CERN and at our institutions worldwide without whom ATLAS could not be operated efficiently.

The crucial computing support from all WLCG partners is acknowledged gratefully, in particular from CERN, the ATLAS Tier-1 facilities at TRIUMF/SFU (Canada), NDGF (Denmark, Norway, Sweden), CC-IN2P3 (France), KIT/GridKA (Germany), INFN-CNAF (Italy), NL-T1 (Netherlands), PIC (Spain), RAL (UK) and BNL (USA), the Tier-2 facilities worldwide and large non-WLCG resource providers. Major contributors of computing resources are listed in Ref.~\cite{ATL-SOFT-PUB-2025-001}.

We gratefully acknowledge the support of ANPCyT, Argentina; YerPhI, Armenia; ARC, Australia; BMWFW and FWF, Austria; ANAS, Azerbaijan; CNPq and FAPESP, Brazil; NSERC, NRC and CFI, Canada; CERN; ANID, Chile; CAS, MOST and NSFC, China; Minciencias, Colombia; MEYS CR, Czech Republic; DNRF and DNSRC, Denmark; IN2P3-CNRS and CEA-DRF/IRFU, France; SRNSFG, Georgia; BMFTR, HGF and MPG, Germany; GSRI, Greece; RGC and Hong Kong SAR, China; ICHEP and Academy of Sciences and Humanities, Israel; INFN, Italy; MEXT and JSPS, Japan; CNRST, Morocco; NWO, Netherlands; RCN, Norway; MNiSW, Poland; FCT, Portugal; MNE/IFA, Romania; MSTDI, Serbia; MSSR, Slovakia; ARIS and MVZI, Slovenia; DSI/NRF, South Africa; MICIU/AEI, Spain; SRC and Wallenberg Foundation, Sweden; SERI, SNSF and Cantons of Bern and Geneva, Switzerland; NSTC, Taipei; TENMAK, T\"urkiye; STFC/UKRI, United Kingdom; DOE and NSF, United States of America.

Individual groups and members have received support from BCKDF, CANARIE, CRC and DRAC, Canada; CERN-CZ, FORTE and PRIMUS, Czech Republic; COST, ERC, ERDF, Horizon 2020 and Marie Sk{\l}odowska-Curie Actions, European Union; Investissements d'Avenir Labex, Investissements d'Avenir Idex and ANR, France; DFG and AvH Foundation, Germany; Herakleitos, Thales and Aristeia programmes co-financed by EU-ESF and the Greek NSRF, Greece; BSF-NSF and MINERVA, Israel; NCN and NAWA, Poland; La Caixa Banking Foundation, CERCA Programme Generalitat de Catalunya and PROMETEO and GenT Programmes Generalitat Valenciana, Spain; G\"{o}ran Gustafssons Stiftelse, Sweden; The Royal Society and Leverhulme Trust, United Kingdom.

In addition, individual members wish to acknowledge support from Chile: Agencia Nacional de Investigaci\'on y Desarrollo (FONDECYT 1230812, FONDECYT 1240864, Fondecyt 3240661, Fondecyt Regular 1240721); China: Chinese Ministry of Science and Technology (MOST-2023YFA1605700, MOST-2023YFA1609300), National Natural Science Foundation of China (NSFC - 12175119, NSFC 12275265); Czech Republic: Czech Science Foundation (GACR - 24-11373S), Ministry of Education Youth and Sports (ERC-CZ-LL2327, FORTE CZ.02.01.01/00/22\_008/0004632), PRIMUS Research Programme (PRIMUS/21/SCI/017); EU: H2020 European Research Council (ERC - 101002463); European Union: European Research Council (BARD No. 101116429, ERC - 948254, ERC 101089007), European Regional Development Fund (SMASH COFUND 101081355, SLO ERDF), European Union, Future Artificial Intelligence Research (FAIR-NextGenerationEU PE00000013); France: Agence Nationale de la Recherche (ANR-21-CE31-0022, ANR-22-EDIR-0002, ANR-24-CE31-0504-01); Germany: Deutsche Forschungsgemeinschaft (DFG - 469666862, DFG - CR 312/5-2); China: Research Grants Council (GRF); Italy: Ministero dell'Università e della Ricerca (NextGenEU 153D23001490006 M4C2.1.1, NextGenEU I53D23000820006 M4C2.1.1, NextGenEU I53D23001490006 M4C2.1.1, SOE2024\_0000023); Japan: Japan Society for the Promotion of Science (JSPS KAKENHI JP22H01227, JSPS KAKENHI JP22H04944, JSPS KAKENHI JP22KK0227, JSPS KAKENHI JP24K23939, JSPS KAKENHI JP24KK0251, JSPS KAKENHI JP25H00650, JSPS KAKENHI JP25H01291, JSPS KAKENHI JP25K01023); Norway: Research Council of Norway (RCN-314472); Poland: Ministry of Science and Higher Education (IDUB AGH, POB8, D4 no 9722), Polish National Science Centre (NCN 2021/42/E/ST2/00350, NCN OPUS 2023/51/B/ST2/02507, NCN UMO-2019/34/E/ST2/00393, UMO-2022/47/O/ST2/00148, UMO-2023/49/B/ST2/04085, UMO-2023/51/B/ST2/00920, UMO-2024/53/N/ST2/00869); Portugal: Foundation for Science and Technology (FCT); Spain: Ministry of Science and Innovation (RYC2019-028510-I, RYC2020-030254-I, RYC2021-031273-I, RYC2022-038164-I); Sweden: Carl Trygger Foundation (Carl Trygger Foundation CTS 22:2312), Swedish Research Council (Swedish Research Council 2023-04654, VR 2021-03651, VR 2022-03845, VR 2022-04683, VR 2023-03403, VR 2024-05451), Knut and Alice Wallenberg Foundation (KAW 2018.0458, KAW 2022.0358, KAW 2023.0366); Switzerland: Swiss National Science Foundation (SNSF - PCEFP2\_194658); United Kingdom: The Binks Trust, Royal Society (NIF-R1-231091); United States of America: U.S. Department of Energy (ECA DE-AC02-76SF00515), Neubauer Family Foundation.

%
%


%
%
%
%
%
%
%
%
%

%
%
%
\printbibliography
\clearpage
\input{atlas_authlist}
%

%
%

%

%
%
%
%

\end{document}

%% file: atlas_authlist.tex

\begin{flushleft} 
\hypersetup{urlcolor=black} 
{\Large The ATLAS Collaboration}

\bigskip

\AtlasOrcid[0000-0002-6665-4934]{G.~Aad}$^\textrm{\scriptsize 103}$,
\AtlasOrcid[0000-0001-7616-1554]{E.~Aakvaag}$^\textrm{\scriptsize 17}$,
\AtlasOrcid[0000-0002-5888-2734]{B.~Abbott}$^\textrm{\scriptsize 122}$,
\AtlasOrcid[0000-0002-0287-5869]{S.~Abdelhameed}$^\textrm{\scriptsize 118a}$,
\AtlasOrcid[0000-0002-1002-1652]{K.~Abeling}$^\textrm{\scriptsize 55}$,
\AtlasOrcid[0000-0001-5763-2760]{N.J.~Abicht}$^\textrm{\scriptsize 49}$,
\AtlasOrcid[0000-0002-8496-9294]{S.H.~Abidi}$^\textrm{\scriptsize 30}$,
\AtlasOrcid[0009-0003-6578-220X]{M.~Aboelela}$^\textrm{\scriptsize 45}$,
\AtlasOrcid[0000-0002-9987-2292]{A.~Aboulhorma}$^\textrm{\scriptsize 36e}$,
\AtlasOrcid[0000-0001-5329-6640]{H.~Abramowicz}$^\textrm{\scriptsize 156}$
% ...
\bigskip
\\

% ...
$^{17}$Department for Physics and Technology, University of Bergen, Bergen; Norway.\\
% ...
$^{30}$Physics Department, Brookhaven National Laboratory, Upton NY; United States of America.\\
% ...
$^{36}$$^{(a)}$Facult\'e des Sciences Ain Chock, Universit\'e Hassan II de Casablanca;$^{(b)}$Facult\'{e} des Sciences, Universit\'{e} Ibn-Tofail, K\'{e}nitra;$^{(c)}$Facult\'e des Sciences Semlalia, Universit\'e Cadi Ayyad, LPHEA-Marrakech;$^{(d)}$LPMR, Facult\'e des Sciences, Universit\'e Mohamed Premier, Oujda;$^{(e)}$Facult\'e des sciences, Universit\'e Mohammed V, Rabat;$^{(f)}$Institute of Applied Physics, Mohammed VI Polytechnic University, Ben Guerir; Morocco.\\
% ...
$^{45}$Physics Department, Southern Methodist University, Dallas TX; United States of America.\\
% ...
$^{49}$Fakult\"{a}t Physik , Technische Universit{\"a}t Dortmund, Dortmund; Germany.\\
% ...
$^{55}$II. Physikalisches Institut, Georg-August-Universit\"{a}t G\"ottingen, G\"ottingen; Germany.\\
% ...
$^{103}$CPPM, Aix-Marseille Universit\'e, CNRS/IN2P3, Marseille; France.\\
% ...
$^{118}$$^{(a)}$New York University Abu Dhabi, Abu Dhabi;$^{(b)}$United Arab Emirates University, Al Ain; United Arab Emirates.\\
% ...
$^{122}$Homer L. Dodge Department of Physics and Astronomy, University of Oklahoma, Norman OK; United States of America.\\
% ...
$^{156}$Raymond and Beverly Sackler School of Physics and Astronomy, Tel Aviv University, Tel Aviv; Israel.\\
% ...
$^{a}$ Also at Affiliated with an institute formerly covered by a cooperation agreement with CERN.\\
$^{b}$ Also at An-Najah National University, Nablus; Palestine.\\
$^{c}$ Also at Borough of Manhattan Community College, City University of New York, New York NY; United States of America.\\
$^{d}$ Also at Center for Interdisciplinary Research and Innovation (CIRI-AUTH), Thessaloniki; Greece.\\
$^{e}$ Also at Centre of Physics of the Universities of Minho and Porto (CF-UM-UP); Portugal.\\
$^{f}$ Also at CERN, Geneva; Switzerland.\\
% ...

\end{flushleft}
